\pdfoutput=1 

\documentclass[a4paper,12pt]{article}
\usepackage{typearea}

\newcommand{\LTadd}[1]{}
\newcommand{\LTskip}[1]{#1}

\usepackage[T1]{fontenc}
\usepackage[utf8]{inputenc}
\usepackage[ngerman,italian,main=english]{babel}
\usepackage{csquotes}
\usepackage{microtype}

\usepackage{mathtools}
\mathtoolsset{centercolon,mathic}
\allowdisplaybreaks[2] 
\usepackage{amssymb}
\usepackage{amsthm}
\usepackage{derivative}
\usepackage{enumitem}

\newcommand{\statement}[1]{\paragraph{#1}\pdfbookmark[1]{#1}{#1}} 

\renewcommand{\Delta}{\varDelta}

\renewcommand{\Phi}{\varPhi}

\renewcommand{\Psi}{\varPsi}

\renewcommand{\Lambda}{\varLambda}

\renewcommand{\Gamma}{\varGamma}

\renewcommand{\Omega}{\varOmega}

\DeclarePairedDelimiter{\intervaloo}{\lparen}{\rparen}
\DeclarePairedDelimiter{\intervaloc}{\lparen}{\rbrack}
\DeclarePairedDelimiter{\intervalco}{\lbrack}{\rparen}
\DeclarePairedDelimiter{\intervalcc}{\lbrack}{\rbrack}

\DeclarePairedDelimiter{\abs}{\lvert}{\rvert}
\DeclarePairedDelimiter{\paren}{\lparen}{\rparen}

\DeclarePairedDelimiter{\norm}{\lVert}{\rVert}

    \newcommand{\VERT}[1]{#1|\mkern-1.5mu#1|\mkern-1.5mu#1|}

    \ExplSyntaxOn
    \makeatletter

    \MT_delim_default_inner_wrappers:n{\tnorm}

    \NewDocumentCommand{\tnorm}{ s o m }{
        \IfBooleanTF{#1}{
        \MT_delim_tnorm_star_wrapper:nnn%
            {\VERT{\bgroup\left}}{#3}{\VERT{\aftergroup\egroup\right}}
        }{
            \IfValueTF{#2}{
                \@nameuse{MT_delim_tnorm_nostarscaled_wrapper:nnn}%
                    {\VERT{\@nameuse {\MH_cs_to_str:N #2 l}}}
                    {#3}
                    {\VERT{\@nameuse {\MH_cs_to_str:N #2 r}}}
            }{
                \MT_delim_tnorm_nostarnonscaled_wrapper:nnn%
                    {\VERT{}}
                    {#3}
                    {\VERT{}}
            }
        }
    }

    \makeatother
    \ExplSyntaxOff

\DeclarePairedDelimiter{\commutator}{\lbrack}{\rbrack}
\DeclarePairedDelimiter{\anticommutator}{\{}{\}}
\DeclarePairedDelimiter{\List}{\{}{\}}

\DeclarePairedDelimiter{\floor}{\lfloor}{\rfloor}

\DeclarePairedDelimiter{\braket}{\langle}{\rangle}
\DeclarePairedDelimiter{\bra}{\langle}{\rvert}
\DeclarePairedDelimiter{\ket}{\lvert}{\rangle}
\DeclarePairedDelimiterXPP{\dist}[1]{d}{\lparen}{\rparen}{}{#1}
\DeclarePairedDelimiterXPP{\diam}[1]{\SYMdiam}{\lparen}{\rparen}{}{#1}

\makeatletter
\let\bBigg@@\bBigg@
\renewcommand{\bBigg@}[2]{{%
  \mathchoice
    {\bBigg@@{#1}{#2}}%
    {\bBigg@@{#1}{#2}}%
    {\big@size=.5\big@size\bBigg@@{#1}{#2}}%
    {\big@size=.3\big@size\bBigg@@{#1}{#2}}}}%
\makeatother

\DeclareDocumentCommand{\trace}{s o e{_} m}{
    \IfValueTF{#3}
        {\Tr_{#3}}
        {\Tr}%
    \IfBooleanTF{#1}
        {\paren*{#4}}
        {
            \IfValueTF{#2}
                {\paren[#2]{#4}}
                {\paren{#4}}%
        }%
}
\DeclareDocumentCommand{\Trace}{s o e{_} m}{
    \IfValueTF{#3}
        {\Tr_{#3}}
        {\Tr}%
    \IfBooleanTF{#1}
        {\paren*{#4}}
        {
            \IfValueTF{#2}
                {\paren[#2]{#4}}
                {\paren{#4}}%
        }%
}

\providecommand\given{}

\newcommand\SetSymbol[1][]{%
    \nonscript\,#1\vert
    \allowbreak
    \nonscript\,
    \mathopen{}}
\DeclarePairedDelimiterX\Set[1]\{\}{%
    \renewcommand\given{%
        \SetSymbol[\delimsize]}
    \nonscript\,
    #1
    \nonscript\,
}

\usepackage[svgnames]{xcolor}

\usepackage[
    colorlinks,
    allcolors=col_1,
    bookmarksnumbered,
    bookmarksopen,
    bookmarksopenlevel=2,
    bookmarksdepth=3,
    pdfusetitle=true,
    breaklinks=true,
]{hyperref}

\usepackage{caption}

\makeatletter
\newcommand*{\diff}{\@ifnextchar^{\DIfF}{\DIfF^{}}}
\def\DIfF^#1{%
    \mathop{\mathrm{\mathstrut d}}\nolimits^{#1}\gobblespace}
\def\gobblespace{\futurelet\diffarg\opspace}
\def\opspace{%
    \let\DiffSpace\!%
    \ifx\diffarg(%
        \let\DiffSpace\relax
    \else
        \ifx\diffarg[%
            \let\DiffSpace\relax
        \else
            \ifx\diffarg\{%
                \let\DiffSpace\relax
            \fi
        \fi
    \fi
    \DiffSpace
} 
\makeatother

\newcommand{\sumstack}[2][]{\ifstrempty{#1}{\sum_{\substack{#2}}}{\smashoperator[#1]{\sum_{\substack{#2}}}}}
\LTadd{\renewcommand{\sumstack}[2][]{\sum_{\substack{#2}}}}

\def\epsi{\varepsilon}

\newcommand{\e}{{\mathrm{e}}}
\newcommand{\I}{\mathrm{i}}
\newcommand{\R}{\mathbb{R}}
\newcommand{\C}{\mathbb{C}}
\newcommand{\N}{\mathbb{N}}
\newcommand{\Z}{\mathbb{Z}}
\newcommand{\Hi}{{\mathcal{H}}}
\newcommand{\alg}{\mathcal{A}}

\newcommand{\unit}{\mathbf{1}}
\newcommand{\cexp}{\mathbb{E}}

\newcommand{\sr}[1]{_{\exp(-#1\,\cdot)}}

\newcommand{\ebeta}[2][-\beta]{\e^{#1#2}}
\newcommand{\etat}{\tilde{\eta}}

\newcommand{\fh}{\hat{f}}

\DeclareMathOperator{\Tr}{Tr}
\newcommand{\SYMdiam}{\operatorname{diam}}
\DeclareMathOperator{\Cov}{Cov}
\DeclareMathOperator{\supp}{supp}

\newcommand{\QBP}{\mathup{QBP}}
\newcommand{\LR}{\mathup{LR}}
\newcommand{\LI}{\mathup{LI}}
\newcommand{\LPPL}{\mathup{LPPL}}
\newcommand{\DC}{\mathup{Cov}}
\newcommand{\interaction}{\mathup{int}}

\newcommand{\quadtext}[1]{\quad\text{#1}\quad}
\newcommand{\qquadtext}[1]{\quad\quadtext{#1}\quad}

\newcommand{\Alignindent}{\hspace*{2em}&\hspace*{-2em}}

\newcommand{\mathup}[1]{\mathrm{#1}}

\newcommand{\tstrut}{\vphantom{\widetilde{X}}}

\let\oldsetminus\setminus
\newbox\mybox
\newcommand\cutsetminus[1]{%
    \setbox\mybox\hbox{\(#1\oldsetminus\)}%
    \ht\mybox=0pt%
    \dp\mybox=0pt%
    \usebox\mybox%
}
\renewcommand\setminus{%
    \mathbin{%
        \mathchoice%
            {\displaystyle\oldsetminus}
            {\textstyle\oldsetminus}
            {\cutsetminus{\scriptstyle}}
            {\cutsetminus{\scriptscriptstyle}}
    }%
}

\setlist[enumerate]{label=\textup{(\alph*)}}
\newlist{theoremenumerate}{enumerate}{1}
\setlist[theoremenumerate]{label=\textup{(\alph*)},ref=\thetheorem\,(\alph*)}

\usepackage{tikz}
\pgfdeclarelayer{background}
\pgfsetlayers{background,main}
\newlength{\ul}
\newlength{\lw}
\newlength{\fattening}
\tikzset{%
    myLine/.style = {
        line width = #1\lw,
    },
    myLine/.default = 1,
    x = \ul,
    y = \ul,
    text = black,
    draw = black,
    shorten/.style = {
        shorten < = #1,
        shorten > = #1
    },
    shorten/.default = 2\lw,
    on layer/.code = {
        \pgfonlayer{#1}\begingroup
        \aftergroup\endpgfonlayer
        \aftergroup\endgroup
    },
    on background/.style = {
        preaction = {
            #1,
            on layer = background,
        },
    },
    only inside/.style = {
        preaction = {
            clip,
            postaction = {
                line width = 2\lw,
                #1
            },
        },
    },
    only inside/.default = {draw},
}
\definecolorset{HTML}{col_}{}{%
    1,0072B2;%
    2,D55E00;%
    3,E69F00
}
\definecolor{anthracite}{HTML}{293133}
\definecolor{gray}{HTML}{6a7479}

\usetikzlibrary{
    arrows.meta,
    calc,
    decorations.pathreplacing,
    shadings,
}

\tikzset{%
    every node/.style = {
        node font = \sffamily,
        inner sep = 4\lw,
    },
    concept/.style = {
        rectangle,
        rounded corners = 5\lw,
        align = center,
        draw = #1,
        fill = #1!45,
        inner sep = 5\lw,
        minimum size = 6mm,
    },
    concept/.default = col_1,
    formula/.style = {
        rectangle,
        rounded corners = 5\lw,
        align = center,
        draw = #1,
        fill = #1!25,
        inner sep = 3\lw,
        minimum size = 6mm,
        font = \footnotesize,
    },
    formula/.default = col_1,
    imply/.style = {
        arrows = {-Implies},
        shorten = 3,
        double equal sign distance,
    },
    brc/.style = {
        draw = #1,
        decoration = {
            brace,
            amplitude = 5\lw,
            mirror,
            raise = 2\lw,
            pre = moveto,
            pre length = 2\lw,
            post = moveto,
            post length = 2\lw
        },
        decorate,
        every node/.append style = {
            below = 6\lw,
            midway,
            color = #1,
        }
    },
    brc/.default = black,
}

\ExplSyntaxOn
\int_new:N \g_block_start_int
\tl_new:N \l_block_label

\NewDocumentCommand \Block { O{} m O{0pt} m m o}{

    \ifstrempty{#1}{}{
        \int_set:Nn \g_block_start_int {#1}
    }

    \IfNoValueTF{#6}{
        \tl_set:Nn \l_block_label {\(#5\)}
    }{
        \tl_set:Nn \l_block_label {#6}
    }

    \node[fill=#4!80!white, rectangle, minimum~width=#2\ul, minimum~height=\ul+#3,anchor=west, outer~sep=0pt] (#5) at (\int_use:N \g_block_start_int -0.5,0) {}; 

    \IfBlankF{#6}{
        \node[anchor=north, text=#4] at (#5.south) {\tl_use:N \l_block_label};
    }

    \foreach \n in {1,...,#2}{ 
        \shade[shading=ball, ball~color=anthracite] ({\n-1+\int_use:N \g_block_start_int},0) circle (.3\ul);
    }

    \int_add:Nn \g_block_start_int {#2}
}
\ExplSyntaxOff

\usepackage[
    bibstyle=phys,
    citestyle=numeric,
    biblabel=brackets,
    mincitenames=1,
    maxcitenames=3,
    minalphanames=3,
    maxalphanames=4,
    minbibnames=3,
    maxbibnames=10,
    arxiv=abs,
    eprint=true,
    doi=false,
    url=false,
    sorting=nty,
    useprefix=true,
    alldates=year,
    urldate=long,
]{biblatex}
\renewbibmacro*{note+pages}{%
    \printfield{note}%
    \setunit{\bibpagespunct}%
    \printfield{pages}%
    \newunit%
}
\makeatletter
\DeclareFieldFormat{eprint:arxiv}{%
    \ifhyperref
    {\href{https://arxiv.org/\abx@arxivpath/#1}{%
            arXiv\addcolon\linebreak[2]#1}}
    {arXiv\addcolon
        \nolinkurl{#1}%
    }
}
\makeatother

\renewbibmacro*{maintitle+title}{%
    \iffieldsequal{maintitle}{title}{%
        \clearfield{maintitle}%
        \clearfield{mainsubtitle}%
        \clearfield{maintitleaddon}%
    }{%
        \iffieldundef{maintitle}{}{%
            \printtext[doi/url-link]{%
                \usebibmacro{maintitle}%
            }%
            \newunit\newblock
            \iffieldundef{volume}{}{%
                \printfield{volume}%
                \printfield{part}%
                \setunit{\addcolon\space}%
            }%
        }%
    }%
    \printtext[doi/url-link]{%
        \usebibmacro{title}%
    }%
    \newunit%
}

\DeclareBibliographyDriver{article}{%
    \usebibmacro{bibindex}%
    \usebibmacro{begentry}%
    \usebibmacro{author/translator+others}%
    \setunit{\labelnamepunct}\newblock
    \usebibmacro{title}%
    \newunit
    \printlist{language}%
    \newunit\newblock
    \usebibmacro{byauthor}%
    \newunit\newblock
    \usebibmacro{bytranslator+others}%
    \newunit\newblock
    \printfield{version}%
    \newunit\newblock
    \printtext[doi/url-link]{%
        \usebibmacro{journal+issuetitle}%
        \newunit
        \usebibmacro{byeditor+others}%
        \newunit
        \usebibmacro{note+pages}%
        \newunit\newblock
        \iftoggle{bbx:isbn}
        {\printfield{issn}}
        {}%
        \setunit{\addspace}%
    }%
    \printfield{year}%
    \setunit{\addspace}%
    \usebibmacro{doi+eprint+url}%
    \newunit\newblock
    \usebibmacro{addendum+pubstate}%
    \setunit{\bibpagerefpunct}\newblock
    \usebibmacro{pageref}%
    \newunit\newblock
    \iftoggle{bbx:related}
        {\usebibmacro{related:init}%
        \usebibmacro{related}}
        {}%
    \usebibmacro{finentry}%
}

\usepackage{xurl} 

\usepackage{orcidlink}

\newcommand{\rrr}{r}
\newcommand{\RRR}{R}
\newcommand{\llll}{\ell}

\theoremstyle{plain}
\newtheorem{theorem}{Theorem}
\newtheorem{proposition}[theorem]{Proposition}
\newtheorem{corollary}[theorem]{Corollary}
\newtheorem*{Icorollary}{Main Result (informal)}
\newtheorem{lemma}[theorem]{Lemma}

\theoremstyle{definition}
\newtheorem{definition}[theorem]{Definition}
\newtheorem{conjecture}[theorem]{Conjecture}

\theoremstyle{remark}
\newtheorem{remark}[theorem]{Remark}

\makeatletter
\AtBeginEnvironment{definition}{\pushQED{\qed}}
\AtEndEnvironment{definition}{\popQED\@endpefalse}
\AtBeginEnvironment{remark}{\pushQED{\qed}}
\AtEndEnvironment{remark}{\popQED\@endpefalse}
\makeatother

\makeatletter
\def\blindfootnote{\gdef\@thefnmark{}\@footnotetext}
\makeatother

\title{From decay of correlations\\ to locality and stability of the Gibbs state}

\author{
    Ángela Capel%
    \texorpdfstring{%
        \,\orcidlink{0000-0001-6713-6760}
        \footnote{
            \parbox[t]{.75\textwidth}{
                \foreignlanguage{ngerman}{Fachbereich Mathematik, Universität~Tübingen,\\
                Auf~der~Morgenstelle~10, 72076~Tübingen,} Germany
            }
        }
        \footnote{
            \parbox[t]{.85\textwidth}{
                \foreignlanguage{english}{Department of Applied Mathematics and Theoretical Physics, University of Cambridge\\
                Wilberforce Road, Cambridge CB3 0WA,} United Kingdom
            }
        }
    }{}%
    \and Massimo Moscolari%
    \texorpdfstring{%
        \,\orcidlink{0000-0001-7574-1055}
        \footnote{
            \parbox[t]{.75\textwidth}{
                \foreignlanguage{italian}{Dipartimento di Matematica, Politecnico di Milano,\\
                Piazza Leonardo da Vinci~32, 20133~Milano,} Italy
            }
        }
    }{}%
    \and Stefan Teufel%
    \texorpdfstring{%
        \,\orcidlink{0000-0003-3296-4261}
        \footnotemark[1]
    }{}%
    \and Tom Wessel%
    \texorpdfstring{%
        \,\orcidlink{0000-0001-7593-0913}
        \footnotemark[1]
    }{}%
}

\date{April 2025}

\begin{document}
\bgroup

\setlength{\footnotesep}{\dimexpr1.2\baselineskip-\dp\strutbox}

\maketitle
\thispagestyle{empty}

\blindfootnote{
    Email:\quad%
    \parbox[t]{.7\textwidth}{
        \hypersetup{hidelinks}
        \href{mailto:angela.capel@uni-tuebingen.de}{angela.capel@uni-tuebingen.de},
        \href{mailto:massimo.moscolari@polimi.it}{massimo.moscolari@polimi.it},
        \newline
        \href{mailto:stefan.teufel@uni-tuebingen.de}{stefan.teufel@uni-tuebingen.de},
        \href{mailto:tom.wessel@uni-tuebingen.de}{tom.wessel@uni-tuebingen.de}
    }
}

\egroup

\begin{abstract}
    We show that whenever the Gibbs state of a quantum spin system satisfies decay of correlations, then it is stable, in the sense that local perturbations affect the Gibbs state only locally, and it satisfies local indistinguishability, i.e.\ it exhibits local insensitivity to system size.
    These implications hold in any dimension, require only locality of the Hamiltonian, and are based on Lieb-Robinson bounds and on a detailed analysis of the locality properties of the quantum belief propagation for Gibbs states.

    To demonstrate the versatility of our approach, we explicitly apply our results to several physically relevant models in which the decay of correlations is either known to hold or is proved by us.
    These include Gibbs states of one-dimensional spin chains with polynomially decaying interactions at any temperature, and high-temperature Gibbs states of quantum spin systems with finite-range interactions in any dimension.
    We also prove exponential decay of correlations above a threshold temperature for Gibbs states of one-dimensional finite spin chains with translation-invariant and exponentially decaying interactions, and then apply our general results.

\end{abstract}

\clearpage
\tableofcontents

\section{Introduction}

One of the characteristic features of quantum many-body systems is the locality of their interactions.
In the last decades, this property has been largely exploited in the characterization of their thermal (Gibbs) states and their ground states.
In this work we are interested in Gibbs states and, in particular, in their locality and in their stability against local perturbations.
Understanding how the locality of Hamiltonians of many-body systems translates to the locality of the Gibbs states is of crucial importance for the preparation and simulation of quantum states~\cite{TOVP2011,BS2017,CKBG2023}.

There are several ways to describe how local a quantum state is.
Here we focus on three main concepts: \emph{local indistinguishability}, the principle that \emph{local perturbations perturb locally (LPPL)} and the well-known \emph{decay of correlations}.
The main purpose of this work is to show how these three locality properties are related to each other and to show that they are all equivalent under certain conditions.
Let us first describe these concepts informally.

Consider an interacting quantum spin system defined on \(\Lambda \Subset \Z^\nu\), \(\nu \in \N\), with a Hamiltonian~\(H_\Lambda\) that is a sum of local terms.
Let~\(\rho_\beta^\Lambda\) denote the Gibbs state of~\(H_\Lambda\) at inverse temperature~\(\beta > 0\),
\begin{equation*}
    \rho_\beta^{\Lambda}
    :=
    \e^{-\beta H_{\Lambda}} \big/ \Trace[\big]_{\Lambda}{\e^{-\beta H_{\Lambda}}}
    .
\end{equation*}
For every \(\Lambda'\subset \Lambda\), we define the local Gibbs state~\(\rho_\beta^{\Lambda'}\) to be the Gibbs state of the Hamiltonian consisting only of terms supported on~\(\Lambda'\).
Then, let~\(A\) be an observable supported in an inner region \(X \subset \Lambda' \subset \Lambda\): local indistinguishability of the Gibbs state (see Definition~\ref{def:local-indistinguishability}) is a quantitative version of the statement that
\begin{equation*}
    \trace[\big]_\Lambda{\rho_\beta^\Lambda \, A}
    \approx
    \trace[\big]_{\Lambda'}{\rho_\beta^{\mathrlap{\Lambda'}\phantom{\Lambda}} \, A} .
\end{equation*}
The concept of local indistinguishability~\cite{BK2019,BCP2022} goes also under the name of locality of temperature~\cite{KGK2014}.
Roughly speaking, local indistinguishability says that local observables cannot distinguish between the Gibbs state of the full Hamiltonian and the Gibbs state of its truncated version.
On the other hand, the concept of LPPL, originally introduced in the context of gapped ground states~\cite{BMN2011,BRD2021,RS2015,HTW2021}, is concerned  with localized perturbations of a quantum system.
Let~\(V\) be a generic perturbation supported on a region \(Y \subset \Lambda\) and let \(\tilde H_\Lambda:=H_\Lambda + V\).
We denote by \(\tilde{\rho}_\beta^\Lambda\) the Gibbs state associated to~\(\tilde H_\Lambda\).
Let~\(A\) be an observable supported on a set~\(X\) \emph{far} from~\(Y\)\!, then LPPL (see Definition~\ref{def:LPPL}) is a quantitative version of the statement that
\begin{equation*}
    \trace*_\Lambda{\rho_\beta^\Lambda \, A}
    \approx
    \trace*_{\Lambda}{\tilde{\rho}_\beta^\Lambda \, A},
\end{equation*}
which means that the expectation values of local observables supported far from the perturbation are not influenced by the presence of the perturbation.
The notion of LPPL thus concerns the Gibbs states of two different Hamiltonians, the perturbed and unperturbed one.

A condition similar to local indistinguishability named \emph{Local Topological Quantum Order (LTQO)} has been previously considered in the literature~\cite{BHM2010,BH2011,MZ2013} in the context of ground states of topologically ordered Hamiltonians.
For frustration-free Hamiltonians, LTQO implies stability of the spectral gap.
In a sense, it states that the effect of boundary conditions is exponentially suppressed in the bulk.
And while generally difficult to prove, LTQO is known to hold for one-dimensional matrix product states with normal tensors~\cite{CMPS2013} and for some projected entangled pair states with commuting parent Hamiltonians~\cite{SCP2010} in two dimensions.

Another standard way to measure the locality of a quantum state~\(\rho\) is in terms of the decay of correlations in such a state.
This amounts to study the behaviour of the covariance~\cite{KGK2014,BK2019,Araki1969,FU2015, PP2023,BCP2022}
\begin{equation*}
    \Cov_{\rho}(X;Y)
    :=
    \sup_{\substack{\mathstrut A\in \alg_X\colon \norm{A}=1, \\[\lw] B\in \alg_Y\colon \norm{B}=1\phantom{,}}}
    \abs[\big]{
        \trace*_{\Lambda}{\rho \, A \, B}
        - \trace*_{\Lambda}{\rho \, A} \, \trace*_{\Lambda} {\rho \, B}
    }
\end{equation*}
with respect to the distance~\(\dist{X,Y}\) between the support of the observables.

Our main result shows that decay of correlations of the Gibbs state (see Definition~\ref{def:decay-of-correlations}) implies the stability of the Gibbs state against local perturbation, namely LPPL\@.
The main ingredient in our proof is the so-called quantum belief propagation (QBP) introduced by Hastings~\cite{Hastings2007}, and with a recent increase of attention due to its various applications, see e.g.~\cite{Kim2012,Kim2013,EO2019,BK2019,KB2019,HMS2020,AAKS2021,KAA2021,ORFW2023}.
We show that quantum belief propagation, together with the well-known Lieb-Robinson bounds~\cite{LR1972,NSY2017,NSY2019}, allows exploiting the local structure of the Hamiltonian in order to prove LPPL\@.

Then, by using QBP again, we show that uniform decay of correlations implies uniform stability to local perturbations (\emph{uniform LPPL}), which in turn implies local indistinguishability of the Gibbs state.
Uniformity here means that a property does not only hold for a given Gibbs state or Hamiltonian, but also for the Gibbs states of the restrictions of the Hamiltonian to smaller domains.

Finally, to close the circle, we also show that local indistinguishability implies uniform decay of correlations.
This is well-known in the case of finite-range Hamiltonians, but the proof for arbitrary local Hamiltonians requires more care and is again based on QBP\@.

Combining these three implications, our main result is that \emph{local indistinguishability}, \emph{ uniform LPPL} and \emph{uniform decay of correlations} are actually three equivalent ways to describe the locality properties of Gibbs states.
See also Figure~\ref{fig:diagram-introduction}.

\begin{Icorollary}
    Let \(H\) be the Hamiltonian of a finite interacting quantum spin system and~\(\beta>0\).
    Then the following three properties are equivalent:
    \begin{enumerate}
        \item\label{item:main-result-intro-DC}
            The Gibbs state at inverse temperature \(\beta\) satisfies uniform decay of correlations (Definition~\ref{def:decay-of-correlations}).
        \item\label{item:main-result-intro-LPPL}
            The Gibbs state at inverse temperature \(\beta\) satisfies uniform LPPL, i.e.\ the expectation values of observables are stable against local perturbations of the Hamiltonian (Definition~\ref{def:LPPL}).
        \item\label{item:main-result-intro-LI}
            The Gibbs state at inverse temperature \(\beta\) satisfies local indistinguishability (Definition~\ref{def:local-indistinguishability}).
    \end{enumerate}
\end{Icorollary}

In the main text, the three implications {\hypersetup{hidelinks}\(\ref{item:main-result-intro-DC}\Rightarrow\ref{item:main-result-intro-LPPL}\), \(\ref{item:main-result-intro-LPPL}\Rightarrow\ref{item:main-result-intro-LI}\) and \(\ref{item:main-result-intro-LI}\Rightarrow\ref{item:main-result-intro-DC}\)} are split into the three Theorems~\ref{thm:DC0toLPPL}, \ref{thm:uniform-lppl-implies-local-indistinguishability} and~\ref{thm:local-indistinguishability-implies-decay-of-correlations}, respectively, and refined results for one-dimensional spin chains can be found in Theorems~\ref{thm:LPPL-1D}, \ref{thm:LI-1D} and~\ref{thm:local-indistinguishability-implies-decay-of-correlations_1D}.

We emphasize that we present a rigorous framework that allows relating LPPL and local indistinguishability to decay of correlations for very general interactions that have a finite interaction norm of the form~\eqref{eq:norm_interactions}.
In this way, we are able to treat finite-range, short-range, and long-range interactions in a unified way.
To guide the reader through our results, we always show, along with the general statements, the implications for short-range interactions as an immediate and demonstrative example.
The core message of our work is that, whenever there are results on the decay of correlations for the Gibbs state, one can immediately obtain quantitative versions of LPPL and local indistinguishability by applying our theorems.

To illustrate this idea, in Section~\ref{sec:applications-of-the-general-results} we collect some physically relevant applications of our general results and prove LPPL and local indistinguishability for systems for which these properties have not been known before.
First, we focus on one-dimensional quantum spin chains: assuming translation invariance, (uniform) exponential decay of correlations is known to hold at every temperature if the interactions are also finite-range~\cite{Araki1969,BCP2022}, and above a critical temperature in the short-range case~\cite{PP2023} for the infinite-chain.
We first extend the latter result to finite chains, and then, by applying our framework, we show LPPL and local indistinguishability.
Quite recently it has been shown~\cite{KK2024} that decay of correlations for short-range interactions holds at any positive temperature if one relaxes the decay to stretched exponential.
With that input, we directly obtain LPPL and local indistinguishability with a stretched exponential decay rate for such systems.
Furthermore, we also apply our results to spin chains with long-range interactions, where polynomial decay of correlations has recently been shown~\cite{KK2024} at any positive temperature.
Finally, in any dimension, for quantum spin systems with finite-range interactions at high enough temperature, it is known that Gibbs states satisfy (uniform) exponential decay of correlation, as proven in~\cite{KGK2014}.
Thus, as a byproduct of our results, we also recover the known results~\cite{KGK2014} of uniform LPPL and local indistinguishability, both with exponential decay.
Compared to the proof in~\cite{KGK2014} our methods have the advantage of relying only on decay of the covariance of the unperturbed Gibbs state, rather than on decay of the generalized covariance of the perturbed Gibbs states, an object about which much less is known.

On the side, we use QBP to recover a known result saying that the Gibbs state is stable (in trace norm) against small (in norm) perturbations.
Moreover, we show that local expectation values are stable against perturbations by sums of small local terms in the underlying Hamiltonian, assuming some algebraic decay of correlations.
A similar result was recently indicated in~\cite{RGK2023}.
Furthermore, we note that in the recent work~\cite{TFC2024}, the stability against sum of local terms perturbations has been analysed in connection to quantum simulations and quantum advantage.
Related and in some respects stronger results for small perturbations of classical systems at low temperature were obtained in~\cite{DFF1996,DFF1996a}.

\subsection{Organization of the paper}

In Section~\ref{sec:mathematical-setting} we introduce the mathematical framework and provide a precise definition of decay of correlations, LPPL and local indistinguishability.
After that, in Section~\ref{sec:applications-of-the-general-results} we conclude the introductory sections by showing several applications of our main results to Gibbs states of one-dimensional quantum spin chains and to high-temperature Gibbs states in any dimensions.
Section~\ref{sec:quantum-belief-propagation} is devoted to the analysis of quantum belief propagation, which is the tool at the core of our proofs.
The proofs of the results of Section~\ref{sec:quantum-belief-propagation} are given in Section~\ref{sec:proof-qbp}.
Section~\ref{sec:LPPL-from-decay-of-correlations} contains the theorems, which allow concluding (uniform) LPPL from (uniform) decay of correlations.
Then, in Section~\ref{sec:LIfromULPPL} and Section~\ref{sec:clustering-from-local-indistinguishability}, we provide the equivalence shown in Figure~\ref{fig:diagram-introduction}.
In Section~\ref{sec:stabilty-against-SLT-perturbations} we show stability of local expectation values against perturbations in the underlying Hamiltonian.
Finally, Section~\ref{sec:1D-spin-chains} is dedicated to the analysis of one-dimensional spin chains.

\begin{figure}
    \centering%
    \setlength{\fboxsep}{0pt}%
    \begin{tikzpicture}[
            myLine,
            x=\ul,y=\ul,
        ]
        \begin{scope}[
                every node/.append style={concept},
                every label/.append style={label distance=-\lw,formula},
            ]
            \node[
                name=Corr,
                label={[name=Corr expl,anchor=north]below:\(
                        \Cov_{\rho_\beta^{\Lambda'}}(X;Y)
                        \leq
                        C \, \abs{X} \, \e^{-a\dist{X,Y}}
                    \)
                },
            ] at (0,0) {\footnotesize{uniform (for \(\Lambda' \subset \Lambda\))}\\[-2pt]decay of correlations};
            \node[
                name=LPPL,
                label={[name=LPPL expl]below:\(
                        \begin{aligned}
                            &
                            \abs[\big]{
                                \trace[\big]{\rho_\beta^{\Lambda'}[H]\,B}
                                -\trace[\big]{\rho_\beta^{\Lambda'}[H+V]\,B}
                            }
                            \\&\quad \leq
                            C \, \norm{B} \, \e^{c\beta\norm{V}} \, \e^{-a\dist{X,Y}}
                        \end{aligned}
                    \)
                },
            ] at (8,-4.5) {\footnotesize{uniform (for \(\Lambda' \subset \Lambda\))}\\[-2pt]local perturbations perturb locally};
            \node[
                name=Indi,
                label={[name=Indi expl]below:\(
                        \begin{aligned}
                            &
                            \abs[\big]{
                                \trace[\big]{\rho_\beta^\Lambda\,B}
                                -\trace[\big]{\rho_\beta^{\Lambda'}\,B}
                            }
                            \\&\quad \leq
                            C \, \norm{B} \, \e^{-a \dist{Y,\Lambda\setminus\Lambda'}}
                        \end{aligned}
                    \)
                },
            ] at (-8,-4.5) {local indistinguishability};
        \end{scope}
        \draw[imply] (Corr) -| node[right,anchor=north west,align=left,font=\footnotesize] {Theorem~\ref{thm:DC0toLPPL}\\ QBP and LR-bounds} (LPPL);
        \draw[imply] (LPPL) -- node[below,align=center,font=\footnotesize] {Theorem~\ref{thm:uniform-lppl-implies-local-indistinguishability}} (Indi);
        \draw[imply] (Indi) |- node[left,anchor=north east,align=right,font=\footnotesize] {Theorem~\ref{thm:local-indistinguishability-implies-decay-of-correlations}} (Corr);
    \end{tikzpicture}
    \caption{
        The diagram shows the main implications discussed in this work for short-range interactions.
        In particular, we show “equivalence” of the three concepts in the picture.
        Note, that the formulas are mainly illustrative for the concepts and in particular the constants change, see Remark~\ref{remark:decay-of-correlations-after-one-circle}.
        A crucial ingredient in all the implications is quantum belief propagation (QBP) coupled with Lieb-Robinson bounds.
        For precise statements we refer to the Theorems.
        In certain physical dimensions and temperature regimes, exponential decay of correlations is known to hold by earlier results, for which all three properties are thus satisfied.
    }
    \label{fig:diagram-introduction}
\end{figure}

\section{Mathematical setup and important concepts}
\label{sec:mathematical-setting}

Consider the regular lattice~\(\Z^\nu\), for fixed~\(\nu \in \N\), equipped with the \(\ell^1\)-metric \(d\colon \Z^\nu \times \Z^\nu \rightarrow \N\).
We denote arbitrary subsets as~\(\Lambda\subset \Z^\nu\) (including equality) and finite subsets by~\(\Lambda' \Subset \Lambda\) (again including equality if~\(\Lambda\) is finite).
The cardinality of a set \(\Lambda\Subset \Z^\nu\) is denoted by~\(\abs{\Lambda}\).
Given any two subsets \(X\), \(Y \subset \Z^\nu\) we denote by \(\dist{X,Y}\) their distance with respect to the metric~\(d\).
Likewise, we denote by \(\diam{X} := \sup_{x, y \in X} d(x,y)\) the diameter of~\(X\).
For a set \(X \subset \Lambda\) and \(\rrr>0\), we denote the \(\rrr\)-neighbourhood
\begin{equation*}
    X_\rrr:= \Set[\big]{x \in \Lambda \given \dist{x,X} \leq \rrr}
    ,
\end{equation*}
where the base set~\(\Lambda\) will be clear from the context.

With every site \(x \in \Z^\nu\) we associate a finite-dimensional local Hilbert space \(\Hi_x \equiv \C^D\) with the corresponding space of linear operators denoted by \(\alg_x:=\mathcal{B} \paren*{\C^D}\).
For each \(\Lambda \Subset \Z^\nu\) we define the Hilbert space \(\Hi_{\Lambda}:=\bigotimes_{x \in \Lambda} \Hi_x\), and denote the algebra of bounded linear operators on~\(\Hi_{\Lambda}\) by \(\alg_{\Lambda} := \mathcal{B}(\Hi_{\Lambda})\).
Due to the tensor product structure, we have \(\alg_{\Lambda}=\bigotimes_{x \in \Lambda} \mathcal{B} (\Hi_x)\).
Hence, for \(X \subset \Lambda \Subset \Z^\nu\), any \(A \in \alg_{X}\) can be viewed as an element of~\(\alg_{\Lambda}\) by identifying~\(A\) with \(A \otimes \unit_{\Lambda \backslash X} \in \alg_{\Lambda}\), where~\(\unit_{\Lambda \backslash X}\) denotes the identity in~\(\alg_{\Lambda \backslash X}\).
This identification is always understood implicitly and for \(B\in \alg_\Lambda\) we denote by~\(\supp(B)\) the smallest \(Y\subset \Lambda\) such that~\(B\in \alg_Y\).
For every \(\Lambda\subset \Z^\nu\), let
\begin{equation*}
    \alg_{\Lambda}
    :=
    \overline{\bigcup_{\Lambda'\Subset\Lambda} \alg_{\Lambda'}}^{\raisebox{-.2ex}{\kern1pt\(\scriptstyle\norm{\cdot}\)}}
\end{equation*}
be the algebra of quasi-local observables, where completion is with respect to the operator norm and is only relevant if~\(\Lambda\) is not finite.

An \emph{interaction} on \(\Lambda\subset \Z^\nu\) is a function
\begin{equation*}
    \Psi \colon \List{X\Subset\Lambda} \to \alg_{\Z^\nu},
    \quad X\mapsto \Psi (X)\in\alg_X
    \quadtext{with} \Psi(X)=\Psi(X)^*.
\end{equation*}
For each \(\Lambda' \Subset \Lambda\), the corresponding local Hamiltonian is then defined as
\begin{equation*}
    H_{\Lambda'}
    :=
    \sum_{X \subset \Lambda'} \Psi(X).
\end{equation*}

There are several types of interactions we will consider in the following.
The first are \emph{finite-range interactions}, for which there exists \(R>0\) and \(J>0\) such that \(\Psi(X) = 0\) whenever \(\diam{X}>R\) and \(\norm{\Psi(X)} \leq J\) for all~\(X\subset \Lambda\).

The other are general decaying interactions, for which we define the interaction norm
\begin{equation}
    \label{eq:norm_interactions}
    \norm{\Psi}_F
    :=
    \sup_{z\in \Lambda} \sumstack{Z\Subset\Lambda \mathpunct{:}\\z\in Z}
    \frac{\abs{Z} \, \norm{\Psi(Z)}}{F\paren[\big]{\diam{Z}}}
    ,
\end{equation}
for some decaying \(F\colon \intervalco{0,\infty} \to \intervaloo{0,\infty}\) with \(F(0) \leq 1\) and require~\(\norm{\Psi}_F<\infty\).
The extra~\(\abs{Z}\) is included in the norm to obtain simpler explicit bounds.
Along all the general statements, we will provide exemplary results for so-called \emph{short-range} or \emph{exponentially decaying} interactions with \(F(r) = \exp(-b \, r)\), for some~\(b>0\).
Other regularly used classes are \emph{stretched exponentially decaying} interactions with \(F(r) = \exp(-b \, r^p)\) for some \(p\in \intervaloo{0,1}\) and so-called \emph{long-range} or \emph{polynomially decaying} interactions with \(F(r) = (r+1)^{-\alpha}\) for some~\(\alpha>0\).
Relevant applications to all those types of interactions can be found in Section~\ref{sec:applications-of-the-general-results}.

Note that on finite~\(\Lambda\), every interaction can be seen as a finite-range interaction with range~\(R \leq \diam{\Lambda}\).
However, the range~\(R\) will enter our bounds explicitly.
Hence, even on finite lattices, it makes sense to consider general decaying interactions, for which only the interaction norm~\eqref{eq:norm_interactions} enters the bounds.

We also emphasize that the various constants, which appear in all definitions and results below, do not depend on~\(\Lambda\).
Thus, all results are uniform in \(\abs{\Lambda}\), and one can take the thermodynamic limit.

For any Hilbert space~\(\Hi\) and self-adjoint operator \(H\in \mathcal{B}(\Hi)\) we denote the Gibbs state at inverse temperature \(\beta\in \intervaloo{0,\infty}\) by
\begin{equation*}
    \rho_\beta^{\Hi}[H]
    :=
    \frac{\e^{-\beta H}}{\trace{\e^{-\beta H}}},
\end{equation*}
where~\(\Tr\) denotes the (unnormalized) trace over~\(\Hi\).
For a lattice system on \(\Lambda\subset \Z^\nu\) and \(\Lambda'\Subset \Lambda\) we abbreviate
\begin{equation*}
    \rho_\beta^{\Lambda'}[K]
    :=
    \rho_\beta^{\Hi_{\Lambda'}}[K]
    \quadtext{for}
    K\in \alg_{\Lambda'}
    \qquadtext{and}
    \rho_\beta^{\Lambda'}
    :=
    \rho_\beta^{\Lambda'}[H_{\Lambda'}]
    .
\end{equation*}
Additionally, whenever we consider a path \(H_{\Lambda'}(s) = H_{\Lambda'} + s\,V \in \alg_{\Lambda'}\), we denote
\begin{equation*}
    \rho_\beta^{\Lambda'}(s)
    :=
    \rho_\beta^{\Lambda'}[H_{\Lambda'}(s)] .
\end{equation*}
Let us now introduce the three concepts for which we want to prove equivalence as described in the introduction and depicted in Figure~\ref{fig:diagram-introduction}.

\subsection{Decay of correlations}

Let \(\Lambda\Subset\Z^\nu\) and~\(\rho\) be a state on~\(\alg_\Lambda\).
Then the \textit{covariance} of two operators \(A\in \alg_X\) and \(B\in \alg_Y\), localized in \(X\), \(Y \subset \Lambda\) is defined as
\begin{equation}
    \label{eq:definition-covariance}
    \Cov_{\rho}(A,B)
    :=
    \trace[\big]{\rho \, A \, B}
    - \trace[\big]{\rho \, A} \, \trace[\big]{\rho \, B}
    .
\end{equation}
To remove the explicit dependence on the operators, we define
\begin{equation*}
    \Cov_{\rho}(X;Y)
    :=
    \sup_{\substack{A\in \alg_X\colon \norm{A}=1, \\[\lw] B\in \alg_Y\colon \norm{B}=1\phantom{,}}}
    \abs[\big]{\Cov_{\rho}(A,B)}
    .
\end{equation*}

One of the main concepts, we will use later is \emph{decay of correlations}, sometimes also called \emph{clustering (of correlations)}.

\begin{definition}[(Uniform) decay of correlations]
    \label{def:decay-of-correlations}
    Let \(\Lambda\Subset\Z^\nu\) and~\(\rho\) be a state on~\(\alg_\Lambda\).
    We say that~\(\rho\) satisfies \emph{decay of correlations} with respect to the continuous functions~\(\zeta_\DC\), \(f_\DC\colon \intervalco{0,\infty} \to \intervalco{0,\infty}\) and \(n \geq 0\) if and only if
    \begin{equation*}
        \Cov_{\rho}(X;Y)
        \leq
        \abs{X}^n
        \, f_\DC\paren[\big]{\abs{Y}}
        \, \zeta_\DC\paren[\big]{\dist{X,Y}}
    \end{equation*}
    for all \(X\), \(Y \subset \Lambda\).

    With a little abuse of notation, we say that an interaction~\(\Psi\) satisfies \emph{uniform decay of correlations} (at inverse temperature~\(\beta\)) on~\(\Lambda\) if and only if the Gibbs states~\(\rho_{\beta}^{\Lambda'}[H_{\Lambda'}]\) satisfy decay of correlations with respect to the same functions and~\(n\) for every~\({\Lambda' \subset \Lambda}\).
\end{definition}

One could equivalently take the minimum with the same bound where~\(X\) and~\(Y\) are exchanged.
But for simplicity in the presentation, we will always write the bound in this way without writing the minimum explicitly.
We choose this rather general definition with arbitrary functions as a compromise between understandable proofs and validity for different results on decay of correlations coming from previous literature~\cite{Araki1969,KGK2014,KK2024}.
A possible extension is to consider a bound where the growth is only in the size of the boundaries~\(\partial X\) and~\(\partial Y\) of the sets.
This setting will be discussed in detail for one-dimensional spin chains, where \(\abs{\partial X} = 2\) for all intervals~\(X\).

We say that~\(\rho\) satisfies exponential decay of correlations, if \(\zeta_\DC(r) \leq C_\DC \, \e^{-c_\DC r}\) for some~\(C_\DC\),~\(c_\DC>0\).

\subsection{Local perturbations perturb locally}

The second concept measures stability of a system against local perturbations, in the sense that expectation values of local observables supported far away from the perturbation change only very little.
We refer to this stability as \emph{local perturbations perturb locally} (LPPL), a term used for the local stability of ground states in previous works~\cite{BMN2011,RS2015,HTW2021,BRD2021}.

\begin{definition}[(Uniform) local perturbations perturb locally (LPPL)]
    \label{def:LPPL}
    Let \(\Lambda \Subset\Z^\nu\) and \(H\in \alg_\Lambda\) be self-adjoint.
    We say that~\(H\) satisfies \emph{LPPL} (at inverse temperature~\(\beta\)) with respect to the continuous functions~\(f_\LPPL\), \(g_\LPPL\), \(\zeta_\LPPL\colon \intervalco{0,\infty} \to \intervalco{0,\infty}\) and \(n \geq 0\), if and only if
    \begin{equation*}
        \abs[\big]{
            \trace[\big]{\rho_\beta^\Lambda[H] \, B}
            - \trace[\big]{\rho_\beta^\Lambda[H+V] \, B}
        }
        \leq
        \norm{B}
        \, \abs{X}^n
        \, f_\LPPL\paren[\big]{\abs{Y}}
        \, g_\LPPL\paren[\big]{\norm{V}}
        \, \zeta_\LPPL\paren[\big]{\dist{X,Y}}
    \end{equation*}
    for all \(X, Y \subset \Lambda\), \(V\in \alg_X\) self-adjoint and~\(B\in \alg_Y\).

    We say that an interaction~\(\Psi\) satisfies \emph{uniform LPPL} (at inverse temperature~\(\beta\)) on~\(\Lambda\) if and only if~\(H_{\Lambda'}\) satisfies LPPL with respect to the same functions and~\(n\) for every~\({\Lambda' \subset \Lambda}\).
\end{definition}

\subsection{Local indistinguishability}

The final property we are interested in is locality of the Gibbs state, in the sense that expectation values of local operators can be well approximated by the expectation values in the Gibbs state on a smaller set \(\Lambda'\subset \Lambda\) as long as they are localized far apart from~\(\Lambda\setminus\Lambda'\).
This property goes under the name \emph{local indistinguishability}~\cite{BK2019,BCP2022}.
We note that it was also called \emph{locality of temperature} in~\cite{KGK2014}.

\begin{definition}[Local indistinguishability]
    \label{def:local-indistinguishability}
    Let \(\Lambda \Subset\Z^\nu\) and~\(\Psi\) be an interaction.
    We say that~\(\Psi\) satisfies \emph{local indistinguishability} (at inverse temperature~\(\beta\)) on~\(\Lambda\) with respect to the continuous functions~\(f_\LI\) and \(\zeta_\LI\colon \intervalco{0,\infty} \to \intervalco{0,\infty}\) if and only if
    \begin{equation*}
        \abs[\big]{
            \trace[\big]{
                \rho_\beta^\Lambda[H_\Lambda] \, B
            }
            -\trace[\big]{
                \rho_\beta^{\Lambda'}[H_{\Lambda'}] \, B
            }
        }
        \leq
        \norm{B}
        \, f_\LI\paren[\big]{\abs{Y}} \,
        \, \zeta_\LI\paren[\big]{\dist{Y,\Lambda\setminus\Lambda'}}
    \end{equation*}
    for all \(Y \subset \Lambda' \subset \Lambda\) and~\(B\in \alg_Y\).
\end{definition}

A similar property for ground states is also of great interest and often called local topological quantum order (LTQO) for historical reasons.
See~\cite[section~2.2.2]{NSY2020} for a discussion of LTQO\@.
To emphasize the above flavour, it sometimes gets more descriptive names like “indistinguishability radius”~\cite{NSY2020}.
In most works, LTQO is taken as an assumption, see e.g.~\cite{MZ2013,BRD2021}.
It is only known to be satisfied in very simple systems, see~\cite{NSY2020} for an overview and~\cite{CMPS2013,SCP2010,HTW2021,BRD2021}.

\begin{remark}
    Note that decay of correlations is a property of a state, LPPL is a property of a Hamiltonian, and local indistinguishability is a property of an interaction.
    Clearly in all three concepts the most relevant information is the decay function: \(\zeta_\DC\) encodes the rate of decay of correlations in a state between different regions in space, \(\zeta_\LPPL\) controls the rate at which the influence of a perturbation decays in the distance to the region where it is supported, and~\(\zeta_\LI\) encodes the rate at which the influence of the boundary on a Gibbs state decays into the bulk of a system.
\end{remark}

\begin{remark}
    \label{rmk:KMS}
    Let us briefly comment on how our results extend to infinite volume systems and how they are related and might be useful for related questions in infinite volume.
    Since all our estimates are uniform in the system size \(\abs{\Lambda}\), they extend to the KMS state for the infinite volume system that is obtained as the weak\(^*\)-limit of the finite volume Gibbs states \(\rho^\Lambda_\beta\) for~\(\Lambda\nearrow \Z^\nu\).
    The existence of this limit is guaranteed, for example, by the local indistinguishability property.
    In particular, our circle of equivalences implies that uniform decay of correlations or uniform LPPL at a certain inverse temperature~\(\beta\) are also sufficient conditions for the existence of this limit.
    Note, however, that even if one had a property like local indistinguishability uniformly for different boundary conditions, uniqueness of the KMS state is not expected to our knowledge~\cite[chapter~6]{BR1981}.

    Related questions concerning the stability of KMS states in infinite volume that have been discussed intensively in the literature are return and approach to equilibrium (see e.g.~\cite{Robinson1973,JPT2024} and references therein).
    Roughly speaking, \emph{return to equilibrium} is the property that a KMS state of a locally perturbed system returns under the unperturbed dynamics to the associated KMS state of the unperturbed system in the limit \(t\to\infty\) and in the weak\(^*\)-topology~\cite{Robinson1973}.
    The idea is that local changes in the KMS state disperse or propagate to spatial infinity under the unperturbed evolution, which is usually assumed to be asymptotically abelian.
    Note that this problem is somewhat independent of LPPL, which claims that a local perturbation leads to a local change in the Gibbs or KMS state.
    It is conceivable that a local perturbation changes the KMS state only locally, but that this change then persists under the unperturbed dynamics.
    On the other hand, it is also possible that a system does not satisfy LPPL, but still exhibits return to equilibrium.
    This is because LPPL implies that the perturbed KMS state is normal with respect to the unperturbed one, while this is not a necessary condition for return to equilibrium~\cite{Robinson1973}.

    The problem of \emph{approach to equilibrium} in its general formulation proposed in~\cite{JPT2024} is completely open.
    The question is whether general initial states approach in the long time average and in the weak\(^*\) sense a superposition of KMS states.
    In~\cite{JPT2024} this problem was studied for initial states that are themselves KMS states for a different system.
    While to our understanding this problem is also independent of the LPPL property, our Theorem~\ref{thm:stability-SLT-perturbations} establishes a form of Lipschitz continuity of the Gibbs state as a function of the defining interaction, which might be a useful ingredient when studying approach to equilibrium.
\end{remark}

\section{Applications of the general results}
\label{sec:applications-of-the-general-results}

Before formally stating and proving our main results, in this section we first present, also as a motivation, a collection of applications to various classes of spin systems.
In all cases, we start from quantitative bounds on the decay of correlations, which have either been shown before or will be shown later.

We focus first on one-dimensional quantum spin systems, for which we consider the following cases separately: translation-invariant short-range interactions at high enough temperature, short-range interactions at any positive temperature, and long-range interactions at any positive temperature.

For all of them, decay of correlations (with different decay rates) is known to hold~\cite{Araki1969,BCP2022,PP2023, KK2024} or is proved in this paper (see Theorem~\ref{thm:DC-short-range-interactions-1D-informal}).
We then turn our attention to finite-range quantum spin systems of arbitrary dimension, for which (uniform) exponential decay of correlations above a threshold temperature is known to be satisfied~\cite{KGK2014}.

\subsection{One-dimensional translation-invariant short-range systems}
\label{sec:results-one-dimensional-translation-invariant-short-range}

In this section, we restrict our attention to translation-invariant short-range spin systems.
For one-dimensional translation-invariant systems, Araki~\cite{Araki1969} proved that the infinite chain satisfies exponential decay of correlations at every positive temperature, and this was subsequently extended to short-range interactions in~\cite{PP2023} above a threshold temperature.
As a consequence of this, the analogous result for finite-range finite chains was recently shown in~\cite{BCP2022}.
In the current manuscript, we extend this to short-range interactions above a threshold temperature.
A precise definition of translation-invariant interactions is given in Section~\ref{sec:1D-spin-chains}, basically it means that \(\Psi(X)\) and \(\Psi(X+n)\) are the same operator on different parts \(\alg_X \simeq \alg_{X+n}\) of the lattice.

\begin{theorem}
    \label{thm:DC-short-range-interactions-1D-informal}
    Let~\(b>0\) and~\(\Psi\) be a translation-invariant interaction with \({\norm{\Psi}\sr{b} < \infty}\) and denote \(\beta^* :=b/\paren{2 \, \norm{\Psi}_{1}}\).
    Then, for all \(\beta \in \intervaloo{0 , \beta^*}\), the Gibbs state satisfies decay of correlations in the sense that there exist~\(C_\DC\), \(c_\DC > 0\) such that for every finite interval \(I \subset \Z\) and subintervals \(X, Y \subset I\), it holds that
    \begin{equation*}
        \Cov_{\rho^I_\beta}(X;Y)
        \leq
        C_\DC \, \e^{- c_\DC \dist{X,Y}}
        .
    \end{equation*}
\end{theorem}

From Theorems~\ref{thm:LPPL-1D} and~\ref{thm:LI-1D}, we conclude the following statement.

\begin{corollary}\label{cor:LPPL-and-LI-translation-invariant-short-range}
    Let~\(b\) and~\(C_\interaction>0\), and~\(\Psi\) be a translation-invariant interaction with \({\norm{\Psi}\sr{b} < \infty}\).
    Denote \(\beta^* :=b/\paren{2 \, \norm{\Psi}_{1}}\).
    Then, for all \(\beta \in \intervaloo{0 , \beta^*}\), there exist constants \(C_\LPPL\), \(c_\LPPL\), \(C_\LI\), \(c_\LI>0\) such that the following statements hold for all intervals \(I\Subset \Z\).
    \begin{theoremenumerate}
        \item
            The Gibbs state satisfies LPPL in the sense that
            \begin{equation*}
                \abs[\big]{
                    \trace[\big]{\rho_\beta^{I}[H_I] \, B}
                    - \trace[\big]{\rho_\beta^{I}[H_I + V] \, B}
                }
                \leq
                C_\LPPL
                \, \e^{ 3 \beta \norm{V} }
                \, \paren[\big]{1+\norm{V}}
                \, \norm{B}
                \, \e^{-c_\LPPL \, \dist{X,Y}}
            \end{equation*}
            for all subintervals \(X,Y_1,Y_2\subset I\), such that \(Y_1 < X < Y_2\), \(Y = Y_1 \cup Y_2\), all self-adjoint perturbations \(V \in \alg_X\) and all observables~\(B = B_1 \otimes B_2\) with \(B_1\in \alg_{Y_1}\) and~\(B_2\in \alg_{Y_2}\).
        \item
            The interaction satisfies local indistinguishability in the sense that
            \begin{equation*}
                \abs[\big]{
                    \trace[\big]{
                        \rho_\beta^{I} \, B
                    }
                    -\trace[\big]{
                        \rho_\beta^{I \setminus X} \, B
                    }
                }
                \leq
                C_\LI
                \, \norm{B}
                \, \e^{-c_\LI \dist{Y,X}}
            \end{equation*}
            for all subintervals \(X,Y_1,Y_2\subset I\), such that \(Y_1 < X < Y_2\), \(Y = Y_1 \cup Y_2\), and all observables~\(B = B_1 \otimes B_2\) with \(B_1\in \alg_{Y_1}\) and~\(B_2\in \alg_{Y_2}\).
    \end{theoremenumerate}
\end{corollary}

The proof of this result is a straightforward application of Theorems~\ref{thm:LPPL-1D} and~\ref{thm:LI-1D} under the conditions of Theorem~\ref{thm:DC-short-range-interactions-1D-informal}.
We refer to Section~\ref{sec:1D-spin-chains} for a detailed exposition of all these results.

\subsection{One-dimensional short-range systems}
\label{sec:results-one-dimensional-short-range}

In this and the next section we restrict to so called \(k\)-local interactions, which satisfy \(\Psi(Z)=0\) if~\(\abs{Z}>k\).
Additionally we will – as before – require a decay of the terms~\(\Psi(Z)\) in~\(\diam{Z}\) by specifying an interaction norm.
And again, we only consider one-dimensional systems~\(\Lambda\Subset\Z\).
For such systems, \textcite{KK2024} recently proved decay of correlations for short-range interactions.

\begin{theorem}[{\Cite[Theorem~1]{KK2024}}]
    \label{thm:DC-KK2024-short-range}
    Let \(b\), \(C_\interaction\) and~\(k>0\).
    Then, for any \(\beta>0\) there exist constants \(C_\DC\), \(c_\DC>0\) such that all \(k\)-local interactions \(\Psi\) with \(\norm{\Psi}\sr{b} < C_\interaction\) satisfy decay of correlations such that
    \begin{equation*}
        \Cov_{\rho_\beta^\Lambda}(X;Y)
        \leq
        C_\DC \, \e^{-c_\DC \, \sqrt{\dist{X,Y}}}
    \end{equation*}
    for all \(\Lambda\Subset\Z\), and intervals \(X\), \(Y\subset \Lambda\).
\end{theorem}

From Theorems~\ref{thm:DC0toLPPL} and~\ref{thm:uniform-lppl-implies-local-indistinguishability} we obtain the following statement.
A short proof is given in Appendix~\ref{app:results-one-dimensional-short-range}.

\begin{corollary}
    \label{cor:LPPL-and-LI-from-KK2024-short-range}
    Let \(b\), \(C_\interaction\) and~\(k>0\).
    Then, for any \(\beta>0\) there exist constants \(C_\LPPL\), \(c_\LPPL\), \(C_\LI>0\) such that the following statements hold for all \(k\)-local interactions \(\Psi\) with \(\norm{\Psi}\sr{b} < C_\interaction\).
    \begin{theoremenumerate}
        \item
            The Gibbs state satisfies LPPL in the sense that
            \begin{equation}
                \label{eq:cor-LPPL-from-KK2024-short-range}
                \abs[\big]{
                    \trace[\big]{\rho_\beta^{\Lambda}[H_\Lambda] \, B}
                    - \trace[\big]{\rho_\beta^{\Lambda}[H_\Lambda + V] \, B}
                }
                \leq
                C_\LPPL
                \, \e^{ 3 \beta \norm{V} }
                \, \paren[\big]{1+\norm{V}}
                \, \norm{B}
                \, \e^{-c_\LPPL \, \sqrt{\dist{X,Y}}}
            \end{equation}
            for all \(\Lambda\Subset\Z\), intervals \(X\), \(Y\subset \Lambda\), \(V\in \alg_X\) self-adjoint and~\(B\in \alg_Y\).
        \item
            The interaction satisfies local indistinguishability in the sense that
            \begin{equation*}
                \abs[\big]{
                    \trace[\big]{
                        \rho_\beta^\Lambda \, B
                    }
                    -\trace[\big]{
                        \rho_\beta^{\Lambda'} \, B
                    }
                }
                \leq
                C_\LI
                \, \norm{B}
                \, \paren[\Big]{1+\sqrt{\dist{Y,\Lambda\setminus\Lambda'}}}
                \, \e^{-c_\LI \sqrt{\dist{Y,\Lambda\setminus\Lambda'}}}
            \end{equation*}
            for all \(\Lambda' \subset \Lambda \Subset \Z\), intervals \(Y \subset\Lambda'\) and \(B\in \alg_Y\), with \(c_\LI = b \, c_\LPPL / \sqrt{b^2 + c_\LPPL^2}\).
    \end{theoremenumerate}
\end{corollary}

\Textcite{KK2024} prove a more general statement, which includes stretched exponentially decaying interactions.
For such interactions, one obtains a similar result by using different Lieb-Robinson bounds to calculate~\(\zeta_\QBP\).
Moreover, they prove decay of correlations for long-range interactions, which we discuss in the next section.

\subsection{One-dimensional long-range systems}
\label{sec:results-one-dimensional-long-range}

Let us consider \(k\)-local interactions (see Section~\ref{sec:results-one-dimensional-short-range}) with polynomial decay in the sense that
\begin{equation}
    \label{eq:def-interaction-norm-long-range-KK2024}
    \tnorm{\Psi}_{F_\alpha}
    :=
    \sup_{x,y\in \Lambda} \sumstack[r]{Z\Subset\Lambda:\\x,y\in Z}
    \frac{\norm{\Psi(Z)}}{F_\alpha\paren[\big]{\dist{x,y}}}
    <
    \infty
    ,
\end{equation}
with \(F_\alpha(r):=(r+1)^{-\alpha}\).
For such interactions, \textcite{KK2024} recently obtained a bound on decay of correlations in one-dimensional systems.

\begin{theorem}[{\Cite[Theorem~1]{KK2024}}]
    \label{thm:DC-KK2024-long-range}
    Let \(\alpha>2\), \(\alpha_\DC < \alpha - 2\), \(C_\interaction\) and~\(k>0\).
    Then, for any \(\beta>0\) there exist constants \(C_\DC\), \(c_\DC>0\) such that all \(k\)-local interactions \(\Psi\) with \(\tnorm{\Psi}_{F_\alpha} < C_\interaction\) satisfy decay of correlations such that
    \begin{equation*}
        \Cov_{\rho_\beta^\Lambda}(X;Y)
        \leq
        C \, F_{\alpha_\DC}\paren[\big]{\dist{X,Y}}
    \end{equation*}
    for all \(\Lambda\Subset \Z\) and intervals \(X\), \(Y\subset \Lambda\).
\end{theorem}

From Theorems~\ref{thm:DC0toLPPL} and~\ref{thm:uniform-lppl-implies-local-indistinguishability} we obtain the following statement.
A short proof is given in Appendix~\ref{app:results-one-dimensional-long-range}.

\begin{corollary}
    \label{cor:LPPL-and-LI-from-KK2024-long-range}
    Let \(C_\interaction>0\), \(k\), \(\beta>0\).
    The following statements hold.
    \begin{theoremenumerate}
        \item 
            For every \(\alpha>2\), \(\alpha_\LPPL<\alpha-2\) there exist a constant \(C_\LPPL>0\) such that for all \(k\)-local interactions \(\Psi\) on \(\Z\) with \(\tnorm{\Psi}_{F_\alpha}<C_\interaction\)
            \begin{equation}
                \label{eq:cor-LPPL-from-KK2024-long-range}
                \abs[\big]{
                    \trace[\big]{\rho_\beta^{\Lambda}[H_\Lambda] \, B}
                    - \trace[\big]{\rho_\beta^{\Lambda}[H_\Lambda + V] \, B}
                }
                \leq
                C_\LPPL
                \, \e^{ 3 \beta \norm{V} }
                \, \paren[\big]{1+\norm{V}}
                \, \norm{B}
                \, F_{\alpha_\LPPL}\paren[\big]{\dist{X,Y}}
            \end{equation}
            for all \(\Lambda\Subset \Z\), intervals \(X\), \(Y \subset \Lambda\), \(V\in \alg_X\) self-adjoint and~\(B\in \alg_Y\).
        \item 
            For every \(\alpha>3\), \(\alpha_\LI<\alpha-3\) there exist a constant \(C_\LI>0\) such that for all \(k\)-local interactions \(\Psi\) on \(\Z\) with \(\tnorm{\Psi}_{F_\alpha}<C_\interaction\)
            \begin{equation*}
                \abs[\big]{
                    \trace[\big]{
                        \rho_\beta^\Lambda \, B
                    }
                    -\trace[\big]{
                        \rho_\beta^{\Lambda'} \, B
                    }
                }
                \leq
                C_\LI
                \, \norm{B}
                \, F_{\alpha_\LI}\paren[\big]{\dist{Y,\Lambda\setminus\Lambda'}}
            \end{equation*}
            for all \(\Lambda'\subset \Lambda\Subset \Z\), intervals \(Y \subset \Lambda\) and~\(B\in \alg_Y\).
    \end{theoremenumerate}
\end{corollary}

\subsection{\texorpdfstring{\(\nu\)}{nu}-dimensional short-range systems at high temperature}
\label{sec:results-high-temperature}

At high enough temperature, Gibbs states of finite-range Hamiltonians in any dimension satisfy uniform exponential decay of correlations~\cite{KGK2014} (see Remark~\ref{rem:high-temperature}).
This behaviour is expected also for systems that have short-range interactions.

\begin{conjecture}
    \label{conj:DC}
    Let~\(C_\interaction\), \(b>0\) and \(\nu\in \N\), then there exists \(\beta^*\), \(C_\DC\), \(c_\DC>0\) and \(n\in \N\) such that the following holds:
    Let~\(\Psi\) be an interaction on~\(\Z^\nu\) such that \(\norm{\Psi}\sr{b}<C_\interaction\).
    Then, for every \(\beta<\beta^*\) and \(\Lambda \Subset \Z^\nu\), the Gibbs state~\(\rho^{\Lambda}_\beta\) satisfies decay of correlations in the sense of Definition~\ref{def:decay-of-correlations} with constants such that
    \begin{equation*}
        \Cov_{\rho^\Lambda_\beta}(X;Y)
        \leq
        C_\DC
        \, \abs{X}^n
        \, \abs{Y}^n
        \, \e^{-c_\DC \dist{X,Y}}
    \end{equation*}
    for all \(X\), \(Y \subset \Lambda\).
\end{conjecture}

If the Conjecture~\ref{conj:DC} is satisfied, then our general Theorems~\ref{thm:DC0toLPPL} and~\ref{thm:uniform-lppl-implies-local-indistinguishability} imply that the Gibbs states of such systems are also stable against local perturbations and satisfy local indistinguishability.
A short proof of how the claimed exponential decay rates are obtained from the general formulas is given in Appendix~\ref{app:results-high-temperature}.

\begin{corollary}
    \label{cor:results-from-conjecture}
    Assume that Conjecture~\ref{conj:DC} holds true.
    Let~\(C_\interaction\), \(b>0\) and \(\nu\in \N\), and let \(\beta^*\) and \(n\) as in Conjecture~\ref{conj:DC}.
    Then there exist constants \(C\), \(c>0\) such that the following holds for all interactions \(\Psi\) on~\(\Z^\nu\) satisfying \(\norm{\Psi}\sr{b}<C_\interaction\):
    \begin{theoremenumerate}
        \item \label{item:prop-results-from-conjecture-DC0LPPL}
            For all \(\Lambda\Subset\Z^\nu\), \(X\),~\(Y\subset \Lambda\), \({V\in \alg_X}\)~self-adjoint, \({B\in \alg_Y}\) and \(\beta<\beta^*\)
            \begin{equation*}
                \abs[\big]{
                    \trace[\big]{\rho_\beta^{\Lambda}[H_\Lambda] \, B}
                    - \trace[\big]{\rho_\beta^{\Lambda}[H_\Lambda + V] \, B}
                }
                \leq
                C
                \, \e^{ 3 \beta \norm{V} }
                \, \paren[\big]{1+\norm{V}}
                \, \norm{B}
                \, \abs{X}^n \, \abs{Y}^n
                \, \e^{-c \dist{X,Y}}
                .
            \end{equation*}

        \item \label{item:prop-results-from-conjecture-local-indistinguishability}
            For all \(Y\subset \Lambda'\subset \Lambda\Subset\Z^\nu\), \(B\in \alg_Y\) and \(\beta<\beta^*\)
            \begin{equation*}
                \abs[\big]{
                    \trace[\big]{
                        \rho_\beta^\Lambda \, B
                    }
                    -\trace[\big]{
                        \rho_\beta^{\Lambda'} \, B
                    }
                }
                \leq
                C
                \, {\beta}
                \, \norm{B}
                \, \abs{Y}^{n+1}
                \, \e^{-c \dist{Y,\Lambda\setminus\Lambda'}}
                .
            \end{equation*}
    \end{theoremenumerate}
\end{corollary}

\begin{remark}
    \label{rem:high-temperature}
    A few remarks are in order:
    \begin{itemize}
        \item Conjecture~\ref{conj:DC} is known to be true for finite-range interactions, as it has been shown in~\cite{KGK2014}, which is also one of the very few previous results on LPPL and local indistinguishability for finite-range interactions.
            Indeed, by using the explicit decay of correlations obtained in~\cite{KGK2014} together with our Theorem~\ref{thm:DC0toLPPL} and Theorem~\ref{thm:uniform-lppl-implies-local-indistinguishability} we recover the result of~\cite{KGK2014}, namely estimates analogous to those in~\ref{item:prop-results-from-conjecture-DC0LPPL} and~\ref{item:prop-results-from-conjecture-local-indistinguishability} but with scaling \(\abs{\partial Y}^2\) instead of the \(\abs{X}^n \, \abs{Y}^n\) and~\(\abs{Y}^{n+1}\), respectively.%
            \footnote{
                First, note that \(\partial Y\) in~\cite{KGK2014} is the boundary in the interaction graph, i.e.\ it scales with the range~\(R\).
                To obtain the claimed scaling, we slightly modify their proof:
                After~\cite[eq.~(47)]{KGK2014}, we apply the same bound, but with \(x_0=1\), to upper bound~\cite[eq.~(47)]{KGK2014} by \(2 \, \abs{\partial Y} \, \e^{-L/\xi(\beta)}/(1-\e^{-1/\xi(\beta)})\) for all \(L > L_0 := \xi(\beta) \, \ln\paren[\big]{\abs{\partial Y} / \paren[\big]{1-\e^{-1/\xi(\beta)}}}\), where \(\xi(\beta)\) as in~\cite[eq.~(10)]{KGK2014}.
                Multiplying this bound by \(\abs{\partial Y}/(1-\e^{-1/\xi(\beta)}) = \e^{L_0/\xi(\beta)} > 1\) makes it valid for all \(L \geq 0\) due to the trivial bound~\(\abs{\Cov_\rho(X;Y)} \leq 2\).
            }
            However, we emphasize that the proof in~\cite{KGK2014} requires not only the control on the decay of correlations but also on the decay of the generalized covariance for all the Gibbs states of the perturbed Hamiltonian \(H(s)\).
            In this respect, our result makes the locality and stability results in principle more accessible.
        \item In dimension \(\nu=1\), Conjecture~\ref{conj:DC} is true for translation-invariant interactions.
            Moreover, the critical temperature \(\beta^*\) vanishes when reducing to translation-invariant finite-range interactions in one dimension due to~\cite{PP2023}.
            This setting is addressed in detail in Section~\ref{sec:1D-spin-chains}.

        \item Theorem~3.2 in~\cite{FU2015} implies Conjecture~\ref{conj:DC}.
            However, after finishing this manuscript, it was pointed out to us~\cite{Bluhm2024} that the proof presented in~\cite{FU2015} only proves a scaling \(\exp\paren{\abs{X}+\abs{Y}}\)~\cite{FU2024} instead of the claimed~\(\abs{X}\,\abs{Y}\).
            The exponential prefactor, which is not considered to be optimal, is insufficient to apply our results.
            \qedhere
    \end{itemize}
\end{remark}

\section{Quantum belief propagation}
\label{sec:quantum-belief-propagation}

This section is devoted to the introduction of the \emph{quantum belief propagation} (QBP), the main tool for our analysis.
The concept was originally introduced by \textcite{Hastings2007}.
More recently, the concept of QBP attracted more attention and has been used for different applications, see for example the works~\cite{Kim2012,Kim2013,EO2019,BK2019,KB2019,HMS2020,AAKS2021,KAA2021,RGK2023,ORFW2023}.
See also the review~\cite{Alhambra2022} and Remark~\ref{rmk:QBP} for a more detailed discussion on the literature.

Here, we offer a rigorous exploration of quantum belief propagation, expanding upon earlier findings.
In particular, we show how to extend the technique from the exponential of a Hamiltonian to its Gibbs state.
A step which was used in some of the mentioned works without justification.
Moreover, we also include short- and long-range interactions instead of finite-range interactions only.
To enhance the text's clarity, we have chosen to postpone the proofs to Section~\ref{sec:proof-qbp}.

The first theorem does not need any locality properties of the Hamiltonians.
Therefore, we state the theorem for general self-adjoint operators on a finite-dimensional Hilbert space.
Notice that in~\cite{EO2019} the authors extend the QBP equation for the perturbed exponential, namely equation~\eqref{eq:ode-ebeta}, to KMS states of general \(W^*\)-dynamical systems.

\begin{proposition}[Quantum belief propagation, general statement]
    \label{prop:qbp-ode}
    Let~\(\Hi\) be a finite-dimensional Hilbert space and~\(H\) and~\(V\) self-adjoint operators on~\(\Hi\).
    We consider the path of Hamiltonians \(H(s) := H + s \, V\).
    Then, the following holds true.
    \begin{theoremenumerate}
        \item \label{prop:qbp-ode-exponential}
            The exponentials \(\ebeta{H(s)}\) satisfy the differential equation
            \begin{equation}
                \label{eq:ode-ebeta}
                \odv*{\ebeta{H(s)}}{s}
                =
                -\frac{\beta}{2} \, \anticommutator[\Big]{\ebeta{H(s)},\Phi^{H(s)}_\beta(V)},
            \end{equation}
            where \(\Phi^{H(s)}_\beta(V)\) is defined by%
            \begin{equation}
                \label{eq:def-quasi-local-generator}
                \Phi^{H(s)}_\beta(W)
                :=
                \int_{-\infty}^\infty \diff t \, f_\beta(t) \, \e^{-\I t H(s)} \, W \, \e^{\I t H(s)}
                ,
            \end{equation}
            for all self-adjoint operators~\(W\) on~\(\Hi\) with~\(f_\beta\) an exponentially decaying \(L^1\)-function, which is explicitly given in~\eqref{eq:definition-f-beta}.
            Clearly \(\norm[\big]{\Phi^{H(s)}_\beta(W)} \leq \norm{W}\).

            Moreover, there exists a path of operators \(s \mapsto \eta(s)\), given in~\eqref{eq:definition-eta-s-time-ordering}, such that\linebreak[2] \(\ebeta{H(s)} = \eta(s) \, \ebeta{H(0)} \, \eta(s)^*\) and \(\norm{\eta(s)} \leq \e^{\frac{\beta}{2} s \norm{V} }\).
        \item \label{prop:qbp-ode-gibbs-state}
            Equation~\eqref{eq:ode-ebeta} implies that the path of perturbed Gibbs state \(s \mapsto \rho_\beta(s):=\rho_\beta^{\Hi}[H(s)]\) satisfies the differential equation
            \begin{equation}
                \label{eq:ode-gibbs-state}
                \odv*{\rho_\beta(s)}{s}
                =
                - \frac{\beta}{2} \, \anticommutator[\Big]{\rho_\beta(s),\Phi_\beta^{H(s)}\paren[\big]{V-\braket{V}_{\rho_\beta(s)}}}.
            \end{equation}

            Moreover, there exists a path of operators \(s \mapsto \etat(s)\) such that\linebreak[2] \(\rho_\beta(s) = \etat(s) \, \rho_\beta(0) \, \etat(s)^*\),\linebreak[2] \(\norm{\etat(s)} \leq \e^{\beta s \norm{V} }\) and
            \begin{equation}
                \label{eq:stability-small-perturbations}
                \norm{\rho_\beta(0)-\rho_\beta(s)}_1
                \leq
                \e^{2\beta s \norm{V}}-1.
            \end{equation}
    \end{theoremenumerate}
\end{proposition}

Proposition~\ref{prop:qbp-ode} not only gives two useful differential equations, but also proves continuity of the Gibbs state in the Hamiltonian and stability of the Gibbs state against small perturbations, see~equation~\eqref{eq:stability-small-perturbations}.
To make the result more explicit, we often use the bound
\begin{equation}
    \label{eq:exp(x)-1-bound}
    \e^{xy}-1 \leq x \, (\e^y-1)
    \qquad\text{for all \(x\in \intervalcc{0,1}\), \(y \geq 0\)}
    ,
\end{equation}
which for example allows to bound \(\e^{2\beta s \norm{V}} - 1 \leq s \, \paren[\big]{\e^{2\beta \norm{V}}-1}\).

Note that the existence of an operator \(\kappa(s)\) such that \(\ebeta{H(s)} = \kappa(s) \, \ebeta{H(0)} \, \kappa(s)^*\) is not surprising, and the same holds for the normalized state.
Indeed, many choices are possible since we only require one state to be mapped to another.
One can also choose \(\kappa(s) = \ebeta{H(s)/2} \, \ebeta[\beta]{H(0)/2}\), see~\cite[Corollary~5.4.2]{BR1981} together with the discussion following it and~\cite[section~III.B]{Alhambra2022}.
This is enough to obtain continuity of the Gibbs state, i.e.\ \(\norm{\rho_\beta(0) - \rho_\beta(s)} \to 0\) for \(s \to 0\) but without the explicit bound in~\eqref{eq:stability-small-perturbations}, see~\cite[Theorem~5.4.4(3)]{BR1981}.

The main advantage of the specific operators~\(\eta(s)\) and~\(\etat(s)\) provided by quantum belief propagation is their bounded norm and the locality, which will be stated in Proposition~\ref{prop:qbp-bounds}.
For the~\(\kappa(s)\) given above, boundedness and locality can be proven above a threshold temperature for finite-range Hamiltonian, see~\cite[eq.~(36),(37)]{Alhambra2022}.
But for smaller temperatures, there are nearest neighbour Hamiltonians~\cite[Main~Theorem]{Bouch2015}, where these proofs fail and where we expect that the norm of~\(\kappa(s)\) diverges in the thermodynamic limit.

\begin{remark}
    \label{remark:gapped-ground-states-spectral-flow}
    A very similar concept appears in the context of Hamiltonians with gapped ground states~\cite{BMN2011}.
    There
    \begin{equation*}
        -\I \, \odv*{P(s)}{s}
        =
        - \commutator[\big]{P(s),\Phi^{H(s)}_\gamma(V)},
    \end{equation*}
    where~\(P(s)\) is the projection onto the ground state of~\(H(s)\) and \(\Phi^{H(s)}_\gamma(V)\) is defined as in~\eqref{eq:def-quasi-local-generator} but with a different function~\(f_\gamma\).
    One then solves the Schrödinger equation and obtains the unitary propagator~\(U(s)\) such that
    \begin{equation*}
        P(s) = U(s) \, P(0) \, U(s)^*,
    \end{equation*}
    which one can also approximate.
    It satisfies \(\norm{U(s)}=\norm{U(s)^*}=1\) and \(U(s)^*=U(s)^{-1}\), which we do not have for the~\(\eta(s)\).
    For the ground state projection, the uniform (in~\(s\)) gap above the ground state is crucial, which is not the case for the Gibbs state.
    However, the estimates we obtain for the Gibbs state are not useful to analyse the \(\beta \to \infty\) regime.

\end{remark}

Proposition~\ref{prop:qbp-ode} becomes even more interesting when we apply it to the lattice setting described in Section~\ref{sec:mathematical-setting}, because locality of the Hamiltonian~\(H_\Lambda\) and localization of~\(V\) will result in quasi-locality of~\(\Phi_\beta^{H_\Lambda(s)}\).
The locality of the Hamiltonian is often measured in terms of Lieb-Robinson bounds~\cite{LR1972,NSY2017,NSY2019}.
These bounds measure how fast (the support of) an operator spreads under the Heisenberg time evolution.

\begin{definition}[Lieb-Robinson bound]
    \label{def:Lieb-Robinson-bound}
    Let~\(\Lambda\Subset\Z^\nu\).
    We say that a Hamiltonian \(H\in\alg_\Lambda\) satisfies a Lieb-Robinson bound with decay \(\zeta_\LR\colon \List{X\subset \Lambda} \times \List{Y\subset \Lambda} \times \intervalco{0,\infty} \to \intervalco{0,\infty}\) if
    \begin{equation*}
        \norm[\big]{\commutator[\big]{\e^{-\I t H} \, A \, \e^{\I t H}, B}}
        \leq
        \norm{A} \, \norm{B} \, \zeta_\LR\paren[\big]{X,Y,\abs{t}}
    \end{equation*}
    for all \(X\), \(Y\subset \Lambda\), \(A\in \alg_X\),\(B\in \alg_Y\) and~\(t\in \R\).
\end{definition}

For short-range interactions, the Lieb-Robinson bound decay can be proven to be
\begin{equation*}
    \zeta_\LR\paren[\big]{X,Y,\abs{t}}
    =
    2
    \, \min\List[\big]{\abs{X},\abs{Y}}
    \, \e^{b \paren*{v \abs{t} - \dist{X,Y}}}
\end{equation*}
with the Lieb-Robinson velocity \(v=2 \, \norm{\Psi}\sr{b} / b\), see Proposition~\ref{prop:short-range-LR-bound} for a precise statement.
Such Lieb-Robinson bounds are known for spin systems~\cite[Theorem~3.1]{NSY2019} and lattice fermions~\cite[Theorem~3.1(i)]{NSY2017}.
Similarly, for long-range interactions, Lieb-Robinson bounds with polynomial decay in the distance~\(\dist{X,Y}\) are known~\cite{EMN2020}, see Proposition~\ref{prop:LR-bound-long-range}.
We keep this very general assumption on the Lieb-Robinson bounds to render our results applicable for a large class of interactions.
Moreover, it allows us to obtain improved results for one-dimensional systems when we restrict~\(X\) to be an interval.

\begin{proposition}[Quantum belief propagation on a lattice]
    \label{prop:qbp-bounds}
    Let \(\Lambda\Subset\Z^\nu\), \(H\in\alg_\Lambda\) self-adjoint, \(X\subset \Lambda\), and~\(V\in \alg_X\) self-adjoint.
    For \(s \in [0,1]\), consider the path of Hamiltonians \(H (s):= H + s \, V\).
    Moreover, assume that all Hamiltonians~\(H (s)\) satisfy a Lieb-Robinson bound with~\(\zeta_\LR\) decay uniformly in~\(s\).
    Let \(\zeta_\QBP\colon \Set{X\Subset\Lambda} \times \intervalco{0,\infty} \to \intervalco{0,\infty}\) with
    \begin{equation}
        \label{eq:definition-zeta-QBP-general}
        \zeta_\QBP(X,r)
        :=
        \min\List[\Big]{
            2,
            \inf_{T \geq 0}
            \norm{\zeta_\LR(X,\Lambda\setminus X_r,{\cdot})}_{L^\infty(\intervalcc{-T,T})}
            + 4 \, \e^{-\frac{\pi}{\beta}T}
        }
        ,
    \end{equation}
    which only depends on~\(\zeta_\LR\) and the inverse temperature~\(\beta\).
    Then the following holds:
    \begin{theoremenumerate}
        \item \label{prop:qbp-bounds-generator}
            For any \(W\in \alg_X\), the operators \(\Phi^{H(s)}_\beta(W)\) defined in~\eqref{eq:def-quasi-local-generator} can be approximated by local operators \(\Phi^{H(s)}_{\beta, \rrr}(W) \in \alg_{X_\rrr}\) supported on the \(\rrr\)-neighbourhood~\(X_\rrr\), such that \(\norm{\Phi^{H(s)}_{\beta, \rrr}(W)} \leq \norm{\Phi^{H(s)}_{\beta}(W)}\) and
            \begin{equation*}
                \norm[\big]{
                    \Phi^{H(s)}_{\beta}(W)
                    - \Phi^{H(s)}_{\beta, \rrr}(W)
                }
                \leq
                \norm{W} \, \zeta_\QBP(X,\rrr)
                .
            \end{equation*}

        \item \label{prop:qbp-bounds-exponential}
            The operators~\(\eta(s)\) defined in Proposition~\ref{prop:qbp-ode-exponential} can be approximated by local operators \(\eta_\rrr(s) \in \alg_{X_\rrr}\) supported on~\(X_\rrr\), such that \(\norm{\eta_\rrr(s)} \leq \norm{\eta(s)}\) and
            \begin{equation*}
                \norm[\big]{\eta(s) - \eta_\rrr(s)}
                \leq
                \tfrac{\beta}{2} \, s \, \norm{V}
                \, \e^{\tfrac{\beta}{2} s \norm{V}}
                \, \zeta_\QBP(X,\rrr)
                .
            \end{equation*}

        \item \label{prop:qbp-bounds-gibbs-state}
            The operators~\(\etat(s)\) defined in Proposition~\ref{prop:qbp-ode-gibbs-state} can be approximated by operators \(\etat_\rrr(s) \in \alg_{X_\rrr}\) supported on~\(X_\rrr\), such that \(\norm{\etat_\rrr(s)} \leq \norm{\etat(s)}\) and
            \begin{equation}
                \label{eq:thm-qbp-approximation-etat}
                \norm[\big]{\etat(s) - \etat_\rrr(s)}
                \leq
                \beta
                \, s \, \norm{V}
                \, \e^{\beta s \norm{V}}
                \, \zeta_\QBP(X,\rrr)
                .
            \end{equation}
    \end{theoremenumerate}
\end{proposition}

When using the approximations given by Proposition~\ref{prop:qbp-bounds}, it is important to note that both maps \(\sigma \mapsto \etat(s)\,\sigma\,\etat^*(s)\) and \(\sigma \mapsto \etat_\rrr(s)\,\sigma\,\etat_\rrr^*(s)\) are completely positive, but in general not trace preserving.
Only for the Gibbs state \(\rho_\beta(s)\), clearly, \(\trace[\big]{\etat(s)\,\rho_\beta(0)\,\etat^*(s)} = \trace[\big]{\rho_\beta(s)} = 1\), and by~\eqref{eq:thm-qbp-approximation-etat} one can conclude \(\trace[\big]{\etat_\rrr(s)\,\rho_\beta(0)\,\etat_\rrr^*(s)} \approx 1\).

For the applications discussed in this work, it is enough to have some local approximation of \(\Phi^{H(s)}_\beta(W)\), which we construct in Section~\ref{sec:proof-locality-of-the-generator} using a conditional expectation, which is basically a partial trace for spin systems.
For numerical implementations, it might be useful to instead use \(\Phi^{H_{X_\rrr}}_\beta(W)\), i.e.~\eqref{eq:def-quasi-local-generator} with the Hamiltonian truncated to~\(X_\rrr\), as an approximation.
Qualitatively, the result would be as in Proposition~\ref{prop:qbp-bounds} by using a bound like~\cite[Theorem~3.4\,(ii)]{NSY2019} instead of~\eqref{eq:bound-for-conditional-expectation} in the proof.

\begin{remark}
    In Proposition~\ref{prop:qbp-bounds} we assume a uniform Lieb-Robinson bound for all Hamiltonians \(H (s) = H + s \, V\).
    This simplifies the statement and allows for better bounds when we later prove local indistinguishability.
    The assumption can be dropped by the stability result for the Lieb-Robinson bound we provide in Lemma~\ref{lem:LR-bound-for-perturbed-Hamiltonians} at the price of an extra factor \(\norm{V}\) in~\(\zeta_{\QBP}\), see Lemma~\ref{lem:zeta-QBP-short-range-in-unperturbed}.
\end{remark}

\begin{lemma}[{\(\zeta_\QBP\) for short-range interactions}]
    \label{lem:zeta-QBP-short-range}
    Let \(b>0\) and \(\Psi\) be an interaction on \(\Lambda\Subset\Z^\nu\) satisfying \(\norm{\Psi}\sr{b}<\infty\), \(X\subset \Lambda\) and~\(V\in \alg_X\) self-adjoint.
    Then for all Hamiltonians \(H(s) = H + s \, V\) the following holds:
    \begin{theoremenumerate}
        \item \label{lem:zeta-QBP-short-range-along-the-path}
            If also \(\norm{\Psi+V}\sr{b}<\infty\), then
            \begin{equation*}
                \zeta_\QBP(X,r)
                \leq
                6 \, \abs{X} \, \e^{-\frac{b}{1 + a \beta}\,r}
                ,
            \end{equation*}
            where \(a = \frac{2}{\pi} \, \max\List[\big]{\norm{\Psi}\sr{b}, \norm{\Psi+V}\sr{b}}\).
        \item \label{lem:zeta-QBP-short-range-in-unperturbed}
            In general, it holds that
            \begin{equation}
                \label{eq:remark-zeta-QBP-perturbed-short-range}
                \zeta_\QBP(X,r)
                \leq
                C_\QBP
                \, \paren[\big]{1+\norm{V}}
                \, \abs{X} \, \e^{-\frac{b}{1 + a \beta}\,r}
                ,
            \end{equation}
            where \(C_\QBP = 6 \, \max\List{1,2/(b\,v_b)}\) and \(a = \frac{2}{\pi} \, \norm{\Psi}\sr{b}\).
    \end{theoremenumerate}
    Moreover, if \(X = B_z(R) := \Set{x\in \Lambda \given \dist{x,z} \leq R}\) is a ball (for some \(z\in \Lambda\) and \(R\geq 1\)) we can replace \(\abs{X}\) in both bounds with \(\tfrac{1+b}{b} \, \abs{\partial X}\), where \(\partial X = B_z(R) \setminus B_z(R-1)\).
    In particular, for intervals \(X\subset \Z\) in dimension \(\nu=1\), we can replace \(\abs{X}\) by~\(2 \, \tfrac{1+b}{b}\).
\end{lemma}

We also provide similar estimates for long-range interactions in Lemma~\ref{lem:zeta-QBP-long-range}, the proofs are given in Section~\ref{sec:proof-bounds-for-zeta-qbp}.

\begin{remark}
    \label{rmk:QBP}
    The original proof of the quantum belief propagation is due to Hastings~\cite{Hastings2007} and was only concerned with a differential equation for the perturbed exponential \(\ebeta{H(s)}\), namely the first part of Proposition~\ref{prop:qbp-ode-exponential} and Proposition~\ref{prop:qbp-bounds-generator} and~\ref{prop:qbp-bounds-exponential}.
    However, note that the equation obtained in~\cite{Hastings2007} was different than the one we show here, which is instead the one that is commonly used nowadays and, as far as we know, appeared first in~\cite{Kim2012,Kim2013}.
    The other works that used QBP~\cite{Kim2012,Kim2013,EO2019,BK2019,KB2019,HMS2020,AAKS2021,KAA2021} were mainly focused on a differential equation for the perturbed exponential, while our main focus is the use of QBP directly on the Gibbs state, namely the differential equation~\eqref{eq:ode-gibbs-state} and its locality properties, for which we provide a thorough discussion in Section~\ref{sec:proof-qbp}.
    A similar approach was used for a slightly different application and finite-range interactions in~\cite{ORFW2023}.
\end{remark}

\section{LPPL from decay of correlations}
\label{sec:LPPL-from-decay-of-correlations}

Using quantum belief propagation, we now show that Gibbs states are stable against local perturbations in the Hamiltonian whenever they satisfy decay of correlations.

\begin{theorem}[LPPL from correlations in the unperturbed state]
    \label{thm:DC0toLPPL}
    Let \(\Lambda\Subset\Z^\nu\), \(H\in\alg_\Lambda\) self-adjoint, \(X\subset \Lambda\), and~\(V\in \alg_X\) self-adjoint.
    For \(s \in [0,1]\), consider the path of Hamiltonians \(H(s):= H + s \, V\) with Gibbs states~\(\rho_\beta(s)\).
    Moreover, assume that all Hamiltonians~\(H(s)\) satisfy a Lieb-Robinson bound with~\(\zeta_\LR\)-decay uniformly in~\(s\), and let~\(\zeta_\QBP\) be the function from Proposition~\ref{prop:qbp-bounds}.
    Then, for all \(Y \subset \Lambda\), \(B \in \alg_Y\) and~\(r \geq 0\), we have
    \begin{equation}
        \label{eq:DC0LPPL}
        \abs[\big]{
            \trace[\big]{\rho_\beta (0) \, B}
            - \trace[\big]{\rho_\beta (1) \, B}
        }
        \leq
        \e^{ 2\beta \norm{V} }
        \, \norm{B}
        \, \paren[\big]{
            \Cov_{\rho_\beta (0)}(X_r;Y)
            +
            4
            \, \beta
            \, \norm{V}
            \, \zeta_\QBP(X, r)
        }.
    \end{equation}
\end{theorem}

\begin{remark}[Simplified short-range version]
    \label{remark:DC0toLPPL-short-range}
    Let \(\Psi\) be a short-range interaction and assume that the Gibbs state \(\rho_\beta(0)\) satisfies exponential decay of correlations with respect to \(\zeta_\DC(r) = C_\DC \, \e^{-c_\DC r}\), \(f_\DC(\abs{Y})=\abs{Y}^m\) and \(n \geq 0\) in the sense of Definition~\ref{def:decay-of-correlations}.
    Then Theorem~\ref{thm:DC0toLPPL} implies that for every \(c < c_0 := c_\DC \, c_\QBP / (c_\DC + c_\QBP)\), where \(c_\QBP\) is the decay exponent in~\eqref{eq:remark-zeta-QBP-perturbed-short-range}, there exists a constant \(C>0\) such that for all \(X\), \(Y\subset \Lambda\), \(V\in \alg_X\) self-adjoint and \(B\in \alg_Y\)
    \begin{equation*}
        \abs[\big]{
            \trace[\big]{\rho_\beta (0) \, B}
            - \trace[\big]{\rho_\beta (1) \, B}
        }
        \leq
        C
        \, \e^{ 3 \beta \norm{V} }
        \, \paren[\big]{1+\norm{V}}
        \, \norm{B}
        \, \abs{X}^{\max\List{1,n}}
        \, \abs{Y}^m
        \, \e^{-c \dist{X,Y}}
        .
        \qedhere
    \end{equation*}
\end{remark}

Notice that we only need to know decay of correlations in the unperturbed state~\(\rho_\beta (0)\) in order to control the bound~\eqref{eq:DC0LPPL}.
This allows to use it in conjunction with decay of correlations in translation-invariant systems in one dimension~\cite{Araki1969,PP2023}.
Although the bound~\eqref{eq:DC0LPPL} holds for all temperatures, it still diverges for \(\beta\to\infty\) and is thus not useful in the zero temperature limit.

\begin{proof}
    For the proof, we drop the subscript~\({}_\beta\).
    Under the assumptions of the proposition, let~\(\etat(s)\) be the operators from Proposition~\ref{prop:qbp-ode-gibbs-state} such that \(\rho(s) = \etat(s) \, \rho(0) \, \etat(s)^*\), and \(\etat_\rrr(s)\) be their local approximations from Proposition~\ref{prop:qbp-bounds-gibbs-state} with \(\rrr<\dist{X,Y}\) such that~\(\commutator{B,\etat_\rrr}=0\).
    We abbreviate \(\rho:=\rho(0)\), \(\etat:=\etat(s)\), and~\(\etat_\rrr:=\etat_\rrr(s)\).
    Then,
    \begin{align*}
        \Alignindent
        \trace[\big]{\rho(s) \, B}
        - \trace[\big]{\rho \, B}
        \\&=
        \Trace[\big]{\etat \, \rho \, \etat^* \, B}
        - \Trace[\big]{\rho \, B}
        \\&=
        \Trace[\big]{(\etat-\etat_\rrr) \, \rho \, \etat^* \, B}
        + \Trace[\big]{\etat_\rrr \, \rho \, (\etat^*-\etat_\rrr^*) \, B}
        + \Trace[\big]{\rho \, \etat_\rrr^* \, \etat_\rrr \, B}
        - \Trace[\big]{\rho \, B}
        .
    \end{align*}
    The first two terms are bounded by \(
        \norm{\etat-\etat_\rrr} \, \paren[\big]{\norm{\etat}+\norm{\etat_\rrr}} \, \norm{B}
    \).
    Thus, for~\(B=\unit\), we obtain \(
        \abs[\big]{\Trace{\rho \, \etat_\rrr^* \, \etat_\rrr}-1}
        \leq
        \norm{\etat-\etat_\rrr} \, \paren[\big]{\norm{\etat}+\norm{\etat_\rrr}}
    \).
    Hence, we can replace \(\Trace[\big]{\rho \, B}\) by \(\Trace{\rho \, \etat_\rrr^* \, \etat_\rrr} \, \Trace[\big]{\rho \, B}\) to recover the covariance and obtain
    \begin{align*}
        \Alignindent
        \abs[\big]{
            \trace[\big]{\rho(s) \, B}
            - \trace[\big]{\rho \, B}
        }
        \\&\leq
        2 \, \norm{\etat-\etat_\rrr} \, \paren[\big]{\norm{\etat}+\norm{\etat_\rrr}} \, \norm{B}
        + \abs[\big]{ \Cov_\rho(\etat_\rrr^* \, \etat_\rrr, B) }
        \\&\leq
        4 \, \norm{B}
        \, \beta \, s \, \norm{V}
        \, \e^{2\beta s \norm{V}}
        \, \zeta_\QBP(X,\rrr)
        + \norm{B} \, \e^{2\beta s \norm{V}}
        \, \Cov_\rho(X_\rrr;Y)
        .
        \qedhere
    \end{align*}
\end{proof}

\begin{remark}
    The same calculations can also be done in the ground state setting (see Remark~\ref{remark:gapped-ground-states-spectral-flow}).
    There one does not need to define a~\(\tilde U(s)\) because~\(U(s)\) is already the mapping of the ground state projection.
    Moreover, one does not need to assume decay of correlations because \(\Cov_P(U^* \, U,B)=0\) anyway since~\(U^* \, U = \unit\).
    And the scaling in~\(\norm{V}\) is also better.
    However, as pointed out above, the gap is necessary and thus LPPL of this form only holds if~\(V\) does not close the spectral gap, which is in general only true for small~\(\norm{V}\).
\end{remark}

Furthermore, if we know how to control the correlations for all Gibbs states along the path, we obtain a better scaling in the norm of the perturbation, as is shown in the next proposition.

\begin{theorem}[LPPL from correlations along the path]
    \label{thm:DCStoLPPL}
    Let \(\Lambda\Subset\Z^\nu\), \(H\in\alg_\Lambda\) self-adjoint, \(X\subset \Lambda\), and~\(V\in \alg_X\) self-adjoint.
    For \(s \in [0,1]\), consider the path of Hamiltonians \(H(s):= H + s \, V\) with Gibbs states~\(\rho_\beta(s)\).
    Moreover, assume that all Hamiltonians~\(H(s)\) satisfy a Lieb-Robinson bound with~\(\zeta_\LR\)-decay uniformly in~\(s\) and let~\(\zeta_\QBP\) be the function from Proposition~\ref{prop:qbp-bounds}.
    Then, for all \(Y \subset \Lambda\), \(B \in \alg_Y\) and all~\(r \geq 0\), we have
    \begin{equation*}
        \abs[\big]{
            \trace[\big]{\rho_\beta (0) \, B}
            - \trace[\big]{\rho_\beta (1) \, B}
        }
        \leq
        \beta \, \norm{V} \, \norm{B} \, \paren[\bigg]{
            \sup_{s\in [0,1]} \Cov_{\rho_\beta (s)}(X_r;Y)
            + 2 \, \zeta_\QBP(X,r)} .
    \end{equation*}
\end{theorem}

\begin{remark}[Simplified short-range version]
    \label{remark:DCstoLPPL-short-range}
    Let \(\Psi\) and \(\Psi+V\) be short-range interactions and assume that all Gibbs states \(\rho_\beta(s)\) satisfy exponential decay of correlations with respect to \(\zeta_\DC(r) = C_\DC \, \e^{-c_\DC r}\), \(f_\DC(\abs{Y})=\abs{Y}^m\) and \(n \geq 0\) in the sense of Definition~\ref{def:decay-of-correlations}.
    Then Theorem~\ref{thm:DCStoLPPL} implies that for every \(c<c_0\), with \(c_0\) as in Remark~\ref{remark:DC0toLPPL-short-range}, there exist a constant \(C>0\) such that for all \(Y\subset \Lambda\) and \(B\in \alg_Y\)
    \begin{equation*}
        \abs[\big]{
            \trace[\big]{\rho_\beta (0) \, B}
            - \trace[\big]{\rho_\beta (1) \, B}
        }
        \leq
        C
        \, \beta \, \norm{V}
        \, \norm{B}
        \, \abs{X}^{\max\List{1,n}}
        \, \abs{Y}^m
        \, \e^{-c \dist{X,Y}}
        .
        \qedhere
    \end{equation*}
\end{remark}

\begin{proof}
    We again drop the subscript~\({} _\beta\).
    Integrating~\eqref{eq:ode-gibbs-state}, we obtain
    \begin{align*}
        \Alignindent
        \trace[\big]{\rho(1) \, B}
        - \trace[\big]{\rho(0) \, B}
        \\&=
        -\frac{\beta}{2} \, \int_0^1
        \Trace[\Big]{\anticommutator[\Big]{\rho(s),\Phi^{H(s)}_\beta(V) - \Trace[\big]{\rho(s) \, \Phi^{H(s)}_\beta(V)}} \, B}
        \diff s
        \\&=
        -\frac{\beta}{2} \, \int_0^1 \Cov_{\rho(s)}\paren[\big]{\Phi^{H(s)}_\beta(V),B} + \Cov_{\rho(s)}\paren[\big]{B,\Phi^{H(s)}_\beta(V)} \diff s.
    \end{align*}
    Approximating \(\Phi^{H(s)}_\beta(V)\) in~\(X_r\) using Proposition~\ref{prop:qbp-bounds-generator} and \(\norm{\Phi^{H(s)}_{\beta,r}(V)} \leq \norm{V}\), gives
    \begin{equation*}
        \abs[\big]{
            \trace[\big]{\rho(1) \, B}
            - \trace[\big]{\rho(0) \, B}
        }
        \leq
        \beta \, \norm{V} \, \norm{B}
        \, \paren[\bigg]{
            \int_0^1 \Cov_{\rho(s)}(X_r;Y) \diff s
            + 2 \, \zeta_\QBP(X,r)
        }.
    \end{equation*}
    Bounding the integral with the supremum concludes the proof.
\end{proof}

A result similar to Theorem~\ref{thm:DCStoLPPL} was obtained by \textcite{KGK2014}.
They also start with a differential equation for~\(\rho_\beta\).
But later they need to use not only decay of correlations for all~\(s\), but decay of the generalized covariance for all~\(s\), which, they can only prove in finite-range systems at high temperature.
Instead, Theorem~\ref{thm:DCStoLPPL} can also be applied if one only has information about decay of correlations.
And in Theorem~\ref{thm:DC0toLPPL} it is enough to know decay of correlations in the unperturbed state~\(\rho_\beta(0)\).
See in particular the applications we give in Section~\ref{sec:applications-of-the-general-results}.
In Section~\ref{sec:1D-spin-chains}, we will discuss consequences of the results for translation-invariant one-dimensional spin chains, where decay of correlations is known for finite-range interactions at any temperature~\cite{Araki1969,BCP2022}.

\begin{remark}
    Since the proofs of Theorems~\ref{thm:DC0toLPPL} and~\ref{thm:DCStoLPPL} are mainly based on quantum belief propagation and Lieb-Robinson bounds, which hold true also for fermionic systems, analogous propositions can be stated and proved in a fermionic setting with minor modifications.
    Similarly, the following Theorems~\ref{thm:uniform-lppl-implies-local-indistinguishability} and~\ref{thm:local-indistinguishability-implies-decay-of-correlations} hold for fermionic systems as well.
    In this paper we focus on quantum spin systems for simplicity.
\end{remark}

\section{Local indistinguishability from uniform LPPL}
\label{sec:LIfromULPPL}

In this section we prove that uniform LPPL implies local indistinguishability if also the interaction decays fast enough.
The main idea is to remove \(\Lambda\setminus\Lambda'\) from~\(\Lambda\) point by point, for which we have to assume LPPL at all intermediate steps.
The idea is inspired by~\textcite{BK2019}, who removed point by point a boundary region \(\partial\Lambda' \subset \Lambda\) to decouple the system in~\(\Lambda'\) from the rest.
With our improved method we can improve the scaling with~\(\Lambda'\) and extend the result from finite- to short- and long-range interactions.
The first result will later be used to remove single vertices.

\begin{lemma}
    \label{lem:local-indistinguishability-remove-single-site-short-range}
    Let \(\Lambda\Subset\Z^\nu\), \(F\) be a decay function, and~\(\Psi\) be an interaction such that~\(\norm{\Psi}_{F}<\infty\).
    Moreover, assume that~\(\rho_\beta^\Lambda\) satisfies LPPL with respect to~\(f_\LPPL\), \(g_\LPPL\),\linebreak[2] \(\zeta_\LPPL\), and \(n \geq 0\) as in Definition~\ref{def:LPPL}.
    Then, for any \(X\subset \Lambda\) it holds that
    \begin{equation*}
        \abs[\big]{
            \trace[\big]{
                \rho_\beta^\Lambda \, B
            }
            -\trace[\big]{
                \rho_\beta^{\Lambda\setminus X } \, B
            }
        }
        \leq
        \norm{B}
        \, \abs{X}^n
        \, f_\LPPL\paren[\big]{\abs{Y}}
        \, g\paren[\big]{\abs{X} \, \norm{\Psi}_F}
        \, \zeta\paren[\big]{\dist{Y,X}}
    \end{equation*}
    for all \(Y \subset \Lambda\setminus X\) and \(B\in \alg_Y\), where
    \begin{equation*}
        g(v)
        :=
        \max\List[\big]{
            g_\LPPL(v),
            \paren[\big]{ \e^{2\beta v} -1 }
        }
        \quadtext{and}
        \zeta(r)
        :=
        \min_{0 \leq R \leq r}
        (2R+1)^{n\nu} \, \zeta_\LPPL\paren{r-R}
        + F(R)
        .
    \end{equation*}
\end{lemma}

\begin{proof}
    In the proof we will compare \(\rho_\Lambda := \rho_\beta^\Lambda[H_\Lambda]\) and \(\rho_{\Lambda\setminus X} := \rho_\beta^\Lambda[H_{\Lambda\setminus X}]\) such that we can always use the same trace.
    By the different normalizations in the Gibbs state, it holds that
    \begin{equation*}
        \trace[\big]_\Lambda{\rho_{\Lambda\setminus X} \, B}
        =
        \trace[\big]_\Lambda{\rho_\beta^\Lambda[H_{\Lambda\setminus X}] \, B}
        =
        \trace[\big]_{\Lambda\setminus X}{\rho_\beta^{\Lambda\setminus X}[H_{\Lambda\setminus X}] \, B}
        =
        \trace[\big]_{\Lambda\setminus X}{\rho_\beta^{\Lambda\setminus X} \, B}
    \end{equation*}
    for all \(B\in \alg_{\Lambda\setminus X}\) and the result follows.

    We split the difference \(H_\Lambda - H_{\Lambda\setminus X} = V_1 + V_2\) into a finite-range part and the rest,
    \begin{equation*}
        V_1
        :=
        \sumstack[lr]{Z\subset \Lambda:\\ X \cap Z \neq \emptyset,\\ \diam{Z}\leq R} \Psi(Z)
        \qquadtext{and}
        V_2
        :=
        \sumstack[lr]{Z\subset \Lambda:\\ X \cap Z \neq \emptyset,\\ \diam{Z}> R} \Psi(Z)
    \end{equation*}
    for some~\(R\geq 0\), and use \(\tilde{\rho} := \rho^\Lambda_\beta[H_\Lambda - V_1]\) as an intermediate step.
    The perturbation~\(V_1\) is supported in~\(X_R\) and bounded \(\norm{V_1} \leq \abs{X} \, \norm{\Psi}_{F}\).
    Hence, by LPPL we find
    \begin{equation*}
        \abs[\big]{
            \trace{ \rho_\Lambda \, B }
            - \trace{ \tilde \rho \, B }
        }
        \leq
        \norm{B}
        \, \abs{X_R}^n
        \, f_\LPPL\paren[\big]{\abs{Y}}
        \, g_\LPPL\paren[\big]{\abs{X}\,\norm{\Psi}_{F}}
        \, \zeta_\LPPL\paren[\big]{\dist{Y,X}-R}
    \end{equation*}
    for all~\(B\in \alg_Y\).
    The remaining perturbation~\(V_2\) is small,
    \begin{equation*}
        \norm{V_2}
        \leq
        \sumstack[lr]{Z\subset \Lambda:\\X \cap Z \neq \emptyset,\\ \diam{Z} > R} \norm{\Psi(Z)}
        \leq
        \abs{X} \, \norm{\Psi}_{F} \, F(R),
    \end{equation*}
    and with the bound~\eqref{eq:stability-small-perturbations} from Proposition~\ref{prop:qbp-ode-gibbs-state} and~\eqref{eq:exp(x)-1-bound} it holds that
    \begin{equation*}
        \abs[\big]{
            \trace{ \tilde \rho \, B }
            - \trace{ \rho_{\Lambda\setminus X} \, B }
        }
        \leq
        \norm{B}
        \, \paren[\big]{
            \e^{2\beta \norm{V_2}}-1
        }
        \leq
        \norm{B}
        \, F(R)
        \, \paren[\big]{
            \e^{2\beta \abs{X} \norm{\Psi}_F} -1
        }
        .
    \end{equation*}
    The result follows by triangle inequality.
\end{proof}

The bound provided in Lemma~\ref{lem:local-indistinguishability-remove-single-site-short-range} on its own is not very good because it scales at least exponentially in~\(\abs{X}\).
This might be enough in one-dimensional systems where one needs to remove only a constant number of sites to decouple two halves of a system.
But in general it is more advantageous to remove~\(X\) site by site.
Therefore, we assume that there exists a sequence~\((\Lambda_i)_{i=0}^N\) such that \(\Lambda_0 = \Lambda\) and \(\Lambda_N=\Lambda'\) along which each~\(\rho_\beta^{\Lambda_i}\) satisfies LPPL\@.

\begin{theorem}
    \label{thm:uniform-lppl-implies-local-indistinguishability}
    Let \(\Lambda\Subset\Z^\nu\), \(F\) a decay function, and~\(\Psi\) be an interaction such that~\(\norm{\Psi}_{F}<\infty\).
    Let \(\Lambda'\subset \Lambda\) and assume that there exists a sequence \((x_i)_{i=1}^N \subset \Lambda\setminus\Lambda'\) defining \(\Lambda_i := \Lambda \setminus \Set{x_j \given j=1,\dots,i}\) such that \(\Lambda_N=\Lambda'\) and~\(\rho_\beta^{\Lambda_i}\) satisfies LPPL with respect to~\(f_\LPPL\), \(g_\LPPL\), \(\zeta_\LPPL\), and \(n \geq 0\) as in Definition~\ref{def:LPPL} for every \(i=0,\dotsc,N-1\).

    Then,
    \begin{equation}
        \label{eq:thm-bound-local-indistinguishability-from-uniform-lppl}
        \abs[\big]{
            \trace[\big]{
                \rho_\beta^\Lambda \, B
            }
            -\trace[\big]{
                \rho_\beta^{\Lambda'} \, B
            }
        }
        \leq
        \norm{B}
        \, f_\LPPL\paren[\big]{\abs{Y}}
        \, g\paren[\big]{\norm{\Psi}_{F}}
        \, \sum_{i=1}^N \zeta\paren[\big]{\dist{Y,x_i}}
    \end{equation}
    for all \(Y \subset \Lambda'\) and \(B\in \alg_Y\), where~\(g\) and~\(\zeta\) as in Lemma~\ref{lem:local-indistinguishability-remove-single-site-short-range}.

    Moreover, if~\(\zeta\) decays fast enough, i.e.\ such that \(\tilde \zeta(0)\) defined below converges, one can bound
    \begin{equation}
        \label{eq:prop-local-indistinguishability-from-uniform-lppl-uniform-zeta-tilde}
        \sum_{i=1}^N \zeta\paren[\big]{\dist{Y,x_i}}
        \leq
        \abs{Y}
        \, \tilde\zeta\paren[\big]{\dist{Y,\Lambda\setminus\Lambda'}}
        \quadtext{where}
        \tilde\zeta(r)
        :=
        2^\nu
        \, \sum_{q=r}^\infty q^{\nu-1} \, \zeta(q)
        .
    \end{equation}
\end{theorem}

\begin{remark}[Simplified short-range version]
    Let \(\Psi\) be a short-range interaction on \(\Lambda\Subset \Z^\nu\) such that \(\norm{\Psi}\sr{b}<\infty\) and assume that all Gibbs states \(\rho_\beta^{\Lambda'}\) with \(\Lambda'\subset \Lambda\) satisfy LPPL with respect to \(\zeta_\LPPL(r) = C_\LPPL \, \e^{-c_\LPPL r}\), \(f_\LPPL(\abs{Y})=\abs{Y}^m\), \(g_\LPPL(v) = \e^{3\beta v}\) and \(n \geq 0\) in the sense of Definition~\ref{def:LPPL}.
    Then, for every \(c< b \, c_\LPPL / (b + c_\LPPL)\) there exists \(C>0\) such that for all \(Y\subset \Lambda'\subset \Lambda\) and \(B\in \alg_Y\)
    \begin{equation*}
        \abs[\big]{
            \trace[\big]{
                \rho_\beta^\Lambda \, B
            }
            -\trace[\big]{
                \rho_\beta^{\Lambda'} \, B
            }
        }
        \leq
        C
        \, \norm{B}
        \, \abs{Y}^{m+1}
        \, \e^{3\beta\norm{\Psi}\sr{b}}
        \, \e^{-c \dist{Y,\Lambda\setminus\Lambda'}}
        .
        \qedhere
    \end{equation*}
\end{remark}

\begin{proof}
    Equation~\eqref{eq:thm-bound-local-indistinguishability-from-uniform-lppl} follows from Lemma~\ref{lem:local-indistinguishability-remove-single-site-short-range} and triangle inequality.
    To prove~\eqref{eq:prop-local-indistinguishability-from-uniform-lppl-uniform-zeta-tilde} we split \(\Lambda\setminus\Lambda'\) into shells
    \begin{equation*}
        S_q := \Set[\big]{z\in \Lambda\setminus\Lambda' \given \dist{Y,z}=q}
    \end{equation*}
    which clearly satisfy \(\abs[\big]{S_q} \leq \abs{Y} \, 2^\nu \, q^{\nu-1}\).
    Then, \(\Set{x_i \given i=1,\dotsc,N} = \bigcup_{q=\dist{Y,\Lambda\setminus\Lambda'}}^\infty S_q\) and~\eqref{eq:prop-local-indistinguishability-from-uniform-lppl-uniform-zeta-tilde} follows.
\end{proof}

\begin{figure}
    \centering%
    \begin{tikzpicture}[
            myLine,
            x=\ul,y=\ul,
            set/.style={circle, draw, minimum width=#1},
        ]
        \setlength{\fattening}{9\ul}
        \draw[on background=clip, clip, save path=\pathLambda] (0,-1) rectangle (20,10) node[anchor=north east] {\(\Lambda\)};
        \path[only inside=draw,save path=\pathLambdaPrime] (2,1) rectangle (19,8) node[anchor=north east] {\(\Lambda'\)};
        \node[shape=circle,draw,minimum width=3\ul] (setY) at (7,4.5) {\(Y\)};
        \draw[latex-latex, shorten] (setY) -- node[left] {\(\dist{Y,\Lambda\setminus\Lambda'}\)} (setY |- 17,8);
        \begin{scope}[color=col_1,fill=col_1!20,even odd rule]
            \path[save path=\shellinner] (setY) circle[radius=0.5\fattening-8\lw];
            \path[save path=\shellouter] (setY) circle[radius=.5\fattening];
            \draw[latex-latex, shorten] (setY) -- node[right] {\(\dist{Y,S_q}=q\)} +(+62:.5\fattening-8\lw);
            \draw (setY) +(+62:.5\fattening-8\lw) node[anchor=180+62,circle] {\(S_q\)};
            \clip[use path=\pathLambda\pathLambdaPrime];
            \clip[use path=\shellouter\shellinner\pathLambdaPrime];
            \filldraw[use path=\shellouter\shellinner\pathLambdaPrime, myLine=2];
        \end{scope}
        \begin{scope}[color=col_2]
            \coordinate (x) at ($(setY) + (-70:.5\fattening-4\lw)$);
            \path[on background={shade, inner color=col_2!30!white}] (x) circle[radius=4\ul];
            \node[save path=\pathXR, set=2.5\ul, label={-10:\(X_R=B_x(R)\)}, on background={fill=col_2!30!white}] (setBxR) at (x) {};
            \begin{scope}[even odd rule]
                \clip[use path=\pathXR];
                \clip[use path=\pathLambda\pathLambdaPrime];
                \clip[use path=\shellouter\shellinner\pathLambdaPrime];
                \path[use path=\shellouter\shellinner\pathLambdaPrime, fill=col_1!20!white!70!col_2, draw=col_1!70!col_2, myLine=2];
            \end{scope}
            \path[use path=\pathXR, draw=col_2, myLine=1];
            \draw[latex-latex, shorten] (setY) -- node[right] {\(\dist{X_R,Y} = \dist{x,Y}-R\)} (setBxR);
            \node[fill,minimum width=2\lw,circle,inner sep=0pt, label={-10:\(x\)}] at (x) {};
        \end{scope}
    \end{tikzpicture}
    \caption{
        Depicted is the main idea for the proof of local indistinguishability from uniform LPPL\@.
        The idea is to remove all points \(x\in \Lambda\setminus\Lambda'\) one by one.
        Therefore, we first apply LPPL to the sum of all interaction terms connecting~\(x\) with its \(R\)-neighbourhood~\(B_x(R)\).
        For short-range interactions, the remaining interaction terms including~\(x\) are exponentially small in~\(R\) and can be removed using QBP\@.
        Furthermore, the points~\(x\) are grouped into shells~\(S_q \ni x\) according to their distance \(q:=\dist{x,Y}\) to~\(Y\).
        We then choose the parameter~\(R\) depending on~\(q\), so that the error for operators \(B\in \alg_Y\) introduced by removing all points in~\(S_q\) decays exponentially in~\(q\).
        This allows to sum the error terms introduced by removing all shells with~\(q\) and still obtain exponential decay in the distance~\(\dist{Y,\Lambda\setminus\Lambda'}\).
    }
    \label{fig:proof-local-indistinguishability}
\end{figure}

To recover the result from \textcite{BK2019}, note that they restrict to finite-range interactions and therefore only need to remove enough points along the boundary of~\(\Lambda'\) to decouple the system in~\(\Lambda'\) from the rest.
Then they use the trivial bound \(\abs{\Lambda\setminus\Lambda'} \, \zeta\paren[\big]{\dist{Y,\Lambda\setminus\Lambda'}}\) for the sum in~\eqref{eq:thm-bound-local-indistinguishability-from-uniform-lppl} and thus obtain a linear scaling in~\(\abs{\partial \Lambda'}\).
Our improvement is to observe~\eqref{eq:prop-local-indistinguishability-from-uniform-lppl-uniform-zeta-tilde} such that the statement is independent of~\(\abs{\Lambda\setminus\Lambda'}\) and applicable to short-range interactions.

More specific results can be found in Section~\ref{sec:1D-spin-chains} for one-dimensional spin chains.

\section{Uniform decay of correlations from local indistinguishability}
\label{sec:clustering-from-local-indistinguishability}

In this section we briefly discuss how to close the circle of implications in Figure~\ref{fig:diagram-introduction}, i.e.\ how to conclude decay of correlations from local indistinguishability.
This is a simple and well known consequence for finite-range Hamiltonians.
We present the statement here for short- and long-range Hamiltonians.
We will also discuss in detail the dependence on the support of the observables.

\begin{theorem}
    \label{thm:local-indistinguishability-implies-decay-of-correlations}
    Let \(\Lambda\Subset\Z^\nu\), \(F\) a decay function, and~\(\Psi\) be an interaction such that \(\norm{\Psi}_{F}<\infty\) and assume that~\(\Psi\) satisfies local indistinguishability at inverse temperature~\(\beta\) with respect to~\(f_\LI\) and~\(\zeta_\LI\) in the sense of Definition~\ref{def:local-indistinguishability}.
    Then, for all disjoint \(X\), \(Y\subset \Lambda\),
    \begin{gather*}
        \Cov_{\rho_\beta^\Lambda}(X;Y)
        \leq
        \max\List[\big]{\abs{X},f_\LI\paren[\big]{\abs{X}+\abs{Y}}}
        \, \zeta\paren[\big]{\dist{X,Y}},
    \shortintertext{where}
        \zeta(r)
        :=
        3
        \inf_{0 \leq \llll < r/2} \paren[\Big]{
            \zeta_\LI(\llll)
            + (\e^{2\beta \norm{\Psi}_F}-1 ) \, (2\llll+1)^\nu \, F(r-2\llll)
        }
        .
    \end{gather*}
\end{theorem}

\begin{figure}
    \centering%
    \begin{tikzpicture}[
            myLine,
            x=\ul,y=\ul,
            set/.style={circle, draw, minimum width=#1},
        ]
        \setlength{\fattening}{3\ul}
        \draw[clip,on background={clip}] (0,0) rectangle (20,10) node[anchor=north east] {\(\Lambda\)};
        \node[set=3\ul] (setX) at (5,2) {\(X\)};
        \node[set=3\ul+\fattening, on background={fill=col_1!20}] (setXell) at (setX) {};
        \node[below] at (setXell.north) {\(X_\llll\)};
        \node[set=4\ul] (setY) at (15,5) {\(Y\)};
        \node[set=4\ul+\fattening, on background={fill=col_1!20}] (setYell) at (setY) {};
        \node[below] at (setYell.north) {\(Y_\llll\)};
        \draw[latex-latex, shorten=3] (setXell) -- (setYell) node[pos=0.12,anchor=north west] {\(\dist{X_\llll,Y_\llll}\)};
        \node at (9,7) {\textcolor{col_1}{\(\Lambda'=X_\llll\cup Y_\llll\)}};
    \end{tikzpicture}
    \caption{
        Depicted is the situation from the proof of Theorem~\ref{thm:local-indistinguishability-implies-decay-of-correlations}.
        By local indistinguishability, the covariance of the Gibbs state on~\(\Lambda\) and \(\Lambda'=X_\llll \cup Y_\llll\) are similar.
        The remaining distance~\(\dist{X_\llll,Y_\llll}\) must be chosen so large that the remaining interactions coupling both regions are small.
        In the case of finite-range interactions, the distance must be chosen larger than the interaction range, so that the regions completely decouple.
    }
    \label{fig:proof-decay-of-correlations}
\end{figure}

\begin{remark}[Simplified short-range version]
    Let \(\Lambda\Subset\Z^\nu\) and \(\Psi\) be a short-range interaction such that \(\norm{\Psi}\sr{b}<\infty\) and assume that it satisfies local indistinguishability at inverse temperature \(\beta\) with respect to \(f_\LI\paren[\big]{\abs{Y}} = \abs{Y}^m\) and \(\zeta_\LI(r) = C_\LI \, \e^{-c_\LI r}\).
    Then, for every \(c < b \, c_\LI / (2b + c_\LI)\), there exists \(C>0\) such that for all \(X\), \(Y\subset \Lambda\),
    \begin{equation*}
        \Cov_{\rho_\beta^\Lambda}(X;Y)
        \leq
        C
        \, \paren[\big]{\abs{X}+\abs{Y}}^{\max\List{1,m}}
        \, \e^{-c \dist{X,Y}}
        .\qedhere
    \end{equation*}
\end{remark}

\begin{proof}
    Let \(A\in \alg_X\) and \(B\in \alg_Y\) with unit norm.
    Then choose \(0\leq\llll<\dist{X,Y}/2\) and \(\Lambda' = X_\llll \cup Y_\llll\).
    We first use local indistinguishability to approximate
    \begin{equation}
        \label{eq:LItoUDC1}
        \abs[\big]{
            \Cov_{\rho_\beta^\Lambda}(A,B)
            - \Cov_{\rho_\beta^{\Lambda'}}(A,B)
        }
        \leq
        3
        \, f_\LI\paren[\big]{\abs{X}+\abs{Y}}
        \, \zeta_\LI\paren[\big]{\llll}.
    \end{equation}
    In the case of finite-range interactions with range~\(R\), one could choose~\(\llll\) so that \(\dist{X_\llll,Y_\llll}>R\) for which \(\rho_\beta^{\Lambda'} = \rho_\beta^{X_\llll} \otimes \rho_\beta^{Y_\llll}\) and \(\Cov_{\rho_\beta^{\Lambda'}}(A,B) = 0\).
    For short-range interactions, however, we have to remove the interactions coupling regions~\(X_\llll\) and~\(Y_\llll\) first.
    Therefore, enumerate \(X_\llll = \List{x_1,x_2,\dotsc,x_N}\) and let
    \begin{equation*}
        V_i
        :=
        \sumstack[lr]{Z\subset \Lambda':\\x_i \in Z,\ x_1,\dotsc,x_{i-1} \notin Z,\\ Z \cap Y_\llll \neq \emptyset} \Psi(Z)
        ,\quadtext{which satisfies}
        \norm{V_i}
        \leq
        \sumstack[lr]{Z\subset \Z^\nu:\\x_i\in Z,\\ \diam{Z} \geq \dist{X_\llll,Y_\llll}} \norm{\Psi(Z)}
        \leq
        \norm{\Psi}_F
        \, F\paren[\big]{\dist{X_\llll,Y_\llll}}
        .
    \end{equation*}
    Moreover, \(H_{\Lambda'} - \sum_{i=1}^N V_i = H_{X_\llll} + H_{Y_\llll}\) and by Proposition~\ref{prop:qbp-ode-gibbs-state} and~\eqref{eq:exp(x)-1-bound}
    \begin{equation*}
        \abs[\big]{
            \Cov_{\rho_\beta^{\Lambda'}[H]}(A,B)
            - \Cov_{\rho_\beta^{\Lambda'}[H-V_i]}(A,B)
        }
        \leq
        3 \, F\paren[\big]{\dist{X_\llll,Y_\llll}} \, \paren[\big]{\e^{2\beta\norm{\Psi}_F}-1},
    \end{equation*}
    for all Hamiltonians \(H\in \alg_{\Lambda'}\) and all \(i=1,\dotsc,N\) as long as~\(\dist{X_\llll,Y_\llll} \geq 0\).
    Choosing \(H = H_{\Lambda'} - \sum_{j=1}^{i-1} V_j\), by triangle inequality and vanishing of \(\Cov_{\rho_\beta^\Lambda[H_{X_\llll}+H_{Y_\llll}]}(A,B)\) it follows that
    \begin{equation}
        \label{eq:LItoUDC2}
        \abs[\big]{
            \Cov_{\rho_\beta^{\Lambda'}}(A,B)
        }
        \leq
        3 \, \abs{X} \, (2\llll+1)^\nu \, F\paren[\big]{\dist{X_\llll,Y_\llll}} \, \paren[\big]{\e^{2\beta\norm{\Psi}_F}-1}
        .
    \end{equation}
    Combining~\eqref{eq:LItoUDC1} and~\eqref{eq:LItoUDC2} concludes the proof.
\end{proof}

This statement closes the circle depicted in Figure~\ref{fig:diagram-introduction}.

\begin{remark}
    \label{remark:decay-of-correlations-after-one-circle}
    Notice that the constants get worse in each step, when going around the circle shown in Figure~\ref{fig:diagram-introduction}.
    Indeed, let \(b > 0\) and~\(\Psi\) be an interaction such that \(\norm{\Psi}\sr{b}<\infty\).
    Moreover, let \(\Lambda\Subset\Z^\nu\) and assume that~\(\rho_\beta^{\Lambda}\) satisfies uniform exponential decay of correlations in the sense that there are constants~\(C\), \(c\) and~\(n\) such that
    \begin{equation*}
        \Cov_{\rho_\beta^{\Lambda'}}(X;Y)
        \leq
        C
        \, \paren[\big]{ \abs{X} + \abs{Y} }^n
        \, \e^{-c\dist{X,Y}}
        \qquad\text{for all \(X,Y\subset \Lambda'\subset \Lambda\).}
    \end{equation*}
    Then there exist constants~\(\tilde{C}>C\) and~\(\tilde{c}<c\), which in particular depend on~\(C\), \(c\), \(\beta\) and~\(\norm{\Psi}\sr{b}\), such that after going once through the statements about LPPL and local indistinguishability as indicated in Figure~\ref{fig:diagram-introduction} one obtains
    \begin{equation*}
        \Cov_{\rho_\beta^{\Lambda'}}(X;Y)
        \leq
        \tilde{C}
        \, \paren[\big]{ \abs{X} + \abs{Y} }^{n+1}
        \, \e^{-\tilde{c}\dist{X,Y}}
        \qquad\text{for all \(X,Y\subset \Lambda'\subset \Lambda\).}
        \qedhere
    \end{equation*}
\end{remark}

\section{Stability against small SLT perturbations from decay of correlations}
\label{sec:stabilty-against-SLT-perturbations}

This section is devoted to a result which is not part of the implications depicted in Figure~\ref{fig:diagram-introduction}.
In Proposition~\ref{prop:qbp-ode} we already observed stability of the Gibbs state against small (in norm) perturbations.
Using the idea from Theorem~\ref{thm:DCStoLPPL}, we can extend this to stability against perturbations which are small in an interaction norm.
We call these \emph{sum-of-local-terms (SLT)} perturbations.
Their norm grows like~\(\abs{\Lambda}\).
Hence, if we aim to find a bound uniform in the system size, they are not small in norm.
In contrast to the bound in~\eqref{eq:stability-small-perturbations}, which is in trace norm, we compare local expectation values of the Gibbs states, which is the natural topology for extended systems.

The idea to this observation comes from~\cite{RGK2023} and we give a rigorous proof here.

\begin{theorem}[Stability against small SLT perturbations]
    \label{thm:stability-SLT-perturbations}
    Let \(C_\interaction>0\), \(n\in \N_0\), \(\beta_0>0\) and
    \begin{enumerate}
        \item \label{enum:theorem-stability-SLT-perturbations-short-range}
            \(\alpha_\DC > \nu\), \(b>0\) and \(F(r):=\e^{-br}\) or
        \item \label{enum:theorem-stability-SLT-perturbations-long-range}
            \(\alpha\), \(\alpha_\DC > (n+1) \, \nu\) and \(F(r) = F_\alpha(r) := (1+r)^{-\alpha}\).
    \end{enumerate}

    Then there exists a constant \(C>0\) such that the following holds:
    Let \(\Lambda\Subset\Z^\nu\), \(\beta\in\intervaloo{0,\beta_0}\), \(\Psi_H\) and~\(\Psi_V\) be interactions such that \(\norm{\Psi_H}_{F}\), \(\norm{\Psi_V}_{F}<C_\interaction\) and denote the Gibbs state of \(\Psi_H + s \, \Psi_V\) by~\(\rho_\beta^\Lambda(s)\).
    Assume that all Gibbs states satisfy decay of correlations with respect to \(\zeta_\DC \leq (1+\mathord{\cdot})^{-\alpha_\DC}\), \(f_\DC\) and~\(n\) as in Definition~\ref{def:decay-of-correlations}, then for all \(Y \subset \Lambda\) and \(B \in \alg_Y\) we have
    \begin{equation*}
        \abs[\big]{
            \trace[\big]{\rho_\beta (0) \, B}
            - \trace[\big]{\rho_\beta (1) \, B}
        }
        \leq
        C \, \beta \, \norm{\Psi_V}_{F} \, \abs{Y}
        \, \paren[\Big]{1+f_\DC\paren[\big]{\abs{Y}}}
        \, \norm{B}
        .
    \end{equation*}
\end{theorem}

\begin{remark}
    Concerning temperature dependence, we observe that \(\rho_{\beta+\Delta\beta}^\Lambda[H] = {\rho_{\beta}^\Lambda[H+\frac{\Delta\beta}{\beta}H]}\).
    Hence, Theorem~\ref{thm:stability-SLT-perturbations} implies that, assuming all Gibbs states for inverse temperatures in \(\intervalcc{\beta,\beta+\Delta\beta}\) satisfy decay of correlations (w.r.t.~the functions above), local expectation values change slowly in~\(\beta\), namely
    \begin{equation*}
        \abs[\big]{
            \trace[\big]{\rho^\Lambda_\beta \, B}
            - \trace[\big]{\rho^\Lambda_{\beta+\Delta\beta} \, B}
        }
        \leq
        C \, \Delta\beta \, \norm{\Phi_H}_F \, \abs{Y}
        \, \paren[\Big]{1+f_\DC\paren[\big]{\abs{Y}}}
        \, \norm{B},
    \end{equation*}
    uniformly in~\(\Lambda\) with the constant from above.
    For only this statement, the proof could be simplified since \(\Phi_\beta^{H(s)}(H) = H\) in~\eqref{eq:ode-ebeta} and~\eqref{eq:ode-gibbs-state}.
\end{remark}

\begin{proof}
    As in the proof of Theorem~\ref{thm:DCStoLPPL}, we drop the subscript~\({}_\beta\) and have
    \begin{align*}
        \Alignindent
        \abs[\big]{
            \trace[\big]{\rho(1) \, B}
            - \trace[\big]{\rho(0) \, B}
        }
        \\&\leq
        \frac{\beta}{2} \, \sup_{s\in \intervalcc{0,1}} \paren[\Big]{
            \abs[\big]{\Cov_{\rho(s)}\paren[\big]{\Phi^{H(s)}_\beta(V),B}}
            + \abs[\big]{\Cov_{\rho(s)}\paren[\big]{B,\Phi^{H(s)}_\beta(V)}}
        }
        ,
    \end{align*}
    where \(V=\sum_{Z\subset \Lambda} \Psi_V(Z)\).
    We now bound the first term, as the second is bounded analogously,
    \begin{align*}
        \Alignindent
        \abs[\big]{\Cov_{\rho(s)}\paren[\big]{\Phi^{H(s)}_\beta(V),B}}
        \\&\leq
        \sum_{k=0}^\infty \sumstack{Z\subset \Lambda:\\\dist{Z,Y}=k}
        \abs[\Big]{\Cov_{\rho(s)}\paren[\Big]{\Phi^{H(s)}_\beta\paren[\big]{\Psi_V(Z)},B}}
        \\&\leq
        \begin{aligned}[t]
            &2 \, \norm{B} \sum_{y\in Y} \, \sumstack{Z\subset \Lambda:\\ y\in Z} \norm{\Psi_V(Z)}
            \\&+ \norm{B} \, \sum_{k=1}^\infty \sumstack{Z\subset \Lambda:\\ \dist{Z,Y}=k}
            \norm{\Psi_V(Z)}
            \min_{0 \leq \rrr}
            \paren[\Big]{
                \abs{Z_\rrr}^n
                \, f_\DC\paren[\big]{\abs{Y}}
                \, \zeta_\DC\paren[\big]{\dist{Z_\rrr,Y}}
                + 2 \, \zeta_\QBP(Z,\rrr)
            }
            ,
        \end{aligned}
    \end{align*}
    where we used the approximation \(\Phi^{H(s)}_{\beta,\rrr}\paren[\big]{\Psi_V(Z)}\) from Proposition~\ref{prop:qbp-bounds-generator}, \(\Cov_{\rho(s)}(A,B) \leq 2 \, \norm{A} \, \norm{B}\), and the assumption on decay of correlations.
    The first summand is bounded by \(2 \, \abs{Y} \, \norm{\Psi_V}_{F} \, \norm{B}\).
    For the second summand, we first replace \(\dist{Z_\rrr,Y}=\max\List[\big]{\dist{Z,Y}-\rrr,0}\).
    Then notice that for every \(Z\subset \Lambda\) with \(\dist{Z,Y}=k\), there exist~\(y\in Y\) and~\(z\in Z\) with~\(\dist{z,y}=k\).
    Moreover, under the assumptions~\ref{enum:theorem-stability-SLT-perturbations-short-range} and~\ref{enum:theorem-stability-SLT-perturbations-long-range}, \(\zeta_\QBP(Z,\rrr) = \abs{Z} \, \zeta_\QBP(\rrr)\) scales linearly in \(\abs{Z}\) for some decaying \(\zeta_\QBP(\rrr)\), by Lemmata~\ref{lem:zeta-QBP-short-range} and~\ref{lem:zeta-QBP-long-range}, respectively.
    Hence, by overcounting each of the terms with~\(k \geq 1\), the second summand is bounded by
    \begin{align*}
        \sum_{y\in Y}
        \sumstack{z\in \Lambda\colon\\\dist{z,y}=k}
        \sumstack{Z\subset \Lambda\colon\\z\in Z}
        \norm{\Psi_V(Z)}
        \, \min_{0 \leq \rrr \leq k}
        \paren[\Big]{
            \abs{Z}^n
            \, (1+2\rrr)^{\nu n}
            \, f_\DC\paren[\big]{\abs{Y}}
            \, \zeta_\DC\paren[\big]{k-\rrr}
            + 2 \, \abs{Z} \, \zeta_\QBP(\rrr)
        }
        \\\leq
        \abs{Y}
        \, 2^\nu \, k^{\nu-1}
        \, C_1
        \, \min_{0 \leq \rrr \leq k}
        \paren[\Big]{
            (1+2\rrr)^{\nu n}
            \, f_\DC\paren[\big]{\abs{Y}}
            \, \zeta_\DC\paren[\big]{k-\rrr}
            + 2 \, \abs{Z} \, \zeta_\QBP(\rrr)
        }
        ,
    \end{align*}
    where
    \begin{equation*}
        C_1
        :=
        \sup_{z\in Z} \, \sumstack{Z\subset \Lambda\mathpunct{:}\\z\in Z} \abs{Z}^{\max\List{1,n}} \, \norm{\Psi_V(Z)}
        <
        C_1' \, \norm{\Psi_V}_F
    \end{equation*}
    and we used \(\abs[\big]{\Set{z\in \Lambda \given \dist{z,y}=k}} \leq 2^\nu k^{\nu-1}\) and \(\abs{Z_\rrr} \leq \abs{Z} \, (1+2\rrr)^\nu\).
    Defining
    \begin{equation*}
        C
        :=
        2
        + 2^\nu
        \, C_1'
        \, \sum_{k=1}^\infty
        \, k^{\nu-1}
        \, \min_{0 \leq \rrr \leq k}
        \paren[\Big]{
            \, (1+2\rrr)^{\nu n}
            \, \zeta_\DC\paren[\big]{k-\rrr}
            + 2 \, \zeta_\QBP(\rrr)
        },
    \end{equation*}
    the bound given in the statement follows when~\(C<\infty\).

    For the case of short-range interactions with assumption~\ref{enum:theorem-stability-SLT-perturbations-short-range}, \(\zeta_\QBP(\rrr) \leq 6 \, \e^{-c_\QBP \, \rrr}\) for some \(c_\QBP > 0\), which can be chosen uniformly in \(\beta\in \intervaloo{0,\beta_0}\) and \(\norm{\Psi_H+s\,\Psi_V}_F \leq 2\,C_\interaction\) by Lemma~\ref{lem:zeta-QBP-short-range-along-the-path}.
    To bound \(C_1\), we use \(\abs{Z} \leq (\diam{Z}+1)^\nu\) to find \(C_1' \leq \max_{r\in \N} (r+1)^{n \nu} \, \e^{-b r} < \infty\), which only depends on \(b\) and~\(n\).
    Finally, \(C\) is bounded if \(\zeta_\DC(k) \leq (1+k)^{-q}\) with~\(q>\nu\), by choosing \(\rrr=k^{(q-\nu)/(2\nu n)}/2\).

    For long-range interactions, \(C_1' \leq 1\) and for every \(\alpha_\QBP<\alpha\) and \(\beta_0\) there exist \(C_\QBP\) such that \(\zeta_\QBP(\rrr) \leq C_\QBP \, F_{\alpha_\QBP}(\rrr)\) by Lemma~\ref{lem:zeta-QBP-long-range-along-the-path}.
    Thus, \(C\) can be bounded after choosing \(\rrr=k/2\) under the assumptions specified in~\ref{enum:theorem-stability-SLT-perturbations-long-range}.
\end{proof}

\section{Results for one-dimensional short-range systems}
\label{sec:1D-spin-chains}

We now restrict our attention to translation-invariant one-dimensional spin chains.

For every \(x \in \Z\), \(n \in \N\) consider \(Y=[x,x+n] \subset \Z\) and for every \(A \in \alg_{\{x\}}\) define \(t_n(A) \in \alg_{\{x+n\}}\) by \(t_n(A)=A \otimes \unit_{Y\setminus \{x+n\}}\), where we made use of the canonical identification of \(\alg_{Y\setminus\{x+n\}}\) as a subalgebra of~\(\alg_{Y}\).
Then, let \(I \Subset \Z\) and consider the set \(I+n =\Set{x \in \Z \, \given \, x-n \in I}\).
Let \(Y \subset \Z\) such that \(I, I+n \subset Y\).
Define the algebra \(*\)-isomorphism \(t^I_n\colon \alg_{I}\to \alg_{I+n}\) by \(
    t_n^I(\otimes_{i \in I}A_i)
    =
    \otimes_{i \in I} t_n(A_{i}) \otimes \unit_{Y\setminus (I+n)}
\).
The \(*\)-isomorphisms \(t_n^I\) induce a \(*\)-algebra automorphism \(\tau_n\) of the algebra of quasi-local observables~\(\alg_{\Z}\).
Analogously, one can define \(\tau_n\) for \(-n \in \N\).
The family \(\{\tau_n\}_{n \in \Z}\) is called the family of lattice translations.
Translation-invariant interactions are interactions that satisfy the additional constraint that for all \(X\Subset \Z\) and \(n \in \Z\)
\begin{equation*}
    \tau_n \, \Psi (X) = \Psi (X+n)
    .
\end{equation*}

In this section we assume~\(\nu=1\) and consider only local translation-invariant interactions on~\(\Z\).
For every finite interval \(I=[x,y] \Subset \Z\) and every inverse temperature~\(\beta>0\), we additionally denote the Gibbs state as a functional \(\varphi^{[x,y]}_\beta\colon \alg_{[x,y]} \to \C\) on the operator algebra, defined by
\begin{equation*}
    \varphi^{[x,y]}_\beta(Q)
    :=
    \Trace[\big]{\rho_\beta^{[x,y]}Q}
    \quadtext{for all}
    Q \in \alg_{[x,y]}
\end{equation*}
and extended to a state on~\(\alg_{\Z}\), e.g.\ by using the Hahn-Banach Theorem as in~\cite[Proposition 2.3.24]{BR1979}.
We denote by~\(\varphi_{\beta}\) the unique KMS state over the infinite chain at inverse temperature~\(\beta>0\)~\cite{Araki1975}.
Following the discussion in e.g.~\cite[Proposition 6.2.15]{BR1981}, for every increasing and absorbing sequence \(I_{n} \nearrow \Z\), the sequence of states~\(\varphi_\beta^{I_{n}}\) is \LTskip{weak\(^\ast\)-}convergent to~\(\varphi_{\beta}\), in particular
\begin{equation*}
    \varphi_{\beta}(Q)
    =
    \lim_{n\to\infty} \varphi_\beta^{I_{n}}(Q)
    \quadtext{for all}
    Q \in \alg_{\Z}
    .
\end{equation*}
In the infinite-chain regime, we define the covariance to measure correlations as an extension of equation~\eqref{eq:definition-covariance} by
\begin{equation*}
    \Cov_{\rho_\beta}(A,B)
    :=
    \varphi_\beta(A\,B)
    - \varphi_\beta(A) \, \varphi_\beta(B)
    ,
\end{equation*}
for all \(A\), \(B \in \alg_{\Z}\).

\Textcite{Araki1969} proved that a translation-invariant, finite-range interaction satisfies \emph{exponential decay of correlations}, and this was recently extended to short-range interactions in~\cite{PP2023}.
That is, there exist constants~\(\mathcal{K}\), \(\alpha > 0\) such that for every~\(x\in \Z\), \(k \in \N\), \(A \in \alg_{\intervaloc{-\infty, x}}\) and \(B \in \alg_{\intervalco{x+k,\infty}}\),
\begin{equation*}
    \abs[\big]{
        \varphi_{\beta}(A \, B)
        - \varphi_{\beta}(A) \, \varphi_{\beta}(B)
    }
    \leq
    \mathcal{K} \, \e^{- \alpha \, k} \, \norm{A} \, \norm{B}
    ,
\end{equation*}
whenever \(\beta < \beta^*\), where the precise form of~\(\beta^*\) is given in Theorem~\ref{thm:DC-short-range-interactions-1D}.
For finite-range interactions, the result was extended to the finite-chain regime in~\cite{BCP2022}, where it was proven that any condition of uniform decay of correlations in the infinite-chain can be transferred to the finite-chain, and vice versa.
It is natural to ask whether this is also correct in the presence of exponentially decaying interactions.
And one of the main results of this section is a positive answer to this question.
The precise formulation of this theorem has already appeared in Section~\ref{sec:results-one-dimensional-translation-invariant-short-range}, but we restate it here for convenience with a new notation.
To simplify notation, we will drop the union sign \(XY := X\cup Y\) from unions of finite intervals~\(X\),~\(Y\subset \Z\), and we will write hereafter \(XY\) whenever \(X<Y\) and \(YX\) for~\(Y<X\).
And since we only deal with short-range interactions, we abbreviate
\begin{equation*}
    \norm{\Psi}_{b}
    :=
    \norm{\Psi}\sr{b}, \qquad b\geq 0
    .
\end{equation*}

We restate Theorem~\ref{thm:DC-short-range-interactions-1D-informal} with this notation.
Its proof is deferred to Section~\ref{sec:exp_decaying_interactions_1D}.

\bgroup
\renewcommand{\thetheorem}{\ref*{thm:DC-short-range-interactions-1D-informal}}
\NextLinkTarget{theorem.6.restated}

\begin{theorem}
    \label{thm:DC-short-range-interactions-1D}
    Let~\(b>0\) and~\(\Psi\) be a translation-invariant interaction such that \({\norm{\Psi}_{b} < \infty}\) and denote \(\beta^* :=b/\paren{2 \, \norm{\Psi}_{0}}\).
    Then, for all \(\beta \in \intervaloo{0 , \beta^*}\), the Gibbs state satisfies decay of correlations in the sense that there exist~\(C_\DC\), \(c_\DC > 0\) such that for every finite interval \(I \subset \Z\) and subintervals \(X, Y \subset I\), it holds that
    \begin{equation*}
        \Cov_{\rho^I_\beta}(X;Y)
        \leq
        C_\DC \, \e^{- c_\DC \dist{X,Y}}
        .
    \end{equation*}
\end{theorem}

\addtocounter{theorem}{-1}
\egroup

Note that, for finite-range interactions, we have \(\beta^* = \infty\), recovering the results of~\cite{Araki1969,BCP2022}.
Additionally, in~\cite{BCP2022}, it was shown that having exponential decay of correlations for finite-range interactions is equivalent to local indistinguishability.
This is extended below to exponentially decaying interactions as well.
Following the lines of the equivalence of notions of locality and decay of correlations presented in Figure~\ref{fig:diagram-introduction}, we first show that exponential decay of correlations implies LPPL, and subsequently prove that the latter implies local indistinguishability.
This is reflected in Figure~\ref{fig:diagram-1D}.
The use of quantum belief propagation and Lieb-Robinson bounds is again pivotal to derive these results.

\begin{figure}
    \centering%
    \setlength{\fboxsep}{0pt}%
    \begin{tikzpicture}[
            myLine,
            x=\ul,y=\ul,
        ]
        \begin{scope}[
                every node/.append style={concept},
                every label/.append style={label distance=-\lw,formula},
            ]
            \node[
                name=Corr,
                label={[name=Corr expl,anchor=north]below:\(
                    \Cov_{\rho_\beta^{I}}(X;Y)
                    \leq
                    C \, \e^{-a\dist{X,Y}}
                    \)
                },
            ] at (0,0) {\footnotesize{finite chain}\\decay of correlations};
            \node[
                name=LPPL,
                label={[name=LPPL expl]below:\(
                        \begin{aligned}
                            &
                            \abs[\big]{
                                \trace[\big]{\rho_\beta^{I}[H]\,B}
                                -\trace[\big]{\rho_\beta^{I}[H+V]\,B}
                            }
                            \\&\quad \leq
                            C \, \norm{B} \, \e^{c\beta\norm{V}} \, \e^{-a\dist{X,Y}}
                        \end{aligned}
                    \)
                },
            ] at (8,-4.5) {local perturbations perturb locally};
            \node[
                name=Indi,
                label={[name=Indi expl]below:\(
                        \begin{aligned}
                            &
                            \abs[\big]{
                                \trace[\big]{\rho_\beta^{I}\,B}
                                -\trace[\big]{\rho_\beta^{I'}\,B}
                            }
                            \\&\quad \leq
                            C \, \norm{B} \, \e^{-a \dist{X,Y}}
                        \end{aligned}
                    \)
                },
            ] at (-8,-4.5) {local indistinguishability};
            \node[
                name=InfSpinChain,
                label={[name=InfSpinChain expl]below:\(
                    \Cov_{\rho_\beta}(X;Y)
                    \leq
                    C \, \e^{-a\dist{X,Y}}
                    \)
                },
            ] at (0,6) {\footnotesize{infinite chain}\\decay of correlations};
            \node[name=Araki,align=center,formula=gray] at (0,11) {\(\nu=1\), \(\beta<\beta^*\),\\translation-invariant Hamiltonian};
        \end{scope}
        \path (Araki) edge[imply] node[right, align=left, font=\footnotesize] {\textcite{Araki1969}\\\textcite{PP2023}} (InfSpinChain);
        \draw[imply, transform canvas={xshift=-1\ul}] (InfSpinChain expl) -- node[left,font=\footnotesize] {Theorem~\ref{thm:DC-short-range-interactions-1D}} (Corr);
        \draw[imply, transform canvas={xshift=1\ul}] (Corr) -- node[right] {} (InfSpinChain expl);
        \draw[imply] (Corr) -| node[right,anchor=north west,font=\footnotesize] {Theorem~\ref{thm:LPPL-1D}} (LPPL);
        \draw[imply] (LPPL) -- node[below,font=\footnotesize] {Theorem~\ref{thm:LI-1D}} (Indi);
        \draw[imply] (Indi) |- node[left, anchor=north east, font=\footnotesize] {Theorem~\ref{thm:local-indistinguishability-implies-decay-of-correlations_1D}} (Corr);
    \end{tikzpicture}
    \caption{
        The diagram shows the main implications for one-dimensional (translation-invariant) spin chains, which are discussed in this section.
        Here, \(I \subset \Z\) is a finite interval, \(X\subset I\) a subinterval and~\(Y\subset I\) a union of two intervals.
        In particular, we show “equivalence” of the four concepts in the picture.
        Note that the constants are not the same, and we refer to the Theorems for precise statements.
        A crucial ingredient in all the implications is quantum belief propagation (QBP) coupled with Lieb-Robinson bounds.
        For finite-range or exponentially decaying interactions, exponential decay of correlations is known to hold by earlier results for the infinite-chain regime at every positive or high enough temperature, respectively, for which all four properties are thus satisfied.
    }
    \label{fig:diagram-1D}
\end{figure}

In contrast to the results at high temperature, Theorem~\ref{thm:DC-short-range-interactions-1D} only provides decay of correlations between two operators each supported on an interval.
As a consequence, we will only prove LPPL for perturbations~\(V\) supported in an interval \(X\subset I\) and observables \(B = B_1 \otimes B_2\) with~\(B_1\) and~\(B_2\) supported on intervals~\(Y_1\) and~\(Y_2\), respectively, where \(Y_1 < X < Y_2\) as in Figure~\ref{fig:1D-spin-chain}.
Thus, we use slightly different definition for the three concepts decay of correlations, LPPL, and local indistinguishability and adjust some of the arguments.
One could extend all to arbitrary operators in \(\alg_{Y_1 \cup Y_2}\) by using the Schmidt decomposition and allowing for an exponential growth in~\(\abs{Y_1 \cup Y_2}\).

\begin{theorem}
    \label{thm:LPPL-1D}
    For every~\(b\), \(C_\interaction\), \(\beta\), \(C_\DC\) and~\(c_\DC>0\), there exist constants \(C_\LPPL\) and~\(c_\LPPL>0\) such that the following holds:
    Let~\(\Psi\) be an interaction on~\(\Z\) satisfying \({\norm{\Psi}_b \leq C_\interaction}\), \(I\Subset \Z\) be a finite interval, and assume that the corresponding Gibbs state \(\rho^I_\beta\) satisfies decay of correlations in the sense given in Theorem~\ref{thm:DC-short-range-interactions-1D} with constants~\(C_\DC\) and~\(c_\DC\).
    Then the Gibbs state satisfies LPPL in the sense that for all subintervals \(X,Y_1,Y_2\subset I\), such that \(Y_1 < X < Y_2\), \(Y = Y_1 \cup Y_2\), all self-adjoint perturbations \(V \in \alg_X\) and all observables~\(B = B_1 \otimes B_2\) with \(B_1\in \alg_{Y_1}\) and \(B_2\in \alg_{Y_2}\), it holds true that
    \begin{equation*}
        \abs[\big]{
            \Trace[\big]{\rho_\beta^{I}[H_I] \, B}
            - \Trace[\big]{\rho_\beta^{I}[H_I+V] \, B}
        }
        \leq
        C_\LPPL \, \norm{B} \, \e^{3 \beta \norm{V}} \, \paren[\big]{1+\norm{V}} \, \e^{- c_\LPPL d(X,Y)}
        .
    \end{equation*}
    If also \(\norm{\Psi + V}_b \leq C_\interaction\), one can drop the factor \(\paren[\big]{1+\norm{V}}\).
\end{theorem}

The geometry described in the statement is depicted in Figure~\ref{fig:1D-spin-chain}.
In Theorem~\ref{thm:LPPL-1D} included is the case where~\(Y_i\) is empty and \(B_i = \unit\) which corresponds to~\(Y\) being a single interval.
Together with Theorem~\ref{thm:DC-short-range-interactions-1D}, Theorem~\ref{thm:LPPL-1D} shows that LPPL holds for local, translation-invariant interactions~\(\Psi\) satisfying \(\norm{\Psi}_b<\infty\) at inverse temperatures \(\beta < \beta^* := b / (2 \, \norm{\Psi}_0)\).

\begin{figure}
    \begin{center}
        \begin{tikzpicture}[
                myLine,
                x=\ul,y=\ul,
            ]
            \Block[0]{25}{white}{I}
            \Block[6]{8}[4\lw]{col_1!60!white}{X_l}[]
            \node[anchor=north east,text=col_1!50] at (X_l.south east) {\(X_\rrr\)};
            \Block[8]{4}{col_1}{X}
            \Block[1]{3}{col_3}{Y_1}
            \Block[16]{6}{col_3}{Y_2}

            \node[above,color=col_3] at (Y_1.north) {\(B_1\)};
            \node[above,color=col_3] at (Y_2.north) {\(B_2\)};
            \node[above,color=col_1] at (X.north) {\(V\)};
        \end{tikzpicture}
        \caption{
            Representation of an interval~\(I\) with subintervals \(X,Y_1,Y_2 \subset I\).
            An example of a perturbation \(V \in \alg_X\) such that the distance between~\(X_\rrr\) and~\(Y = Y_1 \cup Y_2\) is at least~\(\rrr\).
            Here, \(\rrr = \floor{d(X,Y)/2} = 2\).
        }
        \label{fig:1D-spin-chain}
    \end{center}
\end{figure}

\begin{proof}
    Let us denote \(\rrr := \floor{d(X,Y)/2}\) and abbreviate \(\rho := \rho_\beta^{I}[H_I]\) and~\(\tilde{\rho} := \rho_\beta^{I}[H_I+V]\).
    From the last line in the proof of Theorem~\ref{thm:DC0toLPPL} we obtain
    \begin{equation}
        \label{eq:proof_1D_1}
        \abs[\big]{
            \Trace[\big]{\rho \, B }
            - \Trace[\big]{\tilde{\rho} \, B }
        }
        \leq
        \e^{2 \beta \norm{V}} \, \paren[\Big]{
            \,\sup_{\substack{W\in \alg_{X_\rrr}\mathpunct{:}\\\norm{W}=1}}\abs[\big]{\Cov_{\rho}(W,B)}
            + 4 \, \beta \, \norm{V} \, \norm{B} \, \zeta_{\QBP}(X,\rrr)
        }
        ,
    \end{equation}
    in this setting, where we write the covariance with a supremum only in the first argument, to use the product structure of \(B = B_1 \otimes B_2 = B_1 \, B_2\) to obtain
    \begin{equation*}
        \Cov_{\rho}(W, B_1 \, B_2)
        =
        \Cov_{\rho}(W\,B_1,B_2)
        + \trace{\rho \, B_2} \, \Cov_{\rho}(W,B_1)
        - \trace{\rho \, W} \, \Cov_{\rho}(B_1,B_2)
        .
    \end{equation*}
    By Theorem~\ref{thm:DC-short-range-interactions-1D}, there exist constants~\(C_\DC\) and~\(c_\DC\) only depending on~\(\norm{\Psi}_b\) and~\(\beta\) such that
    \begin{equation*}
        \abs[\big]{
            \Cov_{\rho}(W, B_1 \, B_2)
        }
        \leq
        3 \, C_\DC \, \norm{W} \, \norm{B_1} \, \norm{B_2} \, \e^{- c_\DC d(X_\rrr,Y) }.
    \end{equation*}
    And since \(B = B_1 \otimes B_2\) is a tensor product, \(\norm{B_1} \, \norm{B_2} = \norm{B}\), such that
    \begin{equation}
        \label{eq:proof_1D_2}
        \sup_{\substack{W\in \alg_{X_\rrr}\mathpunct{:}\,\norm{W}=1}} \abs[\big]{\Cov_{\rho}(W,B)}
        \leq
        3 \, C_\DC \, \norm{B} \, \e^{-\alpha \dist{X,Y}/2}
        .
    \end{equation}
    For the decay function \(\zeta_\QBP\) we use Lemma~\ref{lem:zeta-QBP-short-range} for intervals~\(X\).
    Thus, for every \(b\), \(C_\interaction\), \(\beta\), \(C_\DC\) and~\(c_\DC\), there exists \(C_\QBP\) and \(c_\QBP\) such that
    \begin{equation}
        \label{eq:proof-1D-zeta-QBP}
        \zeta_{\QBP} (X,\rrr)
        \leq
        C_\QBP \, \paren[\big]{1+\norm{V}}^\gamma \, \e^{- c_\QBP \rrr}
        ,
    \end{equation}
    with \(\gamma=0\) if \(\norm{\Psi+V}_b \leq C_\interaction\) and \(\gamma=1\) otherwise.
    Replacing now~\eqref{eq:proof_1D_2} and~\eqref{eq:proof-1D-zeta-QBP} into~\eqref{eq:proof_1D_1}, we obtain
    \begin{align*}
        \Alignindent
        \abs[\big]{
            \Trace[\big]{\rho \, B }
            - \Trace[\big]{\tilde{\rho} \, B }
        }
        \\&\leq
        \e^{2 \beta \norm{V}}
        \,
        \norm{B}
        \, \paren[\Big]{
            3 \, C_\DC \, \e^{- c_\DC d(X,Y)/2 }
            + 4 \, \beta \, \norm{V} \, C_\QBP \, \e^{- c_\QBP \paren[\big]{d(X,Y)/2 - 1} }
        }
        \\&\leq
        C_\LPPL \, \e^{3 \beta \norm{V}} \, \norm{B} \, \e^{- c_\LPPL d(X,Y) } ,
    \end{align*}
    where \(c_\LPPL := \min \List{c_\DC, c_\QBP}/2\) and \(C_\LPPL := 3 \, C_\DC + 4 \, C_\QBP \, \e^{c_\QBP}\), and we have used that~\(x \leq \e^{x}\).
\end{proof}

Next, we show that this slightly restricted version of LPPL also implies local indistinguishability in one-dimensional spin chains.
Together with Theorems~\ref{thm:DC-short-range-interactions-1D} and~\ref{thm:LPPL-1D}, Theorem~\ref{thm:LI-1D} indeed gives that local indistinguishability holds for local, translation-invariant interactions~\(\Psi\) satisfying \(\norm{\Psi}_b<\infty\) at inverse temperatures \(\beta < \beta^* := b / (2 \, \norm{\Psi}_0)\).

\begin{theorem}
    \label{thm:LI-1D}
    For every \(b\), \(\beta\), \(C_\LPPL\)~and~\(c_\LPPL>0\), there exist constants \(C_\LI\) and~\(c_\LI>0\) such that the following holds:
    Let~\(\Psi\) be an interaction on~\(\Z\) satisfying \({\norm{\Psi}_b<\infty}\), \(I\Subset \Z\) be a finite interval, and assume that the corresponding Gibbs state \(\rho^I_\beta\) satisfies LPPL in the sense given in Theorem~\ref{thm:LPPL-1D} with constants~\(C_\LPPL\) and~\(c_\LPPL\).
    Then the Gibbs state satisfies local indistinguishability in the sense that for all subintervals \(X,Y_1,Y_2\subset I\), so that \(Y_1 < X < Y_2\), \(Y = Y_1 \cup Y_2\), and all observables~\(B = B_1 \otimes B_2\) with \(B_1\in \alg_{Y_1}\) and \(B_2\in \alg_{Y_2}\), it holds true that
    \begin{equation*}
        \abs[\big]{
            \trace[\big]{\rho_\beta^I \, B}
            - \trace[\big]{\rho_\beta^{I \setminus X} \, B}
        }
        \leq
        C_\LI \, \norm{B} \, \e^{- c_\LI d(Y,X)}
        .
    \end{equation*}
\end{theorem}

\begin{proof}
    Let us first explain the conceptual difference to the proof of local indistinguishability in Section~\ref{sec:LIfromULPPL}:
    Since we do not have uniform decay of correlations, we cannot remove~\(X\) site by site.
    And looking back at Lemma~\ref{lem:local-indistinguishability-remove-single-site-short-range} we should not remove~\(X\) in one step, because that leads to an exponential scaling in~\(\abs{X}\).
    Instead, here we only remove the interactions between~\(X\) and~\(I\setminus X\), which leads to an exponential scaling in~\(\abs{\partial X} = 2\).

    If \(X \cap Y \neq \emptyset\) the statement is trivial, otherwise we have
    \begin{equation*}
        \Trace[\big]{\rho_\beta^{I\setminus X} B}
        =
        \Trace[\big]{\rho_\beta^{I\setminus X} \otimes \rho_\beta^X \, B}
        =
        \Trace[\big]{\rho_\beta^{I}[H_{I \setminus X} + H_{X}] \, B}
    \end{equation*}
    for all~\(B \in \alg_Y\).
    Thus, we can apply Theorem~\ref{thm:LPPL-1D} with the perturbation given by \(V:= H_{I} - H_{I\setminus X}-H_X\), i.e.~all interaction terms whose support intersects~\(X\) and~\(I\setminus X\).
    Following the idea of Lemma~\ref{lem:local-indistinguishability-remove-single-site-short-range}, we split~\(V\) into
    \begin{equation*}
        V_0
        :=
        \sumstack[lr]{
            Z \subset I:
            \\ (I \setminus X) \cap Z \neq \emptyset
            \\ X \cap Z \neq \emptyset
            \\ \diam{Z}\leq \RRR
        }
        \Psi (Z)
        \qquadtext{and}
        V'
        :=
        \sumstack[lr]{
            Z \subset I:
            \\ (I \setminus X) \cap Z \neq \emptyset
            \\ X \cap Z \neq \emptyset
            \\ \diam{Z}> \RRR
        }
        \Psi (Z)
        ,
    \end{equation*}
    for some~\(\RRR\in \N\) to be chosen later.
    Since each~\(Z\) in the sum of~\(V_0\) needs to contain a site in \(\Set{x\in X \given \dist{x,I \setminus X} \leq \RRR}\) we have \(\norm{V_0} \leq \abs{\partial X} \, \RRR \, \norm{\Psi}_{0}\) and~\(V_0\in \alg_{X_\RRR}\).
    Thus, by Theorem~\ref{thm:LPPL-1D} we find
    \begin{align*}
        \abs[\big]{
            \Trace[\big]{\rho_\beta^{I} \, B}
            - \Trace[\big]{\rho_\beta^{I}[H_I - V_0] \, B}
        }
        &\leq
        C_\LPPL \, \norm{B} \, \e^{6\beta \norm{\Psi}_0 \, \RRR}
        \, \e^{-c_\LPPL \paren[\big]{d(X,Y)-\RRR}}
        \\&\leq
        C_\LPPL \, \norm{B} \, \e^{-c_\LPPL \dist{X,Y}/2},
    \end{align*}
    by choosing \(\alpha := c_\LPPL \, (6\beta \norm{\Psi}_0 + c_\LPPL)^{-1} / 2\) and \(\RRR := \floor{\alpha \, \dist{X,Y} }\).
    Then, denoting \(q_n := \Set{x\in X \given \dist{x,I\setminus X}=n}\) and \(Q_n := \bigcup_{i=1}^{n-1} q_i\) we split \(V' = \sum_{n=1}^\infty V_n\) with
    \begin{equation*}
        V_n
        :=
        \sumstack[lr]{
            Z\subset I \setminus Q_n\mathpunct{:}\\
            Z \cap q_n \neq \emptyset\\
            Z\cap (I \setminus X) \neq \emptyset\\
            \diam{Z}>\RRR
        }
        \Psi(Z),
    \end{equation*}
    where the sum is actually finite.
    Then, \(\norm{V_n} \leq 2 \, \norm{\Psi}_b \, \e^{-b \max\List{\RRR+1,n}}\).
    Using Proposition~\ref{prop:qbp-ode-gibbs-state} and equation~\eqref{eq:exp(x)-1-bound}, we thus find
    \begin{equation*}
        \abs[\big]{
            \Trace[\big]{\rho_\beta^{I}[H_I - {\textstyle\sum_{i=0}^{n-1} V_i}] \, B}
            - \Trace[\big]{\rho_\beta^{I}[H_I - {\textstyle\sum_{i=0}^{n} V_i}] \, B}
        }
        \leq
        \norm{B}
        \, \e^{-b \max\List{\RRR+1,n}}
        \, \paren[\big]{\e^{4\beta\norm{\Psi}_b}-1}
        .
    \end{equation*}
    Hence, by triangle inequality, \(\sum_{n=\RRR+1}^\infty \e^{-bn} < \e^{-b\RRR}/b\), \(\sup_{\RRR>0} \RRR \, \e^{-b\RRR/2} \leq 2/(b\,\e)\) and \({2/\e+1}<2\) we obtain
    \begin{equation*}
        \abs[\big]{
            \Trace[\big]{\rho_\beta^{I}[H_I - V_0] \, B}
            - \Trace[\big]{\rho_\beta^{I}[H_{I\setminus X} + H_X] \, B}
        }
        \leq
        2 \, b^{-1}
        \, \norm{B}
        \, \e^{-b\RRR/2}
        \, \paren[\big]{\e^{4\beta\norm{\Psi}_b}-1}
        .
    \end{equation*}
    Again by triangle inequality,
    \begin{align*}
        \abs[\big]{
            \Trace[\big]{\rho_\beta^{I} \, B}
            - \Trace[\big]{\rho_\beta^{I\setminus X} \, B}
        }
        & \leq
        C_\LPPL \, \norm{B} \, \e^{-c_\LPPL \dist{X,Y}/2}
        + 2 \, b^{-1}
        \, \norm{B}
        \, \e^{-b\RRR/2}
        \, \paren[\big]{\e^{4\beta\norm{\Psi}_b}-1}
        \\& \leq
        C_\LI \, \norm{B} \, \e^{-c_\LI\,d(Y,X)} ,
    \end{align*}
    with \(C_\LI := C_\LPPL + 2 \, b^{-1} \, \paren[\big]{\e^{4\beta\norm{\Psi}_b}-1} \, \e^{-b/2}\) and \(c_\LI := \min \List{c_\LPPL/2, \alpha\,b/2}\).
\end{proof}

To conclude the circle in Figure~\ref{fig:diagram-1D}, we would need to show that local indistinguishability implies decay of correlations.
This is completely analogous to Theorem~\ref{thm:local-indistinguishability-implies-decay-of-correlations} for any dimension at high enough temperature.

\begin{theorem}
    \label{thm:local-indistinguishability-implies-decay-of-correlations_1D}
    For every \(b\), \(\beta\), \(C_\LI\) and~\(c_\LI>0\), there exist constants~\(C_\DC\) and~\({c_\DC>0}\) such that the following holds:
    Let~\(\Psi\) be an interaction on~\(\Z\) satisfying \({\norm{\Psi}_b<\infty}\),\linebreak[2] \(I\Subset \Z\) be a finite interval, and assume that the corresponding Gibbs state \(\rho^I_\beta\) satisfies local indistinguishability in the sense given in Theorem~\ref{thm:LI-1D} with constants~\(C_\LI\) and~\(c_\LI\).
    Then, the Gibbs state satisfies decay of correlations in the sense that for all disjoint intervals \(X,Y\subset I\), and all \(A \in \alg_X\) and \(B \in \alg_Y\), it holds true that
    \begin{equation}
        \label{eq:thm-DC-from-LI-1D}
        \Cov_{\rho_\beta^I}(X;Y)
        \leq
        C_\DC \, \e^{-c_\DC \dist{X,Y}}.
    \end{equation}
\end{theorem}

\begin{proof}
    Without loss of generality we assume \(X<Y\) and extend them to the boundary of~\(I\) such that \(I = X \chi Y\) where also \(\chi\subset I\) is an interval and~\(X\), \(Y\) and~\(\chi\) are pairwise disjoint.
    This puts us in a situation, where we can proceed almost as in Theorem~\ref{thm:local-indistinguishability-implies-decay-of-correlations}, but with the restricted version of local indistinguishability where we can only remove one interval.

    Take \(A \in \alg_X\) and \(B \in \alg_Y\), let \(\rrr < d(X,Y)/2\) to be chosen later and consider \(I':= X_\rrr \cup Y_\rrr\).
    We first use local indistinguishability to cut out a part between~\(X\) and~\(Y\)
    \begin{equation*}
        \abs[\big]{
            \Cov_{\rho_\beta^I}(A,B)
            - \Cov_{\rho_\beta^{I'}}(A,B)}
        \leq
        3 \, C_\LI \, \norm{A} \, \norm{B} \, \e^{-c_\LI \rrr}
        .
    \end{equation*}
    Then we remove the remaining interactions coupling the regions~\(X_\rrr\) and~\(Y_\rrr\) as in the proof of Theorem~\ref{thm:local-indistinguishability-implies-decay-of-correlations} with a slight modification:
    We choose the enumeration \(X_\rrr = \List{x_1,x_2,\dotsc,x_N}\) decreasing such that \(\norm{V_i} \leq \norm{\Phi}_b \, \e^{-b\paren[\big]{\dist{X_\rrr,Y_\rrr}+(i-1)}}\), then we can resum as in the proof of Theorem~\ref{thm:LI-1D} to obtain a result independent of~\(\abs{X}\):
    \begin{equation*}
        \abs[\big]{\Cov_{\rho_\beta^{I'}}(A,B)}
        \leq
        3 \, \e^b \, b^{-1} \, \norm{A} \, \norm{B} \, \e^{-b \dist{X_\rrr,Y_\rrr}} \, \paren[\big]{\e^{2\beta\norm{\Psi}_b}-1}
        .
    \end{equation*}
    Choosing \(\alpha := b \, (c_\LI + 2b)^{-1}\) and \(\rrr := \floor{\alpha \, \dist{X,Y}}\) gives the bound in~\eqref{eq:thm-DC-from-LI-1D} with \(c_\DC := b \, c_\LI \, (c_\LI + 2b)^{-1}\) and \(C_\DC := 3 \, C_\LI \, \e^{c_\LI} + 3 \, \e^b \, b^{-1} \, \paren[\big]{\e^{2\beta\norm{\Psi}_b}-1}\).
\end{proof}

\medskip

There are various reasons for studying the case of one-dimensional spin chains separately.
On one hand, all results from this section present the obvious advantage with respect to those from Section~\ref{sec:results-high-temperature} in the range of~\(\beta\) for which they hold, since the~\(\beta^*\) in this case reduces to~\(\infty\) for super-exponentially decaying interactions, as opposed to the case of high dimensions.
However, they have the drawback that one needs to assume translation invariance for this to be true.
This is a direct consequence of the regimes where correlations are known to decay exponentially fast in one- and higher-dimensions, respectively.
Other cases in which correlations are known to decay with slower rates for one-dimensional systems, such as those of short-range and long-range interactions at every positive temperature, are discussed in Sections~\ref{sec:results-one-dimensional-short-range} and~\ref{sec:results-one-dimensional-long-range}, respectively.

On the other hand, the study of decay of correlations, with different measures than that given by the covariance, in one-dimensional spin chains with translation-invariant, finite-range interactions, has been incredibly fruitful in the past few years.
Given a finite interval~\(I \subset \Z\), with \(X,Y \subset I\), \(X \cap Y \neq \emptyset\), a state~\(\rho\) on~\(I\), and denoting \(\rho_Z := \trace_{I\setminus Z}{\rho}\) for \(Z \subset I\), some other quantities of relevance in this context are for example the \emph{mutual information}, given by \(I_{\rho} (X,Y) := \trace[\big]{ \rho_{XY} \, (\log \rho_{XY} - \log \rho_X \otimes \rho_Y ) }\), and the \emph{mixing condition}, given by \(\norm[\big]{ \rho_{XY}^{} \, \rho_X^{-1} \otimes \rho_Y^{-1} - \unit_{XY}}\).
It is not difficult to show, see~\cite[Section~3.1]{BCP2022}, that, for any state~\(\rho\) on~\(I\) such that~\(\rho_{XY}\) is full-rank, the following holds:
\begin{equation*}
    \tfrac{1}{2} \Cov_\rho (X;Y)^2 \leq I_\rho (X,Y)
    \leq
    \norm[\big]{ \rho_{XY}^{} \, \rho_X^{-1} \otimes \rho_Y^{-1} - \unit_{XY}}
    .
\end{equation*}
Thus, a~\(\zeta\)-decay with~\(d(X,Y)\) for the mixing condition implies the same for the mutual information and a \(\sqrt{2\zeta}\)-decay for the covariance.
Interestingly, in one-dimensional spin chains with translation-invariant, finite-range interactions, a converse is proven for the Gibbs state \({\rho=\rho^I_\beta}\), and all the latter conditions are shown to have equivalent decays.
We expect that a similar result can be derived in the short-range regime.
See~\cite{BCP2023} for an analogous result in high dimensions, at high-enough temperature.

\subsection{Exponential decay of correlations for short-range interactions}
\label{sec:exp_decaying_interactions_1D}

This subsection is devoted to the proof of Theorem~\ref{thm:DC-short-range-interactions-1D}.
The procedure we will follow is very similar to that of~\cite[Theorem 6.2]{BCP2022}.
First, note that~\(I\) can be written as \(I=Z_1 X Z_2 Y Z_3\) for certain intervals \(Z_1, Z_2, Z_3 \subset I\).
Without loss of generality, let us assume for this proof that both~\(Z_1\) and~\(Z_3\) are empty so that we only have to prove the result for the case in which~\(I= X Z Y\).
If not, we enlarge~\(X\) and~\(Y\), which will only allow for more observables~\(A\) and~\(B\) and yield the same bound.

For the rest of the section, we fix~\(b>0\), a translation-invariant interaction~\(\Psi\) satisfying \(\norm{\Psi}_b < \infty\) and \(\beta^* := b/(2\,\norm{\Psi}_0)\).

We need to rephrase some results from~\cite{PP2023}, where the authors use a different interaction norm%
\footnote{%
    While the symbol is the same, this is obviously a different norm than the one used for long-range interactions in Section~\ref{sec:results-one-dimensional-long-range}.
}
\begin{equation*}
    \tnorm{\Psi}_\lambda
    :=
    \sum_{n=0}^\infty \, \e^{\lambda \, n} \, \sup_{z\in \Z} \, \sumstack[lr]{Z\Subset\Z \mathpunct{:}\\z\in Z,\\\diam{Z}\geq n}
    \, \norm{\Psi(Z)}
    .
\end{equation*}
It can be upper bounded with our norm for every~\(\lambda < b\), i.e.
\begin{equation*}
    \tnorm{\Psi}_\lambda
    \leq
    \sum_{n=0}^\infty \, \e^{\lambda \, n}
    \, \norm{\Psi}_b \, \e^{-b \, n}
    \leq
    \frac{1}{1-\e^{\lambda-b}} \, \norm{\Psi}_b
    ,
\end{equation*}
such that \(\norm{\Psi}_b < \infty\) implies \(\tnorm{\Psi}_\lambda < \infty\) for all~\(\lambda < b\).
The statements we use from~\cite{PP2023} all hold for \(\beta<\beta^*_\lambda := \lambda / (2 \, \norm{\Psi}_0)\) if~\(\tnorm{\Psi}_\lambda < \infty\).
With the above observation, they thus also hold for all~\(\beta<\beta^*\).

Next, for \(\beta>0\), \(a \in \Z\), and~\(p\),~\(q \in \N_0\), we define the expansional
\begin{equation*}
    E_{a,p,q}^\beta
    =
    \e^{- \beta H_{[a-p, a+1+q]}} \, \e^{ \beta H_{[a-p, a]}} \, \e^{ \beta H_{[a+1, a+1+q]}}
    .
\end{equation*}
Then we extract the following Lemma from~\cite[Corollary~3.3 and section~4.1]{PP2023}.

\begin{lemma}
    Let~\(\beta < \beta^*\).
    Then there exist constants~\(\mathcal{G}>1\) and~\(\delta>0\) such that for all \(a \in \Z\), and \(p,q \geq 0\),
    \begin{equation*}
        \norm[\big]{ E_{a,p,q}^\beta }, \norm[\big]{(E_{a,p,q}^\beta)^{-1}}
        \leq
        \mathcal{G}
        ,
    \end{equation*}
    and for all \(q' \geq q\),
    \begin{equation*}
        \norm[\big]{ E_{a,p,q}^\beta - E_{a,p,q'}^\beta}
        \leq
        \mathcal{G} \e^{- \delta q}
        .
    \end{equation*}
\end{lemma}

For \(V, W \Subset \Z\), we introduce the slightly more general expansionals
\begin{equation*}
    E_{V,W}^\beta
    :=
    \e^{- \beta H_{VW}} \, \e^{ \beta H_{V}} \, \e^{ \beta H_{W}}
    .
\end{equation*}
As a consequence of the previous lemma, we can provide the following bounds for these expansionals.

\begin{lemma}
    \label{lem:bound_expansional}
    Let~\(\beta < \beta^*\).
    Then there exist constants \(\mathcal{G}>1\) and~\(\delta>0\) such that for all disjoint intervals \(V,W \Subset \Z\),
    \begin{equation*}
        \norm[\big]{ E_{V,W}^\beta }, \norm[\big]{(E_{V,W}^\beta)^{-1}}
        \leq
        \mathcal{G}
        ,
    \end{equation*}
    and if we append intervals~\(\tilde{V}\) and~\(\tilde{W} \Subset \Z\) to~\(V\) and~\(W\)\!, respectively, it holds that
    \begin{equation*}
        \norm[\big]{ E_{V,W}^\beta - E_{\tilde{V}V,W \tilde{W}}^\beta}
        \leq
        \mathcal{G} \, \e^{- \delta q}
        ,
    \end{equation*}
    as long as \(\abs{V}\),~\(\abs{W} \geq q\).
\end{lemma}

Now, let us enlarge the finite interval \(I = XZY \subset \Z\) to \(\tilde{I}= \tilde{X}XZY \tilde{Y}\) as in Figure~\ref{fig:1D-spin-chain-enlarged}.
The appended systems~\(\tilde{X}\) and~\(\tilde{Y}\) are split into two subsystems each, denoted \(\tilde{X}=X_2 X_1\) and~\(\tilde{Y}=Y_1 Y_2\).
Let us further assume, without loss of generality, that \(\abs{X}\),~\({\abs{Y} \geq \abs{Z}}\), otherwise we can again redefine \(X\), \(Y\) and \(Z\) by moving the left third of~\(Z\) into~\(X\) and the right third of~\(Z\) into~\(Y\), reducing~\(\abs{Z}\) to~\(1/3\) of its original value.
Finally, we also choose \(\abs{X_1}\), \(\abs{X_2}\), \(\abs{Y_1}\), \(\abs{Y_2} > \abs{Z}\).

\begin{figure}
    \begin{center}
        \begin{tikzpicture}[
                myLine,
                x=\ul,y=\ul,
            ]
            \Block[0]{2}{white}{X3}[]
            \Block{3}{col_1!40!white}{X_2}
            \Block{3}{col_1!70!white}{X_1}
            \Block{4}{col_1}{X}
            \Block{3}{col_3!70!white}{Z}
            \Block{5}{col_2}{Y}
            \Block{3}{col_2!70!white}{Y_1}
            \Block{3}{col_2!40!white}{Y_2}
            \Block{2}{white}{Y3}[]
            \draw[brc=col_1!70!white,yshift=-8\lw-\baselineskip] (X_2.south west |- 0,0) -- (X_1.south east |- 0,0) node {\(\tilde{X}\)};
            \draw[brc=col_3!70!white,yshift=-8\lw-\baselineskip] (Y_1.south west |- 0,0) -- (Y_2.south east |- 0,0) node {\(\tilde{Y}\)};
        \end{tikzpicture}
        \caption{
            Splitting of an enlarged interval~\(\tilde{I}\) of an original interval~\(I\) with \(X,Y \subset I\) and \(d(X,Y)>0\), to which we have appended regions~\(\tilde{X}\) and~\(\tilde{Y}\) on the left and the right side, respectively, which are subsequently divided in two subregions each.
        }
        \label{fig:1D-spin-chain-enlarged}
    \end{center}
\end{figure}

This brings us in position of proving exponential decay of correlations in this setting for which we now also fix \(\beta<\beta^*\) and the constants~\(\mathcal{G}\) and~\(\delta\) from Lemma~\ref{lem:bound_expansional}.
We define
\begin{equation*}
    E_{\tilde{X},XZY,\tilde{Y}}^{-\beta / 2}
    :=
    \e^{\frac{\beta}{2} H_{\tilde{X}XZY\tilde{Y}}}
    \, \e^{-\frac{\beta}{2} H_{\tilde{X}}-\frac{\beta}{2} H_{XZY \tstrut}-\frac{\beta}{2} H_{\tilde{Y}}}
    ,
\end{equation*}
which can be easily shown to coincide with
\begin{equation*}
    E_{\tilde{X}XZY,\tilde{Y}}^{-\beta / 2} \, E_{\tilde{X},XZY}^{-\beta / 2}
    .
\end{equation*}
Then, by Lemma~\ref{lem:bound_expansional}
\begin{align*}
    \Alignindent
    \norm[\big]{
        E_{\tilde{X},XZY,\tilde{Y}}^{-\beta / 2}
        - E_{X_1,X \tstrut}^{-\beta / 2}
        \, E_{Y,Y_1 \tstrut}^{-\beta / 2}
    }
    \\&=
    \norm[\big]{
        E_{\tilde{X}XZY,\tilde{Y}}^{-\beta / 2}
        \, E_{\tilde{X},XZY}^{-\beta / 2}
        - E_{X_1,X \tstrut}^{-\beta / 2}
        \, E_{Y,Y_1 \tstrut}^{-\beta / 2}
    }
    \\&\leq
    \norm[\big]{
        E_{\tilde{X}XZY,\tilde{Y}}^{-\beta / 2}
        \, E_{\tilde{X},XZY}^{-\beta / 2}
        - E_{Y,Y_1 \tstrut}^{-\beta / 2}
        \, E_{\tilde{X},XZY}^{-\beta / 2}
    }
    + \norm[\big]{
        E_{Y,Y_1 \tstrut}^{-\beta / 2}
        \, E_{\tilde{X},XZY}^{-\beta / 2}
        - E_{X_1,X \tstrut}^{-\beta / 2}
        \, E_{Y,Y_1 \tstrut}^{-\beta / 2}
    }
    \\&\leq
    \mathcal{G} \, \norm[\big]{
        E_{\tilde{X}XZY,\tilde{Y}}^{-\beta / 2}
        - E_{Y,Y_1 \tstrut}^{-\beta / 2}
    }
    + \mathcal{G} \, \norm[\big]{
        E_{\tilde{X},XZY}^{-\beta / 2}
        - E_{X_1,X \tstrut}^{-\beta / 2}
    }
    \\&\leq
    2 \, \mathcal{G} \, \e^{-\delta d(X,Y)}
    .
\end{align*}
Next, following the calculations of~\cite[Section~6]{BCP2022}, and recalling that \(A\in \alg_X\) and \(B\in \alg_Y\), it is not difficult to realize that
\begin{equation*}
    \varphi^{I}_\beta (A \, B)
    =
    \frac{
        \varphi^{\tilde{I}}_\beta\paren[\Big]{
            \paren[\big]{E_{\tilde{X},XZY,\tilde{Y}}^{-\beta / 2}}
            \, A \, B
            \, \paren[\big]{E_{\tilde{X},XZY,\tilde{Y}}^{-\beta / 2}}^*
        }
    }{
        \varphi^{\tilde{I}}_\beta\paren[\Big]{
            \paren[\big]{E_{\tilde{X},XZY,\tilde{Y}}^{-\beta / 2}}
            \, \paren[\big]{E_{\tilde{X},XZY,\tilde{Y}}^{-\beta / 2}}^*
        }
    }
    .
\end{equation*}
In a slight abuse of notation, let us consider the map
\begin{equation*}
    \begin{array}{ccc}
        \alg_I & \rightarrow & \alg_I \\
        Q & \mapsto & \Trace_I {\rho^I_\beta \, Q}
    \end{array}
\end{equation*}
for the unnormalized partial trace in~\(I\)
\begin{equation*}
    \begin{array}{cccc}
        \Tr_I\colon & \alg_J & \rightarrow & \alg_J \\
        & R \otimes S & \mapsto & \Trace{R} (\unit_I \otimes S)
    \end{array}
\end{equation*}
for any \(I \subset J\), and any \(R \in \alg_I\), \(S \in \alg_{J \setminus I}\).
Then, it is contractive as a consequence of the Russo-Dye theorem, and using this as well as Lemma~\ref{lem:bound_expansional}, we see that
\begin{gather*}
    \mathcal{G}^{-4}
    \leq
    \varphi^{\tilde{I}}_\beta \paren[\Big]{
        \paren[\big]{E_{\tilde{X},XZY,\tilde{Y}}^{-\beta / 2}}
        \, \paren[\big]{E_{\tilde{X},XZY,\tilde{Y}}^{-\beta / 2}}^*
    }
    \shortintertext{and}
    \varphi^{\tilde{I}}_\beta \paren[\Big]{
        \paren[\big]{E_{X_1,X}^{-\beta / 2}}
        \, \paren[\big]{E_{X_1,X}^{-\beta / 2}}^*
        \, \paren[\big]{E_{Y,Y_1}^{-\beta / 2}}
        \, \paren[\big]{E_{Y,Y_1}^{-\beta / 2}}^*
    }
    \leq
    \mathcal{G}^4
    .
\end{gather*}
This is fundamental for proving that~\(\varphi^I_\beta(A \, B)\) can be approximated by
\begin{equation*}
    \frac{
        \varphi^{\tilde{I}}_\beta \paren[\Big]{
            \paren[\big]{E_{X_1,X}^{-\beta / 2}}
            \, A
            \, \paren[\big]{E_{X_1,X}^{-\beta / 2}}^*
            \, \paren[\big]{E_{Y,Y_1}^{-\beta / 2}}
            \, B
            \, \paren[\big]{E_{Y,Y_1}^{-\beta / 2}}^*
        }
    }{
        \varphi^{\tilde{I}}_\beta \paren[\Big]{
            \paren[\big]{E_{X_1,X}^{-\beta / 2}}
            \, \paren[\big]{E_{X_1,X}^{-\beta / 2}}^*
            \, \paren[\big]{E_{Y,Y_1}^{-\beta / 2}}
            \, \paren[\big]{E_{Y,Y_1}^{-\beta / 2}}^*
        }
    }
\end{equation*}
up to an error that decays exponentially with the distance between~\(X\) and~\(Y\)\!.
Indeed, using the following standard inequality for scalars~\(a\), \(a\mathrlap{{}'}\), \(b\), \(b' \in \C\)
\begin{equation}
    \label{eq:inequality_scalars}
    \abs[\Big]{\frac{a}{b} - \frac{a'}{b'}}
    \leq
    \frac{1}{\abs{b}} \, \abs{a - a'}
    + \frac{\abs{a'}}{\abs{b} \, \abs{b'}} \, \abs{b-b'}
    ,
\end{equation}
we can prove
\begin{align}
    \label{eq:1D-approx-finite-Gibbs-state-AB}
    \Alignindent
    \abs*{
        \varphi^I_\beta (A \, B)
        - \frac{
            \varphi^{\tilde{I}}_\beta \paren[\Big]{
                \paren[\big]{E_{X_1,X}^{-\beta / 2}}
                \, A
                \, \paren[\big]{E_{X_1,X}^{-\beta / 2}}^*
                \, \paren[\big]{E_{Y,Y_1}^{-\beta / 2}}
                \, B
                \, \paren[\big]{E_{Y,Y_1}^{-\beta / 2}}^*
            }
        }{
            \varphi^{\tilde{I}}_\beta \paren[\Big]{
                \paren[\big]{E_{X_1,X}^{-\beta / 2}}
                \, \paren[\big]{E_{X_1,X}^{-\beta / 2}}^*
                \, \paren[\big]{E_{Y,Y_1}^{-\beta / 2}}
                \, \paren[\big]{E_{Y,Y_1}^{-\beta / 2}}^*
            }
        }
    } \nonumber
    \\&\leq
    \begin{aligned}[t]
        &
        \mathcal{G}^{4} \, \norm[\Big]{
            \paren[\big]{E_{\tilde{X},XZY,\tilde{Y}}^{-\beta / 2}}
            \, A
            \, B
            \, \paren[\big]{E_{\tilde{X},XZY,\tilde{Y}}^{-\beta / 2}}^*
            - \paren[\big]{E_{X_1,X}^{-\beta / 2}}
            \, A
            \, \paren[\big]{E_{X_1,X}^{-\beta / 2}}^*
            \, \paren[\big]{E_{Y,Y_1}^{-\beta / 2}}
            \, B
            \, \paren[\big]{E_{Y,Y_1}^{-\beta / 2}}^*}
        \\&+
        \mathcal{G}^{12}
        \, \norm{A}
        \, \norm{B}
        \, \norm[\Big]{
            \paren[\big]{E_{\tilde{X},XZY,\tilde{Y}}^{-\beta / 2}}
            \, \paren[\big]{E_{\tilde{X},XZY,\tilde{Y}}^{-\beta / 2}}^*
            - \paren[\big]{E_{X_1,X}^{-\beta / 2}}
            \, \paren[\big]{E_{X_1,X}^{-\beta / 2}}^*
            \, \paren[\big]{E_{Y,Y_1}^{-\beta / 2}}
            \, \paren[\big]{E_{Y,Y_1}^{-\beta / 2}}^*
        }
    \end{aligned}
    \nonumber
    \\&\leq
    \tilde{\mathcal{G}} \, \norm{A} \, \norm{B} \, \e^{-\delta d(X,Y) }
    ,
\end{align}
for \(\tilde{\mathcal{G}} >1\) a constant.
Since we have approximated \(\varphi^I_\beta(A \, B)\) by an expression that is independent of~\(X_2\) and~\(Y_2\), we can take a limit with \(\abs{X_2}\),~\(\abs{Y_2} \to \infty\).
Thus, \(\varphi^I_\beta(A \, B)\) can be also approximated by
\begin{equation*}
    \frac{
        \varphi_\beta \paren[\Big]{
            \paren[\big]{E_{X_1,X}^{-\beta / 2}}
            \, A
            \, \paren[\big]{E_{X_1,X}^{-\beta / 2}}^*
            \, \paren[\big]{E_{Y,Y_1}^{-\beta / 2}}
            \, B
            \, \paren[\big]{E_{Y,Y_1}^{-\beta / 2}}^*
        }
    }{
        \varphi_\beta \paren[\Big]{
            \paren[\big]{E_{X_1,X}^{-\beta / 2}}
            \, \paren[\big]{E_{X_1,X}^{-\beta / 2}}^*
            \, \paren[\big]{E_{Y,Y_1}^{-\beta / 2}}
            \, \paren[\big]{E_{Y,Y_1}^{-\beta / 2}}^*
        }
    }
\end{equation*}
keeping the same error as in the finite chain case.
Now we can apply decay of correlations for the infinite chain as proven in~\cite{PP2023}.

\begin{theorem}[{\cite[Theorem~4.4]{PP2023}}]
    Let~\(\beta<\beta^*\).
    Then there exist constants \(\mathcal{G}>1\) and~\(\delta>0\) such that for all \(a,b\in \Z\) with~\(a<b\), \(A\in \alg_{\intervaloc{-\infty,a}}\) and \(B\in \alg_{\intervalco{b,\infty}}\) it holds that
    \begin{equation*}
        \abs[\big]{
            \varphi_\beta(A \, B)
            - \varphi_\beta(A) \, \varphi_\beta(B)
        }
        \leq
        \norm{A} \, \norm{B} \, \mathcal{G} \, \e^{-\delta\abs{b-a}}.
    \end{equation*}
\end{theorem}

Thus, together with~\eqref{eq:inequality_scalars} and similar inequalities to the ones above for the infinite chain, we can find \(\mathcal{K}>1\) and~\(\gamma>0\) such that
\begin{equation}
    \label{eq:approx_infinite_Gibbs_state_by_parts}
    \norm*{
        \text{\footnotesize
            \(\displaystyle
                \begin{aligned}
                    &
                    \frac{
                        \varphi_\beta \paren[\Big]{
                            \paren[\big]{E_{X_1,X}^{-\beta / 2}}
                            \, A
                            \, \paren[\big]{E_{X_1,X}^{-\beta / 2}}^*
                            \, \paren[\big]{E_{Y,Y_1}^{-\beta / 2}}
                            \, B
                            \, \paren[\big]{E_{Y,Y_1}^{-\beta / 2}}^*
                        }
                    }{
                        \varphi_\beta \paren[\Big]{
                            \paren[\big]{E_{X_1,X}^{-\beta / 2}}
                            \, \paren[\big]{E_{X_1,X}^{-\beta / 2}}^*
                            \, \paren[\big]{E_{Y,Y_1}^{-\beta / 2}}
                            \, \paren[\big]{E_{Y,Y_1}^{-\beta / 2}}^*
                        }
                    }
                    \\&\quad- \frac{
                        \varphi_\beta \paren[\Big]{
                            \paren[\big]{E_{X_1,X}^{-\beta / 2}}
                            \, A
                            \, \paren[\big]{E_{X_1,X}^{-\beta / 2}}^*
                        }
                    }{
                        \varphi_\beta \paren[\Big]{
                            \paren[\big]{E_{X_1,X}^{-\beta / 2}}
                            \, \paren[\big]{E_{X_1,X}^{-\beta / 2}}^*
                        }
                    }
                    \frac{
                        \varphi_\beta \paren[\Big]{
                            \paren[\big]{E_{Y,Y_1}^{-\beta / 2}}
                            \, B
                            \, \paren[\big]{E_{Y,Y_1}^{-\beta / 2}}^*
                        }
                    }{
                        \varphi_\beta \paren[\Big]{
                            \paren[\big]{E_{Y,Y_1}^{-\beta / 2}}
                            \, \paren[\big]{E_{Y,Y_1}^{-\beta / 2}}^*
                        }
                    }
                \end{aligned}
            \)
        }
    }
    \leq
    \mathcal{K} \, \norm{A} \, \norm{B} \, \e^{-\gamma \abs{Z}}
    ,
\end{equation}
so that together with~\eqref{eq:1D-approx-finite-Gibbs-state-AB} the second summand in the norm is an approximation of~\(\varphi_\beta^I(A \, B)\).
By choosing~\(A=\unit\) we also obtain
\begin{equation*}
    \abs*{
        \varphi^I_\beta (A)
        - \frac{
            \varphi_\beta \paren[\Big]{
                \paren[\big]{E_{X_1,X}^{-\beta / 2}}
                \, A
                \, \paren[\big]{E_{X_1,X}^{-\beta / 2}}^*
            }
        }{
            \varphi_\beta \paren[\Big]{
                \paren[\big]{E_{X_1,X}^{-\beta / 2}}
                \, \paren[\big]{E_{X_1,X}^{-\beta / 2}}^*
            }
        }
    }
    \leq
    \mathcal{\tilde{K}} \, \norm{A} \, \norm{B} \, \e^{-\tilde{\gamma} \abs{Z}}
\end{equation*}
from~\eqref{eq:1D-approx-finite-Gibbs-state-AB} and~\eqref{eq:approx_infinite_Gibbs_state_by_parts}, and an analogous statement also for the expectation value of~\(B\) with sets~\(Y,Y_1\).
Combining these three approximations, we conclude that there exist~\(C_\DC>1\) and~\(c_\DC>0\) such that
\begin{equation*}
    \abs[\big]{
        \varphi_\beta^I(A \, B)
        - \varphi_\beta^I(A) \, \varphi_\beta^I(B)
    }
    \leq
    C_\DC \, \norm{A} \, \norm{B} \, \e^{ - c_\DC d(X,Y) }
    ,
\end{equation*}
thus concluding the result.

\section{Properties of quantum belief propagation}
\label{sec:proof-qbp}

In this section we review the proof of the quantum belief propagation for the Gibbs state.
The central point is to analyse how the Gibbs state of a certain Hamiltonian is modified when the system is perturbed.

\subsection{Differential equations}

We begin with the discussion of Proposition~\ref{prop:qbp-ode} where no underlying lattice structure is necessary.
Therefore, let~\(\Hi\) be a finite-dimensional Hilbert space and~\(H\) and~\(V\) be two self-adjoint operators on~\(\Hi\).
Furthermore, let \(H(s) := H + s \, V\).

\subsubsection{Differential equation for the exponential\texorpdfstring{ \(\ebeta{H(s)}\)}{}}
\label{sec:proof-ode-ebeta}

The basic idea is to write a differential equation for~\(\ebeta{H(s)}\).
\Textcite{Hastings2007} proves a version where
\begin{equation*}
    \odv*{\ebeta{H(s)}}{s}
    =
    A \, \ebeta{H(s)} + \ebeta{H(s)} \, A^*
\end{equation*}
for some quasi-local (in the case of an underlying lattice structure)~\(A\), while later works use an anti-commutator form
\begin{equation*}
    \odv*{\ebeta{H(s)}}{s}
    =
    -\frac{\beta}{2} \, \anticommutator[\Big]{\ebeta{H(s)},\Phi^{H(s)}_\beta(V)}
\end{equation*}
which we prove in the following.

We mostly follow the proof from \textcite{Hastings2007} but more explicitly work in the energy eigenbasis of~\(H(s)\).
Let \(\List[\big]{\psi_a(s)}_{a}\) be the eigenbasis of~\(H(s)\) such that \(H(s) = \sum_{a} E_a(s) \, \ket{\psi_a(s)}\bra{\psi_a(s)}\).
Then we can also write~\(V\) in this basis as
\begin{equation*}
    V
    =
    \sum_{a,b} V_{a,b}(s) \, \ket{\psi_a(s)}\bra{\psi_b(s)}.
\end{equation*}
Using Duhamel’s formula we find
\begin{align*}
    \odv*{\ebeta{H(s)}}{s}
    &=
    -\beta \int_0^1 \ebeta{\tau H(s)} \, V \, \ebeta{(1-\tau) H(s)} \diff\tau
    \\&=
    -\beta \sum_{a,b} \int_0^1 \ebeta{\tau H(s)} \, V_{a,b}(s) \, \ket{\psi_a(s)}\bra{\psi_b(s)} \, \ebeta[\beta\tau]{H(s)} \diff\tau \, \ebeta{H(s)}
    \\&=
    -\beta \sum_{a,b} V_{a,b}(s) \, \int_0^1 \e^{\beta \tau \Delta E_{a,b}(s)} \diff\tau \, \ket{\psi_a(s)}\bra{\psi_b(s)} \, \ebeta{H(s)}
    \\&=
    -\beta \sum_{a,b} V_{a,b}(s) \, \paren[\big]{1+\e^{\beta \Delta E_{a,b}(s)}}^{-1} \, \int_0^1 \e^{\beta \tau \Delta E_{a,b}(s)} \diff\tau \, \anticommutator[\big]{\ebeta{H(s)}, \ket{\psi_a(s)}\bra{\psi_b(s)}}
    \\&=
    -\frac{\beta}{2} \, \anticommutator[\Big]{\ebeta{H(s)},\Phi^{H(s)}_\beta(V)},
\end{align*}
with \(\Delta E_{a,b}(s) := E_b(s) - E_a(s)\) and
\begin{equation*}
    \Phi^{H(s)}_\beta(V)
    :=
    \sum_{a,b}\fh_\beta\paren[\big]{\Delta E_{a,b}(s)} \, V_{a,b}(s) \, \ket{\psi_a(s)}\bra{\psi_b(s)}
    =
    \int_{-\infty}^\infty f_\beta(t) \, \e^{-\I t H(s)} \, V \, \e^{\I t H(s)} \diff t
    ,
\end{equation*}
where
\begin{equation*}
    \fh_\beta(\omega)
    :=
    2 \, \paren[\big]{1+\e^{\beta\omega}}^{-1} \, \int_0^1 \e^{\beta \tau \omega} \diff\tau
    =
    \begin{dcases}
        \frac{2}{\beta\omega} \, \frac{\e^{\beta\omega}-1}{\e^{\beta\omega}+1}
        & \omega \neq 0\\
        1 & \omega = 0
    \end{dcases}
    =
    \frac{\tanh \frac{\beta\omega}{2}}{\frac{\beta\omega}{2}}
\end{equation*}
and \(f_\beta(t)\) is its inverse Fourier transform given by, see~\cite[SM~Sec.~5.1]{AAKS2021} and~\cite[appendix~C]{EO2019},
\begin{equation}
    \label{eq:definition-f-beta}
    f_\beta(t)
    =
    \frac{1}{2\pi} \int_{-\infty}^\infty \e^{\I \omega t} \, \fh_\beta(\omega) \diff \omega
    =
    \frac{2}{\beta\pi} \, \log\paren*{\frac{\e^{\pi \abs{t} / \beta} + 1}{\e^{\pi \abs{t} / \beta} - 1}}.
\end{equation}
The inverse Fourier transform satisfies \(\norm{f_\beta}_{L^1(\R)} = 1\) and, since \(\ln(x) \leq x-1\),
\begin{equation}
    \label{eq:bound-f-beta}
    f_\beta(t)
    \leq
    \frac{4}{\beta\pi} \, \frac{1}{\e^{\pi \abs{t} / \beta} - 1},
\end{equation}
which decays exponentially in~\(\abs{t}\).

We can now give the general solution of this equation.
Therefore, let
\begin{align}
    \nonumber
    \eta(s)
    &:=
    \mathcal{T} \exp\paren*{
        -\tfrac{\beta}{2} \, \int_0^s \Phi_\beta^{H(\sigma)}(V) \diff \sigma
    }
    \\&:=
    \label{eq:definition-eta-s-time-ordering}
    \sum_{n=0}^\infty \paren*{-\tfrac{\beta}{2}}^n \int_0^s \diff \sigma_1 \int_0^{\sigma_1} \diff \sigma_2 \, \dotsi \int_0^{\sigma_{n-1}} \diff \sigma_n
    \, \Phi_\beta^{H(\sigma_1)}(V) \, \Phi_\beta^{H(\sigma_2)}(V) \dotsm \Phi_\beta^{H(\sigma_n)}(V),
    \intertext{
        with~\(\mathcal{T}\) being the time-ordering operator.
        Then its adjoint is
    }
    \nonumber
    \eta(s)^*
    &:=
    \overline{\mathcal{T}} \exp\paren*{
        -\tfrac{\beta}{2} \, \int_0^s \Phi_\beta^{H(\sigma)}(V) \diff \sigma
    }
    \\&:=
    \nonumber
    \sum_{n=0}^\infty \paren*{-\tfrac{\beta}{2}}^n \int_0^s \diff \sigma_1 \int_0^{\sigma_1} \diff \sigma_2 \, \dotsi \int_0^{\sigma_{n-1}} \diff \sigma_n
    \, \Phi_\beta^{H(\sigma_n)}(V) \, \Phi_\beta^{H(\sigma_{n-1})}(V) \dotsm \Phi_\beta^{H(\sigma_1)}(V),
\end{align}
with \(\overline{\mathcal{T}}\) being the reverse-time-ordering operator.
By definition, these satisfy
\begin{equation*}
    \odv*{\eta(s)}{s}
    = -\tfrac{\beta}{2} \, \Phi_\beta^{H(s)}(V) \, \eta(s)
    \qquadtext{and}
    \odv*{\eta(s)^*}{s}
    = -\tfrac{\beta}{2} \, \eta(s)^* \, \Phi_\beta^{H(s)}(V).
\end{equation*}
Hence,
\begin{equation*}
    \ebeta{H(s)} = \eta(s) \, \ebeta{H(0)} \, \eta(s)^*.
\end{equation*}
Moreover, we have
\begin{equation*}
    \norm{\eta(s)} \leq \e^{\frac{\beta}{2} s \norm{V}}
    \qquadtext{and}
    \norm{\eta(s)-\unit} \leq \e^{\frac{\beta}{2} s \norm{V}} - 1.
\end{equation*}
Later, in Section~\ref{sec:proof-qbp-locality-properties}, we discuss how to approximate \(\Phi_\beta^{H(s)}(V)\) and~\(\eta(s)\) by strictly local operators if one has an underlying lattice structure.

\subsubsection{Differential equation for the Gibbs state \texorpdfstring{\(\rho_\beta(s)\)}{}}
\label{sec:proof-ode-gibbs-state}

In the previous section, we discussed an evolution for the exponential~\(\ebeta{H(s)}\).
Due to the missing normalization, this cannot be directly applied to the Gibbs state \(\rho_\beta(s) = \ebeta{H(s)}/\trace{\ebeta{H(s)}}\), which was overlooked by some previous works, e.g.~\cite{BK2019}.
Instead, we will discuss how to deal with the full Gibbs state \(\rho_\beta(s)\) in the following.

By Leibniz rule, equation~\eqref{eq:ode-ebeta} allows writing the differential equation for the normalized Gibbs state \(\rho_\beta(s)\) as
\begin{equation}
    \label{eq:proof-ode-gibbs-state-old}
    \odv*{\rho_\beta(s)}{s}
    =
    - \frac{\beta}{2} \anticommutator[\Big]{\rho_\beta(s),\Phi^{H(s)}_\beta(V)}
    + \beta \, \rho_\beta(s) \Trace[\Big]{\rho_\beta(s) \, \Phi^{H(s)}_\beta(V)}.
\end{equation}
It appears, with one missing \(\rho_\beta(s)\), already in~\cite[equation~(32)]{Kim2012} (see also~\cite[Lemma 17)]{Kim2013}) and, tested against local observables, in~\cite[appendix~C]{ORFW2023}.
Note that we have \(\Trace[\big]{\rho_\beta(s) \, \Phi^{H(s)}_\beta(V)} = \Trace[\big]{\rho_\beta(s) \, V} =: \braket{V}_{\rho_\beta(s)}\) due to cyclicity of the trace.
Hence, we can further simplify~\eqref{eq:proof-ode-gibbs-state-old} to obtain
\begin{equation*}
    \odv*{\rho_\beta(s)}{s}
    =
    - \frac{\beta}{2} \anticommutator[\Big]{\rho_\beta(s),\Phi_\beta^{H(s)}\paren[\big]{V-\braket{V}_{\rho_\beta(s)}}}.
\end{equation*}

This equation is not linear in \(\rho_\beta(s)\) anymore, but presuming we know \(\rho_\beta(s)\), it still gives a nice evolution
\begin{equation*}
    \rho_\beta(s) = \etat(s) \, \rho_\beta(0) \, \etat(s)^*
    ,
\end{equation*}
with
\begin{equation*}
    \etat(s)
    :=
    \mathcal{T} \exp\paren*{
        -\tfrac{\beta}{2} \, \int_0^s \Phi_\beta^{H(\sigma)}\paren[\big]{V-\braket{V}_{\rho_\beta(\sigma)}} \diff \sigma
    }
    .
\end{equation*}
This~\(\etat(s)\) is very similar to~\(\eta(s)\) from~\eqref{eq:definition-eta-s-time-ordering} and thus has similar properties.
In particular,
\begin{equation*}
    \norm{\etat(s)}
    \leq
    \e^{\frac{\beta}{2} s \sup_{\sigma \leq s}\norm{V-\braket{V}_{\rho_\beta(\sigma)}}}
    \leq
    \e^{\beta s \norm{V}}
    \qquadtext{and}
    \norm{\etat(s)-\unit} \leq \e^{\beta s \norm{V}} - 1.
\end{equation*}
As a simple consequence,
\begin{align*}
    \norm[\big]{\rho_\beta(0) - \rho_\beta(s)}_1
    &=
    \sup_{\norm{A}=1} \abs[\Big]{\trace[\Big]{\paren[\big]{\rho_\beta(0) - \rho_\beta(s)} \, A}}
    \\&=
    \sup_{\norm{A}=1} \norm[\big]{\rho_\beta(0)}_1 \, \norm[\big]{A-\etat^*(s) \, A \, \etat(s)}
    \\&\leq
    \sup_{\norm{A}=1} \paren[\Big]{
        \norm[\big]{A \, \paren[\big]{\unit - \etat(s)}}
        + \norm[\big]{\paren[\big]{\unit - \etat^*(s)} \, A \, \etat(s)}
    }
    \\&\leq
    \e^{2\beta\norm{V}}-1
    \\&\leq
    2\beta \, \norm{V}
    \, \e^{2\beta\norm{V}}
    .
\end{align*}

The formula for~\(\etat(s)\) given above is particularly useful to obtain the same locality results as for~\(\eta(s)\), see the next section.
However, since \(\braket{V}_{\rho_\beta(\sigma)}\) is just a number (with an implicit unit~\(\unit\) in the definition of~\(\etat(s)\)), we can also factor
\begin{equation*}
    \etat(s) = \exp\paren*{
        -\tfrac{\beta}{2} \int_0^s \braket{V}_{\rho_\beta(\sigma)} \diff \sigma
    }
    \, \eta(s).
\end{equation*}
Observing
\begin{equation*}
    \etat(s) \, \rho_\beta(0) \, \etat^*(s)
    =
    \rho_\beta(s)
    =
    \frac{\trace[\big]{\ebeta{H(0)}}}{\trace[\big]{\ebeta{H(s)}}} \, \eta(s) \, \rho_\beta(0) \, \eta^*(s)
    ,
\end{equation*}
we can thus conclude
\begin{equation*}
    \frac{\trace[\big]{\ebeta{H(0)}}}{\trace[\big]{\ebeta{H(s)}}}
    =
    \exp\paren*{
        -\beta \int_0^s \braket{V}_{\rho_\beta(\sigma)} \diff \sigma
    }
    .
\end{equation*}
This might help to understand the difference between the two differential equations~\eqref{eq:ode-ebeta} and~\eqref{eq:ode-gibbs-state} for~\(\ebeta{H(s)}\) and~\(\rho_\beta(s)\), respectively.
In particular, \(\eta(s) \, \rho_\beta(0) \, \eta^*(s)\) differs from \(\rho_\beta(s)\) by exactly this factor.
However, for explicit computations of \(\rho_\beta(s)\) the original approach of~\cite{Hastings2007} to compute \(\eta(s) \, \ebeta{H(0)} \, \eta^*(s)\) and normalize might be more practical.

\subsection{Locality properties}
\label{sec:proof-qbp-locality-properties}

In this section we discuss, how to obtain Proposition~\ref{prop:qbp-bounds} from Proposition~\ref{prop:qbp-ode}, which we discussed in the previous section.
Therefore, we now restrict to the Hilbert spaces~\(\Hi_\Lambda\) with underlying lattice structure as discussed in Section~\ref{sec:mathematical-setting}.

We now fix the interaction~\(\Psi\) of the Hamiltonian~\(H_\Lambda\) and the perturbation \(V\in \alg_X\) with~\(X\subset \Lambda\).
We will only write the proof of Proposition~\ref{prop:qbp-bounds-exponential}.
Part~\ref{prop:qbp-bounds-gibbs-state} then follows by choosing \(\tilde{V} := V - \braket{V}_{\rho_\beta(s)}\) and noting that we can write \(\braket{V}_{\rho_\beta(s)} = \braket{V}_{\rho_\beta(s)} \, \unit_X \in \alg_X\) such that also~\(\tilde{V} \in \alg_X\).

\subsubsection{Lieb-Robinson bound for the perturbed Hamiltonian\texorpdfstring{ \(H_\Lambda+s \, V\)}{}}

To prove locality of the generator \(\Phi^{H(s)}_\beta(V)\) and the exponential~\(\eta(s)\) in the next sections, we need Lieb-Robinson bounds for the Hamiltonian \(H(s) = H + V\).
Hence, we need to extend the Lieb-Robinson bounds for~\(H(0)\) to~\(s>0\).

\begin{lemma}[Lieb-Robinson bound for perturbed Hamiltonians]
    \label{lem:LR-bound-for-perturbed-Hamiltonians}
    Let \(\Lambda\subset \Z^\nu\) and \(H\in \alg_\Lambda\) self-adjoint, and assume that \(H\) satisfies a Lieb-Robinson bound with decay~\(\zeta_{\LR}^{H}\) as in Definition~\ref{def:Lieb-Robinson-bound}.
    Moreover, let \(V\in \alg_X\) self-adjoint, then \(H+V\) satisfies the Lieb-Robinson bound
    \begin{equation}
        \label{eq:lem-LR-bound-for-perturbed-Hamiltonians}
        \begin{aligned}
            \Alignindent
            \norm[\big]{\commutator[\big]{\e^{-\I t (H+V)} \, A \, \e^{\I t (H+V)}, B}}
            \\&\leq
            \norm{A} \, \norm{B}
            \, \paren[\bigg]{
                \zeta_\LR^{H}(Y,Z,\abs{t})
                + 2 \, \norm{V}
                \, \min_{W\in \{Z,Y\}}
                \int_0^{\abs{t}} \zeta_\LR^{H}(X,W,s) \diff s
            }
        \end{aligned}
    \end{equation}
    for all \(A\in \alg_Y\), \(B\in \alg_Z\) and~\(t\in \R\).
\end{lemma}

We will apply the above result in the cases \(A=V\) or \(A=\tilde{V}\) for which \(X=Y\) and the minimum is attained at~\(W=Z\).
Moreover, since Lieb-Robinson bounds are increasing in~\(s\), the integral in~\eqref{eq:lem-LR-bound-for-perturbed-Hamiltonians} can be bounded by \(\abs{t} \, \zeta_\LR^{H}(X,W,\abs{t})\) so that~\eqref{eq:lem-LR-bound-for-perturbed-Hamiltonians} is bounded by
\begin{equation*}
    \norm{A} \, \norm{B}
    \, \paren[\big]{
        1 + 2 \, \norm{V} \, \abs{t}
    }
    \, \zeta_\LR^{H}\paren[\big]{X,Y,\abs{t}}
\end{equation*}
in this case.

\begin{proof}
    We abbreviate \(\tau_t(A) := \e^{-\I t H} \, A \, \e^{\I t H}\) and \(\tilde\tau_t(A) := \e^{-\I t (H+V)} \, A \, \e^{\I t (H+V)}\).
    We will later prove
    \begin{equation}
        \label{eq:proof-LR-bound-perturbed-Hamiltonian-approximation-of-dynamics}
        \norm{\tilde\tau_t(A)-\tau_t(A)}
        \leq
        \norm{V} \, \norm{A} \, \int_0^{\abs{t}} \zeta_\LR^H(X,Y,s) \diff s,
    \end{equation}
    which allows to bound
    \begin{align*}
        \norm[\big]{\commutator[\big]{\tilde\tau_t(A), B}}
        &\leq
        \norm[\big]{\commutator[\big]{\tau_t(A), B}}
        + \norm[\big]{\commutator[\big]{\tilde\tau_t(A)-\tau_t(A), B}}
        \\&\leq
        \norm{A} \, \norm{B}
        \, \paren[\Big]{
            \zeta_\LR^{H}(Y,Z,\abs{t})
            + 2 \, \norm{V} \, \int_0^t \zeta_\LR^H(X,Y,\abs{s}) \diff s
        }.
    \end{align*}
    The minimum can be concluded using \(\norm[\big]{\commutator[\big]{\tilde\tau_{t}(A), B}} = \norm[\big]{\commutator[\big]{A, \tilde\tau_{-t}(B)}}\) and the above argument with the roles of~\(A\) and~\(B\) exchanged.

    To prove~\eqref{eq:proof-LR-bound-perturbed-Hamiltonian-approximation-of-dynamics} we follow very closely the proof of~\cite[Theorem~3.4\,(i)]{NSY2019}.
    By fundamental theorem of calculus,
    \begin{equation*}
        \tilde\tau_t(A)-\tau_t(A)
        =
        \int_0^t \odv*{\paren[\Big]{\tilde\tau_s \circ \tau_{t-s} (A)}}{s} \diff s
        =
        \I \, \int_0^t \tilde\tau_s\paren[\Big]{ \commutator[\big]{ V, \tau_{t-s} (A)}} \diff s,
    \end{equation*}
    which gives~\eqref{eq:proof-LR-bound-perturbed-Hamiltonian-approximation-of-dynamics} by using the Lieb-Robinson bound for \(\norm[\big]{\commutator[\big]{ V, \tau_s (A)}}\).
\end{proof}

In the special case of a short-range Hamiltonian~\(H\) satisfying the Lieb-Robinson bound given in Proposition~\ref{prop:short-range-LR-bound}, we can actually carry out the integration in~\eqref{eq:lem-LR-bound-for-perturbed-Hamiltonians} and obtain the following result.

\begin{corollary}
    \label{cor:LR-bound-for-perturbed-short-range-Hamiltonian}
    Let~\(\Psi\) be a short-range interaction with \(\norm{\Psi}\sr{b}<\infty\) for some~\(b>0\).
    Then, for all \(X\), \(Y\subset \Lambda\Subset\Z^\nu\), \(V\in \alg_X\) self-adjoint, \(A \in \alg_X\), \(B\in \alg_Y\) and \(t\in \R\)
    \begin{equation*}
        \norm[\big]{\commutator[\big]{\e^{-\I t (H + V)} \, A \, \e^{\I t (H + V)},B}}
        \leq
        2 \, \norm{A} \, \norm{B}
        \, \paren[\bigg]{1+\frac{2\,\norm{V}}{b\,v_b}}
        \, \e^{b v_{b} \abs{t}} \, \sum_{x\in X} \e^{-b\dist{x,Y}}
        ,
    \end{equation*}
    with~\(v_b\) as in Proposition~\ref{prop:short-range-LR-bound}.
\end{corollary}

Lemma~\ref{lem:LR-bound-for-perturbed-Hamiltonians} justifies to assume that~\(H(s)\) satisfies a Lieb-Robinson bound with decay~\(\zeta_\LR\) uniformly in~\(s\).
In particular, the perturbation~\(V\) only changes the constant in the Lieb-Robinson bound, but not the Lieb-Robinson velocity.
This Lieb-Robinson bound allows to approximate the Heisenberg time evolution \(\e^{-\I t H(s)} \, W \, \e^{\I t H(s)}\) of any \(W\in \alg_X\) with a local operator in~\(\alg_{X_r}\).

\subsubsection{Locality of the “generator”\texorpdfstring{ \(\Phi_\beta^{H(s)}(V)\)}{}}
\label{sec:proof-locality-of-the-generator}

Given the Lieb-Robinson bound, there exists a conditional expectation~\cite[Lemma~4.1]{NSY2019} \(\cexp_{X_r}\colon \alg_\Lambda \to \alg_{X_r}\), which for us is just the normalized partial trace since we are in finite dimensions, such that
\begin{equation}
    \label{eq:bound-for-conditional-expectation}
    \norm[\big]{
        (\unit-\cexp_{X_r})
        \paren[\big]{\e^{-\I t H(s)} \, W \, \e^{\I t H(s)}}
    }
    \leq
    \norm{W} \, \zeta_\LR(X, \Lambda \setminus X_r, t)
    \quadtext{for all \(W\in \alg_X\).}
\end{equation}
It allows decomposing \(\Phi^{H(s)}_\beta(V) = \Delta_r(s) + \overline\Delta_r(s)\) into two parts
\begin{align*}
    \Delta_r(s)
    &:=
    \int_{-\infty}^\infty \diff t \, f_\beta(t) \, \cexp_{X_r}\paren[\big]{\e^{-\I t H(s)} \, V \, \e^{\I t H(s)}}
    \quad\text{and}
    \\
    \overline\Delta_r(s)
    &:=
    \int_{-\infty}^\infty \diff t \, f_\beta(t) \, (\unit-\cexp_{X_r})\paren[\big]{\e^{-\I t H(s)} \, V \, \e^{\I t H(s)}}.
\end{align*}
By definition \(\Delta_r(s) \in \alg_{X_r}\) is strictly local and bounded by \(\norm{\Delta_r(s)} \leq \norm{V}\).
Using the Lieb-Robinson bound and the conditional expectation we can bound the remainder by
\begin{align*}
    \norm{\overline\Delta_r(s)}
    &\leq
    \norm{V} \, \int_{\abs{t} \leq T} f_\beta(t) \, \zeta_\LR(X,\Lambda\setminus X_r,t) \diff t
    +
    2 \, \norm{V} \int_{\abs{t} \geq T} f_\beta(t) \diff t
    \\&\leq
    \norm{V} \, \paren[\Big]{
        \norm{f_\beta}_{L^1(\R)} \, \norm{\zeta_\LR(X,\Lambda\setminus X_r,{\cdot})}_{L^\infty(\intervalcc{-T,T})}
        + \frac{32}{\pi^2} \, \e^{-\frac{\pi}{\beta}T}
    }.
\end{align*}
Here, we bounded the second integral assuming \(T \geq \frac{\beta}{\pi} \, \ln 2\), by using~\eqref{eq:bound-f-beta} and substituting \(u=\frac{\pi}{\beta} \, t - \ln 2\)
\begin{equation*}
    \int_{\abs{t} \geq T} f_\beta(t) \diff t
    \leq
    \frac{8}{\pi^2} \, \int_{\frac{\pi}{\beta}T-\ln 2}^\infty \frac{1}{2 \, \e^u - 1} \diff u
    \leq
    \frac{16}{\pi^2} \, \e^{-\frac{\pi}{\beta}T}.
\end{equation*}
Additionally, we have the trivial bound \(\norm{\overline\Delta_r(s)} \leq 2 \, \norm{V} \, \norm{f_\beta}_{L^1(\R)} = 2 \, \norm{V}\), such that one can now optimize the bound for a given~\(\zeta_\LR\).
More precisely, we obtain
\begin{equation*}
    \norm{\overline\Delta_r(s)}
    \leq
    \norm{V} \, \zeta_\QBP(X,r)
\end{equation*}
with
\begin{equation*}
    \zeta_\QBP(X,r)
    :=
    \min\List[\Big]{
        2,
        \inf_{T \geq \frac{\beta}{\pi} \, \ln 2}
        \norm{\zeta_\LR(X,\Lambda\setminus X_r,{\cdot})}_{L^\infty(\intervalcc{-T,T})}
        + 4 \, \e^{-\frac{\pi}{\beta}T}
    }
    ,
\end{equation*}
where we bounded \(32/\pi^2 < 4\).
Then, \(4 \, \e^{-\frac{\pi}{\beta}T} > 2\) for \(T < \frac{\beta}{\pi} \, \ln 2\), which together with the trivial bound allows us to take the infimum over all \(T \geq 0\) instead of \(T \geq \frac{\beta}{\pi} \, \ln 2\).

We bound \(\zeta_\QBP\) explicitly for short- and long-range interactions in Section~\ref{sec:proof-bounds-for-zeta-qbp}.

\subsubsection{Locality of the exponential\texorpdfstring{ \(\eta(s)\)}{}}
\label{sec:proof-locality-of-eta-and-etat}

The local approximation of~\(\Phi^{H(s)}_\beta(V)\) from the previous section also allows us to approximate~\(\eta(s)\) by a local version
\begin{equation}
    \label{eq:definition-eta-ell-s}
    \eta_\rrr(s)
    :=
    \mathcal{T} \exp\paren*{
        -\tfrac{\beta}{2} \, \int_0^s \Delta_\rrr(\sigma) \diff \sigma
    }.
\end{equation}
It is easy to show by induction that for all operators \(A_1,\dotsc,A_n\) and \(B_1,\dotsc,B_n\) it holds that
\begin{equation*}
    A_1 \, A_2 \dotsm A_n
    =
    B_1 \, B_2 \dotsm B_n
    + \sum_{j=1}^n A_1 \dotsm A_{j-1} \, (A_j-B_j) \, B_{j+1} \dotsm B_n.
\end{equation*}
Hence,
\begin{equation}
    \label{eq:auxQBPL}
    \norm{
        A_1 \, A_2 \dotsm A_n
        - B_1 \, B_2 \dotsm B_n
    }
    \leq
    n \, \paren[\bigg]{\sup_{\,j\in \List{1,n}} \max\List[\big]{\norm{A_j},\norm{B_j}}}^{n-1} \, \sup_{j\in\List{1,n}}\norm{A_j-B_j}.
\end{equation}
Using~\eqref{eq:auxQBPL} for \(A_j := \Phi^{H(\sigma_j)}_\beta(V)\) and \(B_j := \Delta_\rrr(\sigma_j)\) together with~\eqref{eq:definition-eta-s-time-ordering} we find
\begin{align*}
    \norm[\big]{\eta(s)-\eta_\rrr(s)}
    &\leq
    \sum_{n=0}^\infty \paren*{\tfrac{\beta}{2}}^n \, \frac{s^n}{n!} \, n \, \norm{V}^{n-1}
    \, \norm{V} \, \zeta_\QBP(X,\rrr)
    \\&\leq
    \tfrac{\beta}{2} \, s \, \norm{V} \, \e^{\tfrac{\beta}{2} s \norm{V}} \, \zeta_\QBP(X,\rrr).
\end{align*}
And by the definition~\eqref{eq:definition-eta-ell-s} we have the same bound \(\norm{\eta_\rrr(s)} \leq \e^{\frac{\beta}{2} s \norm{V}}\) as for~\(\eta(s)\).

\subsection{Specific decay \texorpdfstring{\(\zeta_\QBP\)}{zeta\_QBP} for short- and long-range interactions}
\label{sec:proof-bounds-for-zeta-qbp}

We now obtain the specific~\(\zeta_\QBP\) stated in Section~\ref{sec:quantum-belief-propagation} for short- and long-range interactions.
Moreover, we will explain how to optimize the bound on \(\zeta_\QBP(X,r)\) for balls \(X\), which we use for intervals in the one-dimensional setting.
In all cases, we start with a discussion of available Lieb-Robinson bounds.

\subsubsection{Short-range interactions}

For short-range interactions, Lieb-Robinson bounds have been proven in several different forms, for example in~\cite{LR1972,HK2006,NSY2019,Maier2022}.
A more complete discussion about the history is given in~\cite{NSY2019}.
For our specific interaction norm, we adopt the proof of~\cite[Theorem~7.3.1]{Maier2022} which is based on the proof of~\cite[Theorem~3.1]{NSY2019}.

\begin{proposition}[{Lieb-Robinson bound}]
    \label{prop:short-range-LR-bound}
    Let \(\Lambda\Subset\Z^\nu\) and~\(\Psi\) be a short-range interaction on~\(\Lambda\) with \(\norm{\Psi}\sr{b}<\infty\) for some~\(b>0\).
    Then, for all \(X\), \(Y\subset \Lambda\), operators \(A\in \alg_X\), \(B\in \alg_Y\), and~\(t\in \R\)
    \begin{equation*}
        \norm[\big]{\commutator[\big]{\e^{-\I t H_\Lambda} \, A \, \e^{\I t H_\Lambda}, B}}
        \leq
        2 \, \norm{A} \, \norm{B}
        \, \e^{b v_{b} \abs{t}} \, \sum_{x\in X} \e^{-b\dist{x,Y}},
    \end{equation*}
    where \(v_b := 2 \, \norm{\Psi}\sr{b} / b\) is the Lieb-Robinson velocity.
\end{proposition}

\begin{proof}
    The interaction norm used in~\cite{Maier2022} does not have the extra factor~\(\abs{Z}\).
    In the proof of the Lieb-Robinson bound, they need to bound a factor~\(\abs{Z}\), for which they use a fraction of the decay in~\(\diam{Z}\) and \(\sup_{Z \Subset \Lambda} \abs{Z} \, \e^{-c \diam{Z}} < \infty\), instead, we use the extra \(\abs{Z}\) in the interaction norm.

    More specifically, instead of~\cite[eq.~(7.3.5)]{Maier2022}, we start with the induction hypothesis
    \begin{equation*}
        a_k(X,Y)
        \leq
        \norm{\Psi}\sr{b}^k \, g_b(X,Y)
        ,
        \qquadtext{where}
        g_b(X,Y)
        :=
        \sum_{x\in X} \e^{-b\dist{x,Y}},
    \end{equation*}
    where \(a_k(X,Y)\) is defined in~\cite[p.~313]{Maier2022}.
    The proof for \(a_1(X,Y)\) is the same, while for the induction step, we find
    \begin{align*}
        a_{k+1}(X,Y)
        &\leq
        \norm{\Psi}\sr{b}^k
        \, \sum_{x\in X}
        \, \sumstack{Z\subset \Lambda\mathpunct{:}\\x\in Z} \frac{\norm{\Psi(Z)}}{\e^{-b\diam{Z}}}
        \, \sum_{z\in Z} \e^{-b\diam{Z}} \, \e^{-b\dist{z,Y}}
        \\&\leq
        \norm{\Psi}\sr{b}^k
        \, \sum_{x\in X}
        \, \e^{-b\dist{x,Y}}
        \, \sumstack{Z\subset \Lambda\mathpunct{:}\\x\in Z} \frac{\norm{\Psi(Z)}}{\e^{-b\diam{Z}}} \, \abs{Z}
        \\&\leq
        \norm{\Psi}\sr{b}^{k+1}
        \, g_b(X,Y)
        .
    \end{align*}
    With this small modification, the proof in~\cite{Maier2022} yields the claim.
\end{proof}

We are now ready to obtain \(\zeta_\QBP\) for short-range interactions.

\begin{proof}[Proof of Lemma~\ref{lem:zeta-QBP-short-range}]
    Note that \(
        \norm{\Psi+s\,V}\sr{b}
        \leq
        \max\List[\big]{\norm{\Psi}\sr{b}, \norm{\Psi+V}\sr{b}}
    \) by convexity.
    Hence, to prove Lemma~\ref{lem:zeta-QBP-short-range-along-the-path} we can choose the Lieb-Robinson velocity \(v_b = 2 \, \max\List[\big]{\norm{\Psi}\sr{b}, \norm{{\Psi+V}}\sr{b}} / b\).
    We begin with the proof for general sets and bound
    \begin{equation}
        \label{eq:proof-zeta-qbp-bound-sum-X-general-sets}
        \sum_{x\in X} \e^{-b\dist{x,Y}} \leq \abs{X} \, \e^{-b\dist{X,Y}}
        ,
    \end{equation}
    which yields \(
        \zeta_\LR(X,\Lambda\setminus X_r, t)
        \leq
        C_\LR \, \abs{X} \, \e^{b(v_b \abs{t}-r)}
    \) with \(C_\LR=2\).
    This~\(\zeta_\LR\) attains its supremum for \(t\in \intervalcc{-T,T}\) at~\(t=T\).
    Hence, the infimum in~\eqref{eq:definition-zeta-QBP-general} is attained at
    \begin{equation*}
        T
        =
        \frac{b \, r - c}{\frac{\pi}{\beta} + b \, v_b}
        \qquadtext{with}
        c
        :=
        \ln\paren[\bigg]{
            C_\LR \, \abs{X} \, b \, v_b \, \beta / (4 \, \pi)
        }
        .
    \end{equation*}
    To obtain a simpler result, we just choose \(T := b \, r/(\frac{\pi}{\beta}+b \, v_b)\)
    to get
    \begin{equation}
        \label{eq:zeta-QBP-short-range-for-T-larger-constant}
        \zeta_\QBP(X,r)
        \leq
        (C_\LR+4) \, \abs{X} \, \e^{-\frac{b}{1 + b v_b \beta / \pi\xmathstrut{.2}} \, r}
        ,
    \end{equation}
    where \(C_\LR + 4 = 6\).

    To obtain~\eqref{eq:remark-zeta-QBP-perturbed-short-range}, we start with the Lieb-Robinson bound for the perturbed Hamiltonian from Corollary~\ref{cor:LR-bound-for-perturbed-short-range-Hamiltonian}, which means to replace~\(C_\LR\) by \(2 \, \paren[\big]{1+2/(b\,v_b)\,\norm{V}}\).
    In~\eqref{eq:zeta-QBP-short-range-for-T-larger-constant} we then bound \(
        2 \, \paren[\big]{1+2/(b\,v_b)\,\norm{V}} + 4
        \leq
        6 \, \max\List{1,2/(b\,v_b)} \, \paren[\big]{1+\norm{V}}
    \).

    If \(X = B_{z}(R)\) is a ball, we can improve the bound in~\eqref{eq:proof-zeta-qbp-bound-sum-X-general-sets} by summing over shells \(S_k = \Set{x\in X \given \dist{z,x}=k}\) which all satisfy \(\abs{S_k} \leq \abs{\partial X}\) such that
    \begin{equation*}
        \sum_{x\in X}\e^{-b\dist{x,\Lambda\setminus X_r}}
        \leq
        \sum_{k=0}^R \abs{S_k} \, \e^{-b (r+R-k)}
        \leq
        \abs{\partial X} \paren[\bigg]{\e^{-b r} + \int_r^\infty \e^{-b q} \diff q}
        \leq
        \tfrac{1+b}{b} \, \abs{\partial X} \, \e^{-b r}
        .
    \end{equation*}
    Hence, we just replace \(\abs{X}\) with \(\tfrac{1+b}{b} \, \abs{\partial X}\) in \(\zeta_\QBP\) to obtain the improved result for balls.
    With the same proof it is also clear that this replacement works for every interval \(X\subset \Z\), while strictly speaking, only intervals with an odd number of sites are balls.
\end{proof}

\subsubsection{Long-range interactions}

Also for long-range interactions various Lieb-Robinson bounds are available.
Some of them only apply for a restricted set of times \(t\) or only for two-body interactions.
In particular, there was a focus on proving so-called linear light cones in the last years.
The bounds have the property that for each \(\epsi>0\) there exists \(v>0\) such that \(\zeta_\LR\paren[\big]{X,Y,\dist{X,Y}/v} < \epsi\) in the limit \(\dist{X,Y} \to \infty\), see e.g.~\cite{KS2020,TGB2021} and the discussion of light-cones in~\cite{EMN2020}.
For our applications, it turns out that we need a good decay in~\(\dist{X,Y}\) and a bound which holds for all \(t\), while the growth in \(t\) is not too important.
This is due to the fact, that we only need to use the Lieb-Robinson bound until~\(T\) in~\eqref{eq:definition-zeta-QBP-general} and can then use the exponential decay in~\(T\) in the second term to our advantage.

One bound, which provides a good decay and allows for general interaction, is proven in~\cite{EMN2020}, which is based on~\cite{MKN2016}.
We give a slightly improved version here.
From now on, we will fix \(F_\alpha(r) := (1+r)^\alpha\).

\begin{proposition} \label{prop:LR-bound-long-range}
    Let \(\alpha>\nu\) and \(\sigma\in \intervaloo[\big]{(\nu+1)/(\alpha+1),1}\).
    Then there exist constants \(C\) and \(c>0\) such that for all \(\Lambda\Subset \Z^\nu\) and interactions \(\Psi\) on \(\Lambda\) with \(\norm{\Psi}_{F_\alpha}<\infty\) the following Lieb-Robinson bound hold:
    For all \(X\), \(Y\subset \Lambda\), \(A\in \alg_X\), \(B\in \alg_Y\)
    \begin{equation}
        \label{eq:prop-LR-bound-long-range}
        \norm[\big]{\commutator[\big]{\e^{-\I t H_\Lambda} \, A \, \e^{\I t H_\Lambda}, B}}
        \leq
        C \, \norm{A} \, \norm{B}
        \sum_{x\in X}
        \, \paren[\Big]{
            \e^{v \abs{t} - r^{1-\sigma}}
            + v\,\abs{t} \, \paren[\big]{1 + (v\,\abs{t})^{\nu/(1-\sigma)}} \, F_{\sigma \alpha}(r)
        }
    \end{equation}
    where \(r=\dist{x,Y}\) and \(v := c \, \norm{\Psi}_{F_\alpha}\).
\end{proposition}

\begin{proof}
    First note, that the assumptions in~\cite[eq.~(3),(5)]{EMN2020} are satisfied if \(\alpha>\nu\) and \(\norm{\Psi}_{F_\alpha}<\infty\).
    Then~\cite[Theorem~1]{EMN2020} proves the statement for a single point \(X=\List{x}\).
    From~\cite[Lemma~4]{TCE2020} one obtains the statement for \(\abs{X}>1\) with the sum over~\(x\in X\).

    Strictly speaking, the bound in~\cite{EMN2020} is only provided for \(\ell^2\)-distance on \(\Z^\nu\), which agrees with our setting only for~\(\nu=1\).
    However, the proof can be modified to obtain the same result also with \(\ell^1\)-distance and directly with a linear scaling in~\(\abs{X}\).
    Details will be given in a future work, where
    we also consider fermions, for which~\cite[Lemma~4]{TCE2020} is not applicable.
\end{proof}

Before we estimate \(\zeta_\QBP\), let us further upper bound~\eqref{eq:prop-LR-bound-long-range} to obtain the simpler
\begin{equation}
    \label{eq:LR-bound-long-range-simplified}
    \zeta_\LR(X,Y,t)
    \leq
    C
    \, \sum_{x\in X} \, \paren[\Big]{
        \e^{v t - \dist{x,Y}^{1-\sigma}}
        + \paren[\big]{1 + (v\,t)^{1+\nu/(1-\sigma)}} \, F_{\sigma \alpha}\paren[\big]{\dist{x,Y}}
    }
\end{equation}
after including a factor~\(2\) in~\(C\).
Then, for the dynamics of \(H_\Lambda+V\) with \(V\in \alg_X\) we obtain
\begin{equation}
    \label{eq:LR-bound-long-range-simplified-perturbed}
    \zeta_\LR(X,Y,t)
    \leq
    C
    \, \paren[\big]{1+\norm{V}}
    \, \sum_{x\in X} \, \paren[\Big]{
        \e^{v t - \dist{x,Y}^{1-\sigma}}
        + \paren[\big]{1 + (v\,t)^{2+\nu/(1-\sigma)}} \, F_{\sigma \alpha}\paren[\big]{\dist{x,Y}}
    }
\end{equation}
using Lemma~\ref{lem:LR-bound-for-perturbed-Hamiltonians}, again after adjusting~\(C\).

\begin{lemma}[{\(\zeta_\QBP\) for long-range interactions}]
    \label{lem:zeta-QBP-long-range}
    Let \(\alpha>\nu\), \(\alpha_\QBP < \alpha\), \(\beta_0>0\) and~\(C_\interaction>0\).
    Then there exists a constant \(C_\QBP>0\) such that the following holds:

    Let \(\Psi\) be an interaction on \(\Lambda\Subset\Z^\nu\), \(X\subset \Lambda\), \(V\in \alg_X\) self-adjoint and \(\beta\in \intervaloo{0,\beta_0}\).
    Then for all Hamiltonians \(H(s) = H + s \, V\) the following holds:
    \begin{theoremenumerate}
        \item \label{lem:zeta-QBP-long-range-along-the-path}
            If \(\norm{\Psi}_{F_\alpha}\), \(\norm{\Psi+V}_{F_\alpha}<C_\interaction\), then
            \begin{equation*}
                \zeta_\QBP(X,r)
                \leq
                C_\QBP \abs{X} \, F_{\alpha_\QBP}(r)
                .
            \end{equation*}
        \item \label{lem:zeta-QBP-long-range-in-unperturbed}
            If only \(\norm{\Psi}_{F_\alpha} < C_\interaction\), then
            \begin{equation*}
                \zeta_\QBP(X,r)
                \leq
                C_\QBP
                \, \paren[\big]{1+\norm{V}}
                \, \abs{X} \, F_{\alpha_\QBP}(r)
                .
            \end{equation*}
    \end{theoremenumerate}
    Moreover, if \(X = \Set{x\in \Lambda \given \dist{x,z} \leq R}\) is a ball (for some \(z\in \Lambda\) and \(R>0\)) and \(\alpha_\QBP < \alpha - 1\) we can replace \(\abs{X}\) in both bounds with~\(\abs{\partial X}\).
    This replacement in particular also works for all intervals \(X\subset \Z\), where \(\abs{\partial X}=2\).
\end{lemma}

\begin{proof}
    For the proof, we will assume the \(\zeta_\LR(X,Y,t)\) from~\eqref{eq:LR-bound-long-range-simplified-perturbed} but without the \(\paren[\big]{1+\norm{V}}\).
    It clearly upper bounds~\eqref{eq:LR-bound-long-range-simplified} and we can later add the factor \(\paren[\big]{1+\norm{V}}\) as in the proof of Lemma~\ref{lem:zeta-QBP-short-range}.
    This way, we only need to bound~\eqref{eq:definition-zeta-QBP-general} once.

    As in the proof of Lemma~\ref{lem:zeta-QBP-short-range}, we start with the result for general \(\abs{X}\), where we replace the sum with \(\abs{X}\) and \(r\) with~\(\dist{X,Y}\).
    Then,
    \begin{align*}
        \Alignindent
        \norm{\zeta_\LR(X,\Lambda\setminus X_r,{\cdot})}_{L^\infty(\intervalcc{-T,T})}
        \\&\leq
        C \, \abs{X}
        \, \paren[\Big]{
            \e^{v T - r^{1-\sigma}}
            + \paren[\big]{1 + (v\,T)^{2+\nu/(1-\sigma)}} \, F_{\sigma \alpha}\paren{r}
        }
        \\&\leq
        C \, \abs{X}
        \, \paren[\Big]{
            \e^{r^{p(1-\sigma)} - r^{1-\sigma}}
            + F_{\sigma \alpha-2 p (1-\sigma) - \nu p}\paren{r}
        }
    \end{align*}
    after choosing \(p\in \intervaloo{0,1}\), \(T=r^{p(1-\sigma)}/v\) and adjusting the constant by a factor~\(2\) after using \(1+r^x \leq 2\,(1+r)^x\).
    We now choose \(\sigma\) and \(p\) such that \(\alpha_\QBP = \sigma \alpha-2 p (1-\sigma) - \nu p\), which is possible, because we assumed \(\alpha_\QBP < \alpha\) and can choose \(\sigma\) arbitrarily close to~\(1\) and \(p\) arbitrarily close to~\(0\), which will only change the constant~\(C\) according to Proposition~\ref{prop:LR-bound-long-range}.
    For every choice of \(\sigma\) and \(p\), we can also upper bound
    \(\e^{r^{p(1-\sigma)} - r^{1-\sigma}} \leq C \, F_{\alpha_\QBP}(r)\) for some~\(C>0\).
    Thus, we obtain
    \begin{equation*}
        \norm{\zeta_\LR(X,\Lambda\setminus X_r,{\cdot})}_{L^\infty(\intervalcc{-T,T})}
        \leq
        C \, \abs{X}
        F_{\alpha_\QBP(r)}
        ,
    \end{equation*}
    where~\(T\) is chosen as above and the constant \(C\) only depends on \(\alpha\), \(\alpha_\QBP\) and \(\nu\) as in the Lieb-Robinson bound.
    For the second summand in~\eqref{eq:definition-zeta-QBP-general} we bound \(
        4 \, \e^{-\frac{\pi}{\beta}T}
        =
        4 \, \e^{-\frac{\pi}{\beta v}r^{p(1-\sigma)}}
        \leq
        C \, F_{\alpha_\QBP}(r)
    \) for some~\(C>0\), where this \(C\) additionally depends on~\(\beta_0\) and~\(C_\interaction\).
    Combining the two, proves Lemma~\ref{lem:zeta-QBP-long-range-along-the-path}.
    And as said above, adding the additional factor \(\paren[\big]{1+\norm{V}}\) also proves Lemma~\ref{lem:zeta-QBP-long-range-in-unperturbed}.

    For balls \(X = B_{z}(R)\) we use the same strategy as in the proof of Lemma~\ref{lem:zeta-QBP-short-range}.
    Here, we bound
    \begin{equation*}
        \sum_{x\in X} F_{\alpha}\paren[\big]{\dist{x,\Lambda\setminus X_r}}
        \leq
        \sum_{k=0}^R \abs{S_k} \, F_{\alpha}\paren{r+R-k}
        \leq
        \abs{\partial X} \, \sum_{k=r}^\infty F_\alpha(k)
        \leq
        \tfrac{\alpha}{\alpha-1} \, \abs{\partial X} \, F_{\alpha-1}(r)
        ,
    \end{equation*}
    using
    \begin{equation}
        \label{eq:proof-bound-sum-F_alpha}
        \sum_{k=r}^\infty F_\alpha(k)
        \leq
        F_\alpha(r) + \int_r^\infty F_\alpha(x) \diff x
        =
        \tfrac{\alpha}{\alpha-1} \, F_{\alpha-1}(r)
        \quadtext{for}
        \alpha>1
    \end{equation}
    for the second summand in~\eqref{eq:LR-bound-long-range-simplified} and~\eqref{eq:LR-bound-long-range-simplified-perturbed}.
    For the first summand one obtains a polynomial correction in~\(r\), see e.g.~\cite[Lemma~3]{EMN2020} which will be absorbed by the exponential decay in the end.
    Hence, after adjusting the constants, the result also holds with \(\abs{\partial X}\) instead of \(\abs{X}\) for \(\alpha_\QBP < \alpha - 1\).
    As in Lemma~\ref{lem:zeta-QBP-short-range}, the replacement also works for all intervals~\(X\subset \Z\).
\end{proof}

\statement{Acknowledgments}
We thank Álvaro Alhambra, Andreas Blum, Marius Lemm, Alberto Ruiz de~Alarcón, Michael \LTskip{Sigal}, Daniel \LTskip{Ueltschi} and Simone \LTskip{Warzel} for fruitful discussions and Curt von \LTskip{Keyserlingk} for pointing us to~\cite{RGK2023}.

This work was funded by the \foreignlanguage{ngerman}{Deutsche Forschungsgemeinschaft} (DFG, German Research Foundation) –
470903074; 
465199066
.
The work of M.\,M. has been supported by a fellowship of the Alexander von Humboldt Foundation during his stay at the University of Tübingen, where this work initiated.
M.\,M. gratefully acknowledges the support of \foreignlanguage{italian}{PNRR Italia Domani} and Next Generation EU through the \LTskip{ICSC} National Research Centre for High Performance Computing, Big Data and Quantum Computing and the support of the MUR grant \foreignlanguage{italian}{Dipartimento di Eccellenza}~2023–2027.

\appendix

\section{Details on Section~\ref{sec:applications-of-the-general-results}}
\label{app:proof-application-of-gerenal-results}

In this section we collect short proofs for the statements in Section~\ref{sec:applications-of-the-general-results}.

\subsection{One-dimensional short-range systems}
\label{app:results-one-dimensional-short-range}

We begin with the proof for short-range interactions.

\begin{proof}[Proof of Corollary~\ref{cor:LPPL-and-LI-from-KK2024-short-range}]
    LPPL follows from Theorem~\ref{thm:DC0toLPPL} by choosing \(r=\dist{X,Y}/2\), using Theorem~\ref{thm:DC-KK2024-short-range} to bound the covariance and using the specific form of \(\zeta_\QBP\) from Lemma~\ref{lem:zeta-QBP-short-range-in-unperturbed} for intervals.
    For more details, one can follow the proof of Theorem~\ref{thm:LPPL-1D}.

    \medskip

    For local indistinguishability we first observe, that~\eqref{eq:cor-LPPL-from-KK2024-short-range} also holds without the factor \(\paren[\big]{1+\norm{V}}\), if \(\norm{\Psi+V}\sr{b} < C_\interaction\), see Lemma~\ref{lem:zeta-QBP-short-range-along-the-path}.
    This is the form of LPPL that we need in the proof of Lemma~\ref{lem:local-indistinguishability-remove-single-site-short-range}.
    Notice moreover, that we indeed provide uniform LPPL for all \(\Lambda \Subset \Z\), with respect to \(f_\LPPL = C_\LPPL\), \(g_\LPPL(v) = \e^{3\beta v}\), \(\zeta_\LPPL(r) = \e^{-c_\LPPL \sqrt{r}}\) and \(n=0\) (in a restricted sense, where \(X\) and \(Y\) are intervals).
    Hence, we can apply Theorem~\ref{thm:uniform-lppl-implies-local-indistinguishability} by removing the sites from~\(\Lambda\setminus\Lambda'\).
    Therefore, we calculate, see Lemma~\ref{lem:local-indistinguishability-remove-single-site-short-range},
    \(
        g(v) = \e^{3\beta v}
    \)
    and bound
    \begin{equation*}
        \zeta(r)
        \leq
        \paren[\big]{
            \e^{-c_\LPPL \sqrt{(1-\alpha) \, r}}
            + \e^{-b \sqrt{\alpha \, r}}
        }
        \leq
        2 \, \e^{-c_\LI \, \sqrt{r}}
        ,
    \end{equation*}
    where we bounded \(\e^{-b x} \leq \e^{-b \sqrt{x}}\) for \(x \geq 1\) and chose \(R = \alpha \, r\), \(\alpha = c_\LPPL^2 / (b^2 + c_\LPPL^2)\) and \(c_\LI = b \, \sqrt{\alpha}\).
    It remains to bound the sum in~\eqref{eq:thm-bound-local-indistinguishability-from-uniform-lppl}, which in this case is
    \begin{equation*}
        \sumstack[r]{x\in \Lambda\setminus\Lambda'} \zeta\paren[\big]{\dist{Y,x_i}}
        \leq
        2 \, \sum_{k=r}^\infty \zeta(k)
        \leq
        4 \, \e^{-c_\LI \sqrt{r}}
        +
        4 \int_{r}^{\mathrlap{\infty}} \e^{-c_\LI\sqrt{q}} \diff q
        \leq
        \frac{8}{c_\LI^2} \, \paren[\big]{c_\LI\,\sqrt{r} + 1 + c_\LI^2} \, \e^{-c_\LI \, \sqrt{r}}
        ,
    \end{equation*}
    where we abbreviated \(r=\dist{Y,\Lambda\setminus\Lambda'}\).
    Absorbing all the constants in \(C_\LI\) proves the statement.
\end{proof}

\begin{remark}
    With a refined proof as in Section~\ref{sec:1D-spin-chains}, one could also obtain LPPL for observables \(B = B_1 \, B_2\) with \(B_1\in \alg_{Y_1}\), \(B_2\in \alg_{Y_2}\) and \(Y_1\), \(Y_2\subset \Lambda\) intervals such that \(Y_1 < X < Y_2\).
\end{remark}

\subsection{One-dimensional long-range systems}
\label{app:results-one-dimensional-long-range}

We proceed with the proof for long-range interactions.

\begin{proof}[Proof of Corollary~\ref{cor:LPPL-and-LI-from-KK2024-long-range}]

    The interaction norm used in~\cite{KK2024}, which we give in~\eqref{eq:def-interaction-norm-long-range-KK2024} and which is used in the formulation of Theorem~\ref{thm:DC-KK2024-long-range} does not agree with the interaction norm~\(\norm{\cdot}_{F_\alpha}\), which we use in the rest of the paper.
    Clearly
    \begin{equation*}
        \tnorm{\Psi}_{F_\alpha}
        =
        \sup_{x,y\in \Lambda} \sumstack[r]{Z\Subset\Lambda:\\x,y\in Z}
        \frac{\norm{\Psi(Z)}}{F_\alpha\paren[\big]{\dist{x,y}}}
        \leq
        \sup_{x\in \Lambda} \sumstack[r]{Z\Subset\Lambda:\\x\in Z}
        \frac{\norm{\Psi(Z)}}{F_\alpha\paren[\big]{\diam{Z}}}
        =
        \norm{\Psi}_{F_\alpha}
        ,
    \end{equation*}
    so their result also applies for our norm.
    But it will be advantageous in terms of possible \(\alpha\) to bound the \(\norm{\Psi}_{F_{\alpha'}}\) norm with the \(\tnorm{\Psi}_{F_\alpha}\) for some \(\alpha'\) and \(\alpha\) and formulate the assumptions using~\(\tnorm{\Psi}_{F_\alpha}\).
    Therefore, observe the following: For each \(Z\Subset\Z\) one can find \(x\), \(y\in \Z\) such that \(\diam{Z} = \dist{x,y} \leq \dist{x,z} + \dist{y,z}\) for every \(z\in Z\) by triangle inequality.
    Thus, for every \(z\in Z\) one can find \(w\in Z\) (which is \(x\) or \(y\)) such that \(\diam{Z} \leq 2 \, \dist{z,w}\).
    This allows to obtain the following bound for \(k\)-body interactions
    \begin{align*}
        \norm{\Psi}_{F_{\alpha'}}
        &=
        \sup_{z\in \Z} \sumstack{Z\Subset\Z \mathpunct{:}\\z\in Z}
        \frac{\abs{Z} \, \norm{\Psi(Z)}}{F_{\alpha'}\paren[\big]{\diam{Z}}}
        \\&\leq
        2^{\alpha'} \, k \, \sup_{z\in \Z} \sum_{w\in \Z} F_{\alpha-\alpha'}\paren[\big]{\dist{z,w}} \, \sumstack{Z\Subset\Z \mathpunct{:}\\z,w\in Z}
        \frac{\norm{\Psi(Z)}}{F_\alpha\paren[\big]{\dist{z,w}}}
        \\&\leq
        C_{\alpha,\alpha'} \, \tnorm{\Psi}_{F_\alpha}
    \end{align*}
    with
    \begin{equation*}
        C_{\alpha,\alpha'}
        =
        2^{\alpha'} \, k \, \sup_{z\in \Z} \sum_{w\in \Z} F_{\alpha-\alpha'}\paren[\big]{\dist{z,w}}
        <
        \infty
        \qquad\text{if \(\alpha>\alpha'+1\).}
    \end{equation*}

    With the estimates on the interaction norms, for every \(\alpha > 2\) and \(\alpha_\QBP < \alpha' - 1 < \alpha - 2\) we have
    \begin{equation*}
        \zeta_\QBP(X,r)
        \leq
        C_\QBP
        \, \paren[\big]{1+\norm{V}}
        \, \abs{X} \, F_{\alpha_\QBP}(r)
    \end{equation*}
    from Lemma~\ref{lem:zeta-QBP-long-range-in-unperturbed} for intervals \(X\subset \Lambda\).
    We now choose \(\alpha_\QBP = \alpha_\DC = \alpha_\LPPL\) and apply Theorem~\ref{thm:DC0toLPPL} with \(r=\dist{X,Y}/2\).
    Using \(x<\e^x\) to absorb the \(\beta\,\norm{V}\) in the exponential, we obtain the statement on LPPL.

    \medskip

    For local indistinguishability we again observe, that~\eqref{eq:cor-LPPL-from-KK2024-long-range} also holds without the factor \(\paren[\big]{1+\norm{V}}\), if \(\norm{\Psi+V}_{F_\alpha} < C_\interaction\), see Lemma~\ref{lem:zeta-QBP-long-range-along-the-path}.
    As in the proof of Corollary~\ref{cor:LPPL-and-LI-from-KK2024-short-range}, we can apply Theorem~\ref{thm:uniform-lppl-implies-local-indistinguishability}, since we provide uniform LPPL for all \(\Lambda \Subset \Z\), with respect to \(f_\LPPL = C_\LPPL\), \(g_\LPPL(v) = \e^{3\beta v}\), \(\zeta_\LPPL(r) = F_{\alpha_\LPPL}(r)\) and \(n=0\) (in a restricted sense, where \(X\) and \(Y\) are intervals).
    After choosing \(R=\dist{X,Y}/2\), we obtain, see Lemma~\ref{lem:local-indistinguishability-remove-single-site-short-range},
    \(
        g(v)=\e^{3\beta v}
    \)
    and bound
    \(
        \zeta(r)
        \leq
        2^{\alpha_\LPPL} \, F_{\alpha_\LPPL}(r)
    \).
    For Theorem~\ref{thm:uniform-lppl-implies-local-indistinguishability} it is left to bound
    \begin{equation*}
        \sum_{x\in \Lambda\setminus\Lambda'} F_{\alpha_\LPPL}\paren[\big]{\dist{Y,x}}
        \leq
        2 \, \sumstack[lr]{k=\dist{Y,\Lambda\setminus\Lambda'}}^\infty F_{\alpha_\LPPL}(k)
        \leq
        \tfrac{\alpha_\LPPL}{\alpha_\LPPL-1} \, F_{\alpha_\LPPL-1}\paren[\big]{\dist{Y,\Lambda\setminus\Lambda'}}
        ,
    \end{equation*}
    where we used~\eqref{eq:proof-bound-sum-F_alpha}.
    Putting everything together and absorbing the constants in~\(C_\LI\) proves the claim.
\end{proof}

\subsection{\texorpdfstring{\(\nu\)}{nu}-dimensional short-range systems at high temperature}
\label{app:results-high-temperature}

\begin{proof}[Proof of Corollary~\ref{cor:results-from-conjecture}]
    LPPL follows from Theorem~\ref{thm:DC0toLPPL} by choosing \(r=\dist{X,Y}\), using Conjecture~\ref{conj:DC} to bound the covariance and using \(\zeta_\QBP\) from Lemma~\ref{lem:zeta-QBP-short-range-in-unperturbed}.
    See also Remark~\ref{remark:DC0toLPPL-short-range}.

    \medskip

    To obtain local indistinguishability, we first use Theorem~\ref{thm:DCStoLPPL} together with \(\zeta_\QBP\) from Lemma~\ref{lem:zeta-QBP-short-range-along-the-path}, to obtain a better LPPL for the perturbations considered in the proof of Lemma~\ref{lem:local-indistinguishability-remove-single-site-short-range}, similar to what is given in Remark~\ref{remark:DCstoLPPL-short-range}.
    This way, we obtain a linear scaling \(\beta \, \norm{V}\) in the LPPL bound with a constant that can be chosen uniformly for~\(\beta<\beta^*\).
    To keep this scaling also for local indistinguishability, we bound
    \(\e^{2\beta v}-1 \leq \beta \, \tfrac{1}{\beta^*} \, (\e^{2\beta^* v} - 1)\) to obtain \(g\paren[\big]{\norm{\Psi}\sr{b}} \leq C \, \beta \, \norm{\Psi}\sr{b}\) for some~\(C\), which only depends on \(\beta^*\) and~\(C_\interaction\), in Theorem~\ref{thm:uniform-lppl-implies-local-indistinguishability}.
    Since \(\zeta_\LPPL\) and \(F\) decay exponentially, \(\tilde{\zeta}(r)\) in Theorem~\ref{thm:uniform-lppl-implies-local-indistinguishability} converges and decays exponentially.
    Thus, the result follows.
\end{proof}

\printbibliography[heading=bibintoc]

\end{document}